\documentclass[12pt]{iopart}

\usepackage{graphicx}
\usepackage{url}
\begin{document}

\title{A comparison of 3D particle, fluid and hybrid simulations for negative streamers}

\author{Chao Li$^{1,2}$, Jannis Teunissen$^1$, Margreet Nool$^1$, Willem Hundsdorfer$^{1,3}$, Ute Ebert$^{1,2}$}
\address{$^1$Centrum Wiskunde \& Informatica (CWI), P.O.Box 94079, 1090 GB Amsterdam, The Netherlands,}
\address{$^2$Dept.\ Physics, Eindhoven Univ.\ Techn., The Netherlands,}
\address{$^3$Department of Science, Radboud University Nijmegen, The Netherlands.}
\ead{ebert@cwi.nl}

\begin{abstract}
Modeling individual free electrons can be important in the simulation of discharge streamers.
Stochastic fluctuations in the electron density accelerate the branching of streamers.
And in negative streamers, energetic electrons can even `run away' and contribute to processes such as terrestrial gamma-ray and electron flashes.
To track energies and locations of single electrons in relevant regions, we have developed a 3D hybrid model that couples a particle model for single electrons in the region of high fields and low electron densities with a fluid model in the rest of the domain.
Here we validate our 3D hybrid model on a 3D (super-)particle model for negative streamers without photo-ionization in overvolted gaps.
We show that the extended fluid model approximates the particle and the hybrid model well until stochastic fluctuations become important,
while the classical fluid model underestimates velocities and ionization densities.
We compare density fluctuations and the onset of branching between the models, and we compare the front velocities with an analytical approximation.
\end{abstract}

%\pacs{52.80.-s, 52.80.Mg, 52.65.Kj, 52.65.Pp}

\submitto{\PSST--- revised manuscript, 20 July 2012}

\maketitle

\section{Introduction}

Streamers are growing ionized fingers that appear when ionizable matter is suddenly exposed to high voltages.
Streamers pave the path for lightning leaders and precede sparks, and they occur without the subsequent stages in the form of enormous sprite discharges high above thunderclouds.
Streamers are also used in diverse industrial applications.
As reviewed, e.g., in~\cite{Ebert2006/PSST,Ebert2010/JGR,Ebert2011/Nonlinearity}, the evolution of a single streamer consists of phenomena on several length scales:
the ionizing and exciting collisions of fast electrons with molecules,
the emergence of an ionization front with an electric screening layer,
and the emergence of a streamer finger surrounded by such a screening layer and ionization front.
The dynamical instability of a thin screening layer can make a streamer branch \cite{Ebert2006/PSST, Ebert2011/Nonlinearity, PRL2002, Caro2006, Kao, Luque2010/JCPprep, Luque2011}.
In negative streamers with high field enhancement energetic electrons can run away from the front and emit hard electromagnetic radiation;
taking the further interaction of these run-away electrons with the atmosphere into account, this is a possible explanation~\cite{Moss2006,Li2007,Cha2008,Li2009,Cha2010,Celestin2011} of terrestrial gamma-ray flashes~\cite{Fishman1994}, electron beams~\cite{Dwyer2008} or even electron positron beams~\cite{Briggs2011} emitted from active thunderstorms.

Streamer propagation is mostly investigated with a density or fluid approximation for the electrons and ions, which continues to be very challenging due to the widely separated scales; for recent articles we refer to~\cite{babaeva2006, panchesny2008, kacem2012, Duarte2012, Kolobov2012, Papadakis2008} and for a recent review to~\cite{Luque2010/JCPprep}.
However, there are stages of evolution where the statistics of single electrons matters, either due to their nonthermal energy distribution with long tails at very high energies~\cite{Cha2008,Li2011:JCP}, or due to their stochastic presence in non-ionized regions.
Examples include electron run-away from streamers, ionization avalanches created by single electrons that have now been observed experimentally in very clean gases~\cite{Nijdam2010/JPD,Wormeester2010/JPD,Wormeester2011/JJAP}, or density fluctuations that can accelerate streamer branching, as was shown in recent simulations~\cite{Luque2011}; this last study investigated positive streamers
in air with photo-ionization in a background field below the ionization threshold.

To track the energy and density fluctuations accurately, a Monte Carlo particle model for streamer simulations tracks single free electrons as they move and randomly collide with neutrals; neutral molecules are not simulated but provide a background that electrons stochastically collide with, see for example~\cite{Cha2008, Li2011:JCP, Rose2011}.
Here we look at very short timescales on which ions can be assumed to be immobile.
The model contains the energy and location of each electron as well as the spatial distributions of ions and of the different types of excited states.
Therefore it also can accurately simulate rare events like electron run-away or avalanche formation from single electrons.
But computer memory strongly constrains the number of electrons that can be tracked.
Streamers usually form when the total number of free electrons reaches $10^7-10^9$ in air at standard temperature and pressure~\cite{Mee1940,Mon2006:1,Li2008:2}, and during streamer growth the electron number continues to
increase.
This makes computations with real electrons very expensive or even impossible, and typically super-particles representing many real particles are used to accelerate computations.
However, super-particles can introduce unphysical fluctuations and numerical heating, as shown in \cite{Li2008:2} and below.
They also corrupt the statistics of rare events.
This statistical problem motivated us to develop our ``spatially hybrid model''.

In a streamer discharge, most electrons reside in high densities in the low field region in the streamer interior.
This region is typically in the ``hydrodynamic'' regime, that can be well described by a fluid model.
Relatively few electrons are in the region of strong field enhancement at the streamer tip or outside the streamer, and only those electrons should be tracked individually with a single particle model (as opposed to super-particles).
Therefore we have developed a code that is hybrid in space~\cite{Li2007, Li2011:JCP, Li2010:1}, applying a fluid approximation in the streamer interior and a single particle model at the streamer tip and in the essentially non-ionized region around it.
We call it the ``spatially hybrid model'' to distinguish it from other hybrid models, see for example \cite{Kushner2009}.
The model follows single electrons and their fluctuations in the dynamically relevant region.

In the present paper, we test the consistency and correct implementation of the particle and the hybrid model on propagating negative streamers in air (while neglecting photo-ionization), and we illustrate the influence of the (super-)particle fluctuations on the destabilization of the streamer ionization front.
For comparison, we also present simulations of the same system with a classical and with our extended fluid model, and we compare the computing times.
In the conclusion we also discuss the effect of photo-ionization that is excluded in the present work for technical reasons.
Therefore the present results are actually more appropriate for discharges in gases without photo-ionization, like pure nitrogen.

\section{Description of particle, fluid and hybrid model}

\subsection{Particle model}
The Monte Carlo particle model is of PIC-MCC type.
It describes the motion and collisions of free electrons in a streamer discharge in air without photo-ionization.
``Particle in cell'' (PIC) means that the electric charge of electrons and ions is mapped to an electric charge density on a numerical grid; this charge density changes the electric potential (from which the electric field is calculated) according to the Poisson equation.
``Monte Carlo collision'' (MCC) means that collisions of the free electrons with the neutral background molecules occur randomly, so neutral molecules do not need to be simulated.
Ions are treated as immobile.
The Monte-Carlo procedure and the differential cross-sections for the particle model are described in detail in section 2.1 of~\cite{Li2011:JCP}.

The particles represent single electrons during the initial stages of the simulation, as indicated in the text.
However, when the particle number becomes too large for the computer memory, super-particles are introduced that represent several real particles.
While particles carry their generic physical distributions and fluctuations of density and energy, the fluctuations of super-particles are unphysically increased and can generate artifacts when fluctuation effects or rare events become important.
To avoid this problem, we have developed the spatially hybrid model.
The different models are compared in sections \ref{sec:methodresults} and \ref{sec:discussion}.

\subsection{Classical fluid model, bulk and flux coefficients}
The classical fluid model for streamers has a long tradition in streamer modeling, much longer than the more microscopic particle model.
Originally it is a phenomenological model based on the essential physical mechanisms and conservation laws.
It can be traced back at least to the 1930s.
The classical fluid model approximates the electron dynamics by a reaction-drift-diffusion equation for the electron density, and the reaction and transport coefficients are assumed to depend on the local electric field, in the so-called ``local field approximation''.
The model is completed with the reaction equation for the ions and with the Poisson equation for the electric potential.

In order to approximate the particle dynamics well, the coefficients in the fluid model should be derived from the particle model.
This can be done either through averaging over the Boltzmann equation for the electron distribution in configuration space, or through evaluating swarm simulations in a Monte Carlo particle model; in a swarm simulation the evolution of an approximately Gaussian electron distribution in a constant electric field is traced by a particle model.
For the present paper the transport coefficients and the reaction rates for the fluid model are derived by fitting the fluid coefficients to the swarm dynamics in a Monte Carlo particle model, as evaluated in~\cite{Li2011:JCP}.
(Please note that \cite{Li2011:JCP} contains some corrections to the reaction and transport rates described in~\cite{Li2007,Li2010:1}.) The coefficients in~\cite{Li2011:JCP} were derived up to a field of 250~kV/cm in air at standard temperature and pressure, and for stronger electric fields fit formulas
from the same article are used.

Furthermore, in a reactive plasma one needs to distinguish between bulk and flux coefficients~\cite{Robson2005}.
While bulk coefficients characterize the dynamics of a swarm as a whole including its reactions, flux coefficients characterize the dynamics of individual electrons within a swarm.
Robson {\it et al.}~\cite{Robson2005} express the general opinion that bulk data should not be used in low-temperature plasma simulations, see also figure~2 in~\cite{Robson2005}.

However, essential physics is missing in the classical or minimal model as we found in~\cite{Li2010:1}, and as will be discussed much more extensively and systematically in a forth coming article~\cite{Sasa}.
For example, in strong electric fields, the ionization rate cannot be computed accurately from just the local electric field and the local electron density.
Therefore in~\cite{Li2010:1} we extended the generation term in the fluid model with a gradient expansion term while in~\cite{Sasa} a higher order model is derived in a systematic manner by averaging over more moments of the Boltzmann equation.
An important conclusion is that accurate results cannot be expected from the classical fluid model, independently of whether flux or bulk coefficients are used, as the functional form of the equations is not sufficiently general.

For comparing the classical fluid model with the other models, we have chosen to use bulk coefficients.
From studies of planar (1D) fronts, we know that bulk coefficients in the classical fluid model will lead to approximately correct front velocities.
This is the case because the leading edge of an ionization front, that pulls the front along, propagates under swarm-like conditions; for details, we refer to~\cite{Li2007} or to a full mathematical analysis of pulled front dynamics to~\cite{Ebert2000}.
A particle swarm should be parameterized with bulk coefficients in the classical model (recall that the bulk coefficients are constructed to give agreement for particle swarms).
Therefore, the streamer ionization front should also be modeled with bulk coefficients in the classical model.
The front velocity is then close to that of a particle model in the same electric field, although the ionization density behind the front is too low~\cite{Li2007}.
If flux coefficients would be used, the ionization density behind the front would be a bit higher.
(Both sets of coefficients produce the same amount of ionization per unit time in a given electric field.
The flux mobility is typically lower, therefore a larger amount of electrons per unit length is produced, and the electron density is higher.)
But the front velocity would be significantly too low with flux coefficients, and therefore we use bulk coefficients.
We stress again that accurate results can not be expected from the classical fluid model with either set of coefficients.

\subsection{Extended fluid model}

The ionization term in the classical fluid model for streamers is calculated in local field and local density approximation, but the comparison with particle models shows that the ionization densities in the streamer interior are too low behind a planar front in a fixed electric field~\cite{Li2007,Li2010:1, Nai1997}.
This is because the mean electron energy varies even within a swarm in a constant electric field: at the front edge of the swarm, the electrons have higher energies and are more likely to ionize the neutrals, while the electrons at the back end of the swarm are slower on average and less likely to ionize.
By including the first term of a gradient expansion in the electron density in the impact ionization rate, both particle swarms and planar ionization fronts are approximated well~\cite{Li2010:1}.
The extra term was derived in a model with flux coefficients, therefore we need to use flux coefficients for consistency with the model derivation.

\subsection{Hybrid model}

The hybrid model connects the particle model with the extended fluid model through a moving model interface with a buffer zone, as described extensively in~\cite{Li2011:JCP} and briefly recalled in the introduction.
When the flux of electrons across the interface between particle and fluid model is calculated, the same definition of coefficients should be used on both sides in order to be physically consistent; otherwise the physical inconsistency becomes visible in the form of a discontinuity of the electron density at the model interface~\cite{Li2010:1}.
For further details and the numerical implementation, we refer to~\cite{Li2007, Li2011:JCP, Li2010:1}.
The important feature of the spatially hybrid model is that it follows the particle dynamics only in the dynamically relevant region, and that it therefore can continue to track the single electron fluctuations much longer than the (super-)particle model.

\section{Simulation methods and results}
\label{sec:methodresults}

\subsection{The simulated system}

We simulate the evolution of a negative streamer in air without photo-ionization, at standard temperature and pressure.
The streamer propagates in a background field of -100~kV/cm, or 372~Td; this field is well above the break-down value.
The simulation volume is 1.17~mm long in the $z$-direction parallel to the electric field and extends up to $\pm 0.29$~mm outwards from the axis in the $x$ and $y$ direction where homogeneous Neumann boundary conditions are applied to the electric potential.

The initial distribution of electrons and ions is generated in the following manner: first 500 electrons and ions are placed at a distance of 0.115~mm from the cathode on the z-axis and followed by the particle model for 60~ps.
At this time, there are about 2500 electrons and ions with spatial distributions close to a Gaussian; the electrons are at this time all in the interval of 0.120 to 0.136~mm from the cathode.
These electron and ion distributions are then used as an initial condition for the simulations in all four models; for the fluid models, the swarm is mapped to densities on the numerical grid.

\subsection{Numerical implementation}

The models were already described in the previous section, and references to more detailed discussions were given there as well.
Electric field and electron and ion densities are calculated on a uniform grid of $256\times 256 \times 512$ points with $\Delta x $ = $\Delta y $ = $\Delta z $ = 2.3 $\mu$m, using the numerical schemes described in section 2.2 of~\cite{Li2011:JCP}.
The time step is $\Delta t= 0.3$~ps.
The Poisson equation for the electric field is solved in all models at each time step with the same fast elliptic solver {\sc fishpack}~\cite{fishpack90}.

In both particle and hybrid simulation, the particle model with single electrons is used in the early stages, until the number of electrons reaches $2\times10^7$; this occurs at about 0.46 ns.
At this time, the particle model switches to super-particles, while the hybrid model switches to the full hybrid scheme: the fluid model is applied inside the streamer channel where the electric field is less than 0.95 $E_b$ or where the electron density is larger than 0.7 $n_{e,max}$, and the particle model in the remaining part of space; here $E_b$ stands for the background field and $n_{e,max}$ is the maximal electron density in the complete simulation volume.
When the hybrid model is activated, electrons inside the streamer channel are removed from the particle list, and the particle model is only applied at the streamer head where it continues to trace all single electrons.
In the particle model, on the other hand, super-particles are introduced at the time 0.46~ns by removing at random
every second electron, and by doubling the weight of the remaining particles.

\subsection{Overview of simulation results for the four models}

\begin{figure}
\centering
% First row
\includegraphics[width=.12\textwidth,viewport=25 110 150 400, clip]{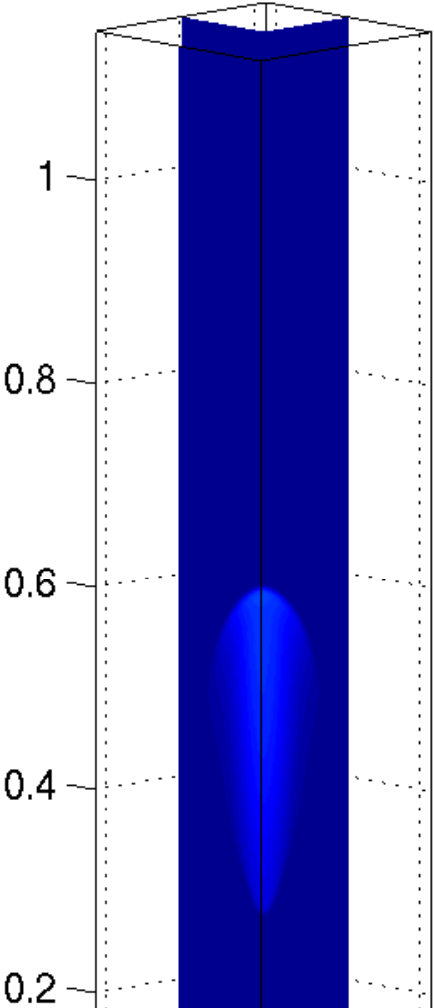}
\includegraphics[width=.12\textwidth,viewport=25 110 150 400, clip]{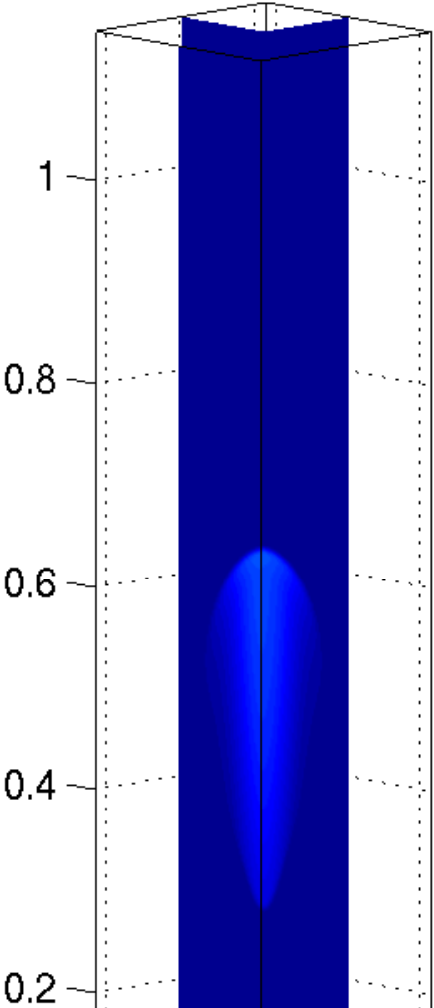}
\includegraphics[width=.12\textwidth,viewport=25 110 150 400, clip]{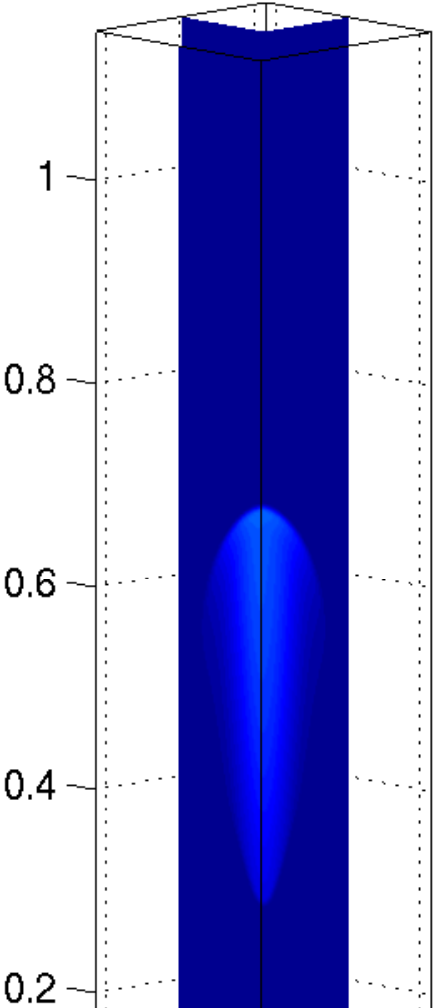}
\includegraphics[width=.12\textwidth,viewport=25 110 150 400, clip]{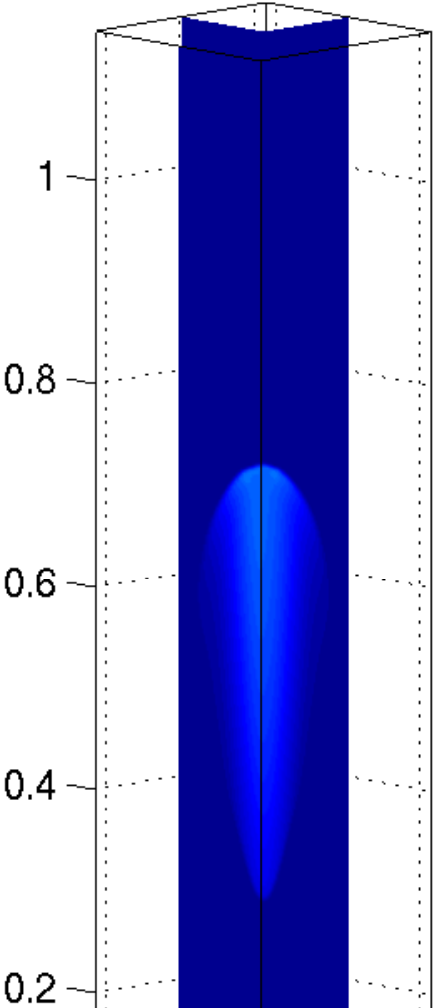}
\includegraphics[width=.12\textwidth,viewport=25 110 150 400, clip]{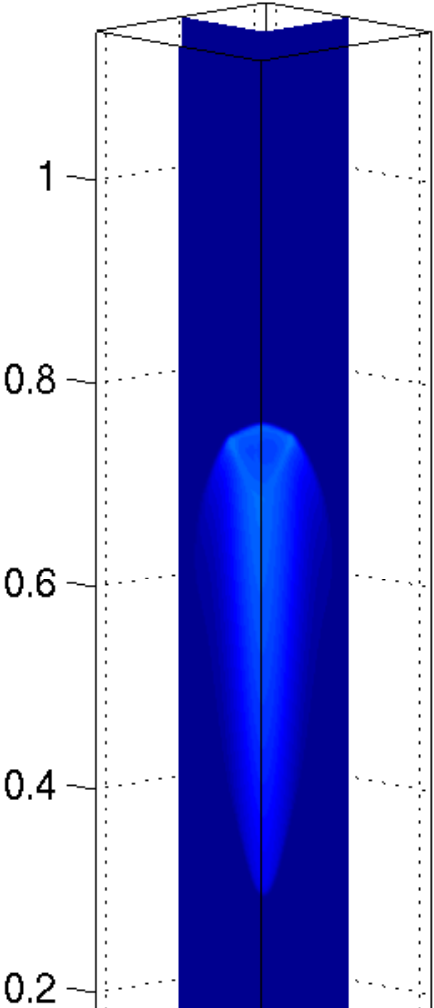}
\includegraphics[width=.12\textwidth,viewport=25 110 150 400, clip]{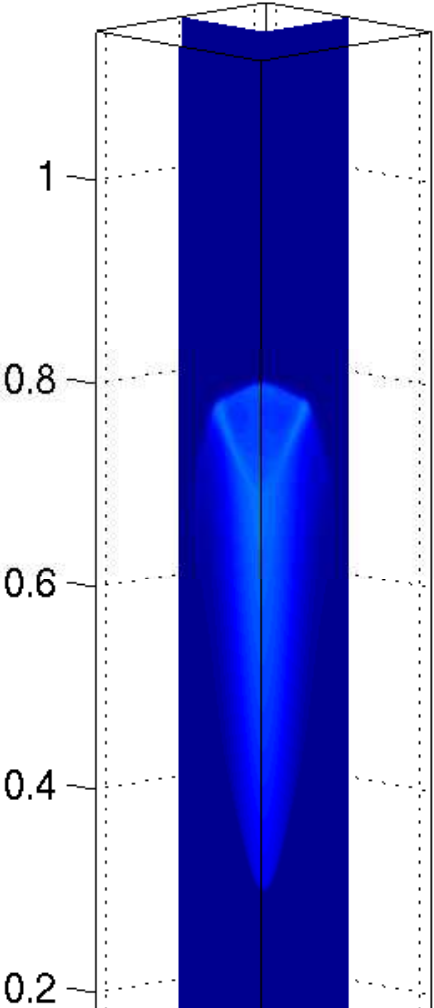}
\includegraphics[width=.12\textwidth,viewport=25 110 150 400, clip]{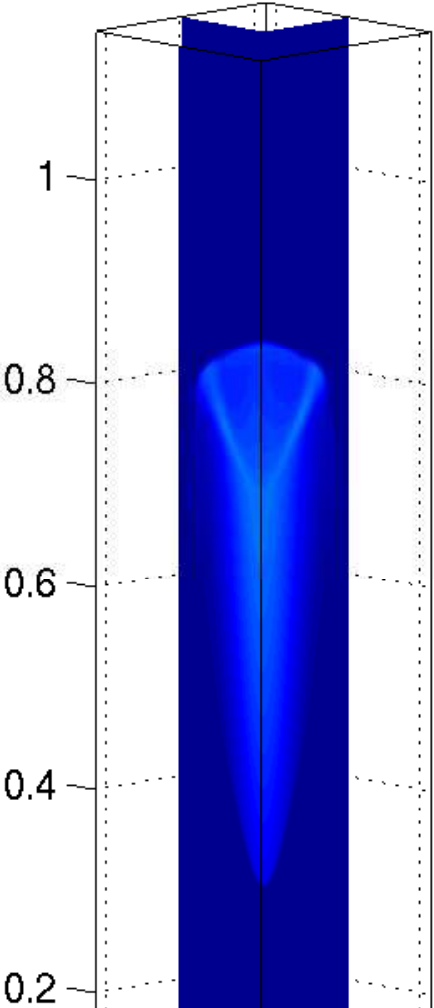}
\includegraphics[width=.04\textwidth,viewport=225 110 275 400, clip]{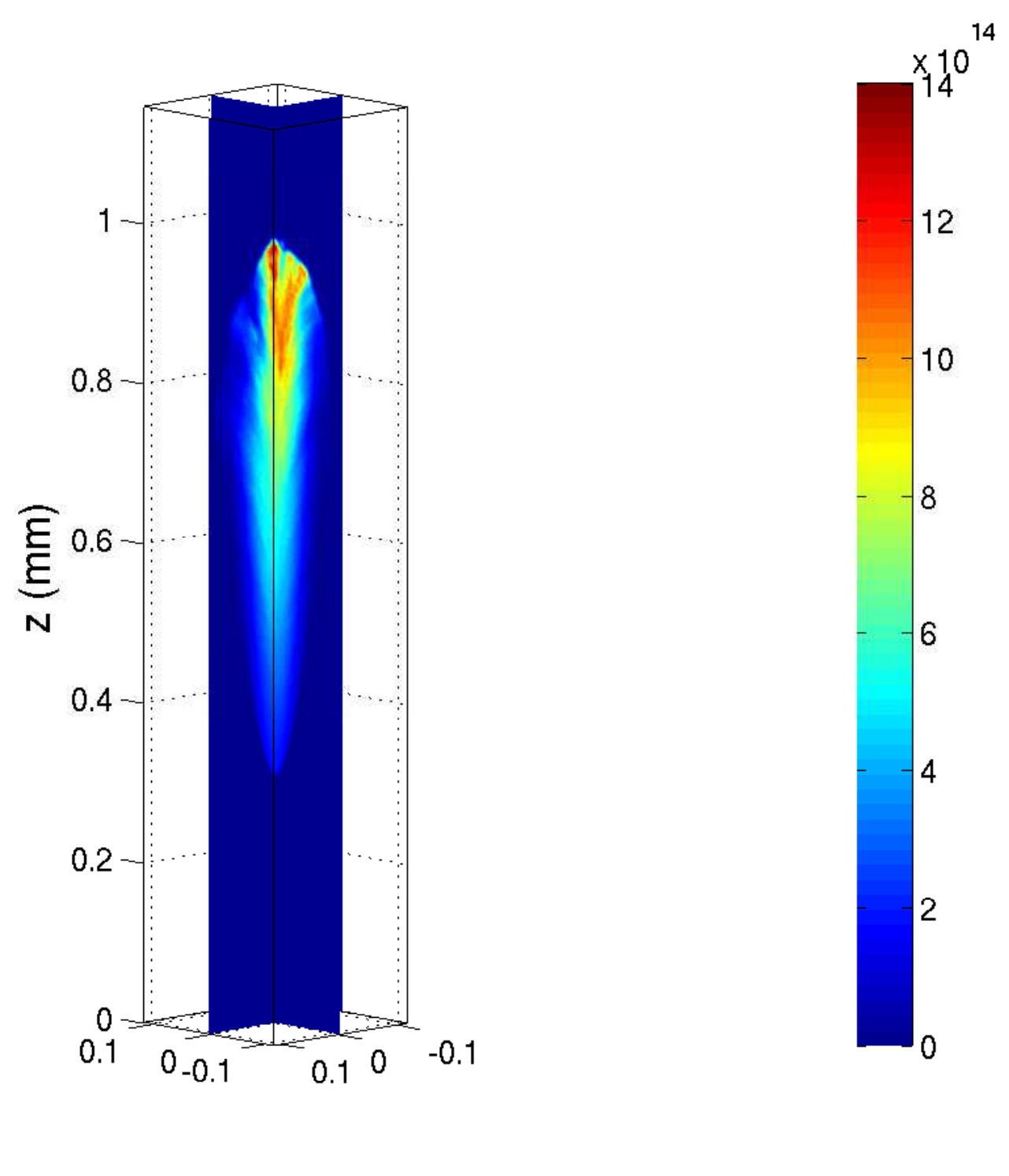} %White space
\\
% Second row
\includegraphics[width=.12\textwidth,viewport=25 110 150 400, clip]{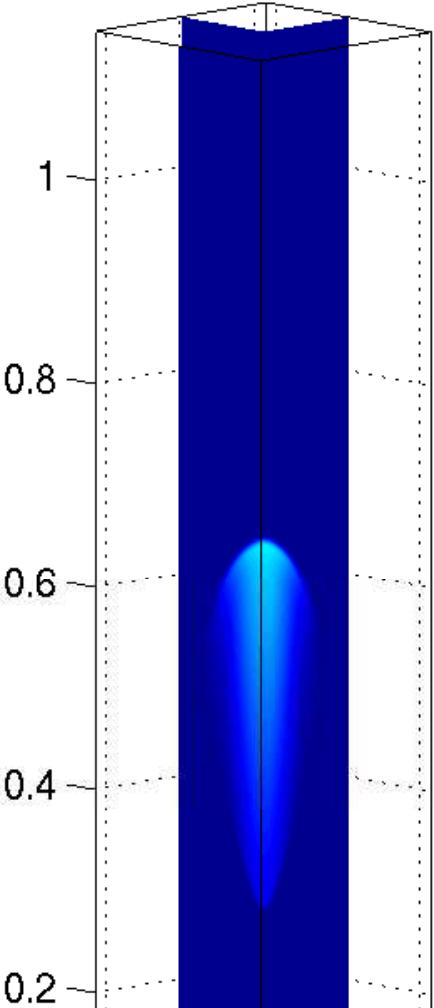}
\includegraphics[width=.12\textwidth,viewport=25 110 150 400, clip]{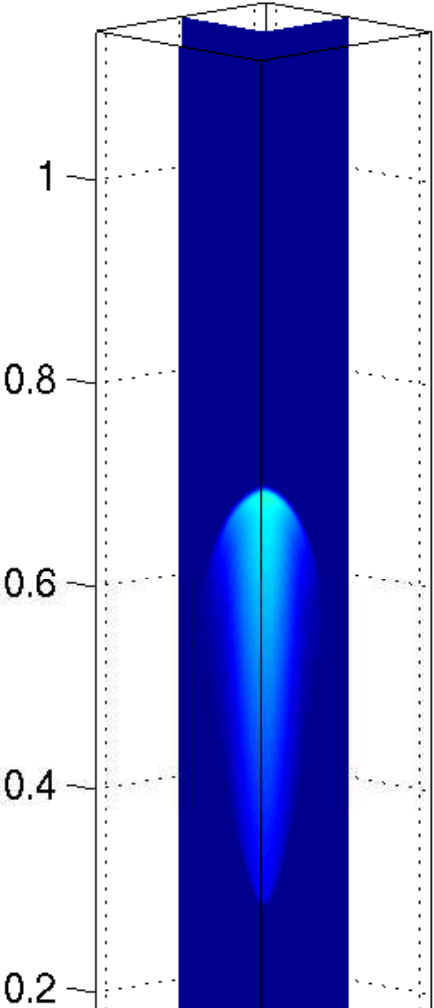}
\includegraphics[width=.12\textwidth,viewport=25 110 150 400, clip]{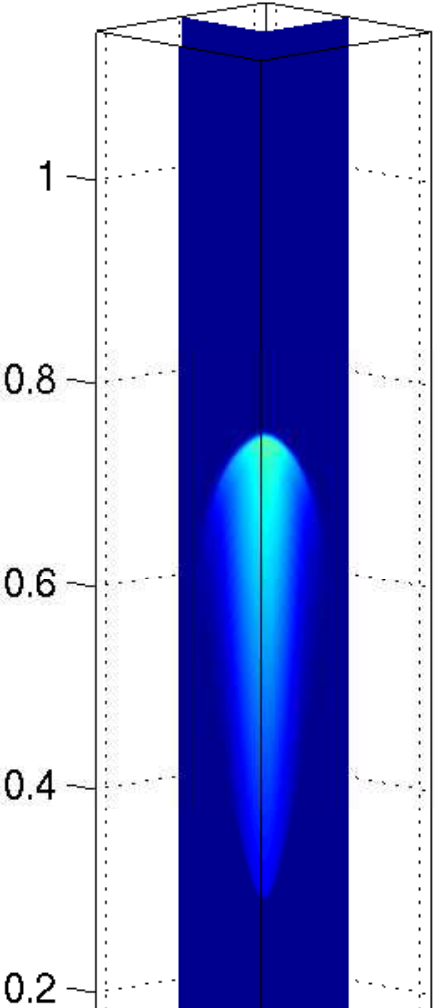}
\includegraphics[width=.12\textwidth,viewport=25 110 150 400, clip]{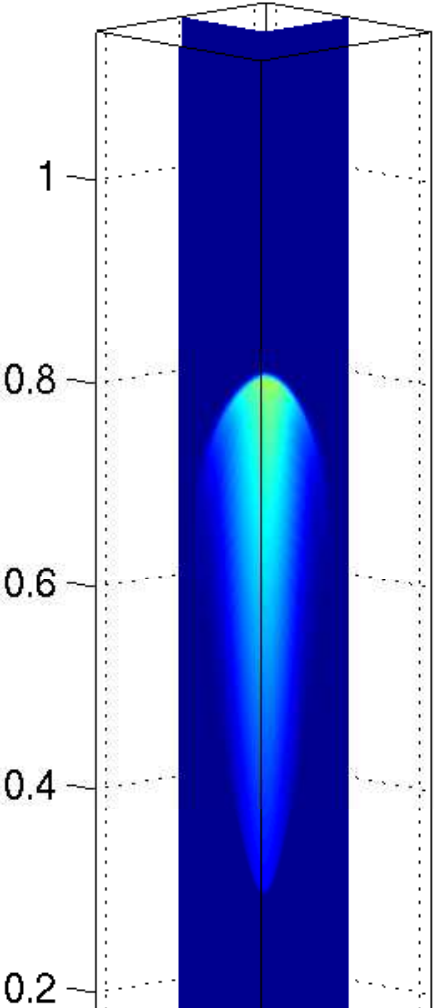}
\includegraphics[width=.12\textwidth,viewport=25 110 150 400, clip]{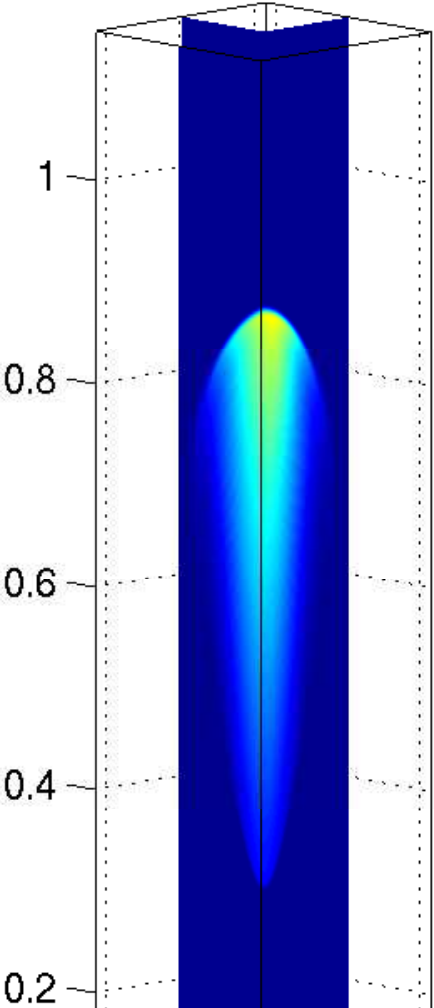}
\includegraphics[width=.12\textwidth,viewport=25 110 150 400, clip]{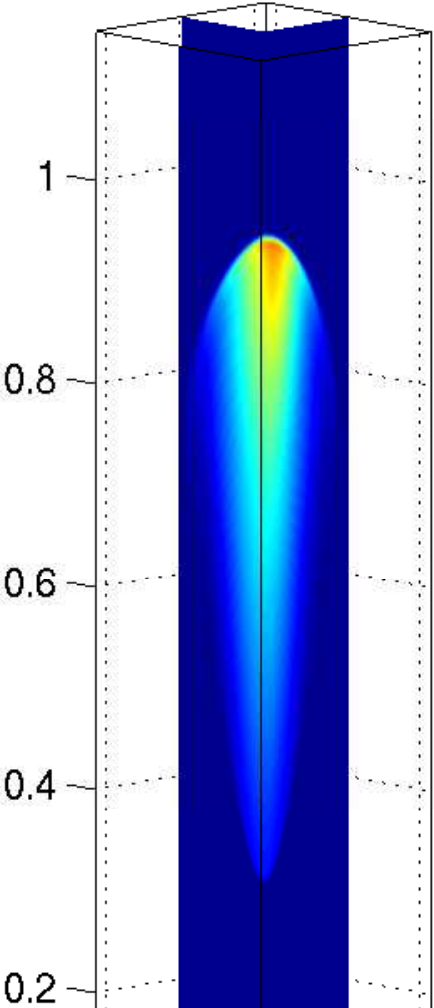}
\includegraphics[width=.12\textwidth,viewport=25 110 150 400, clip]{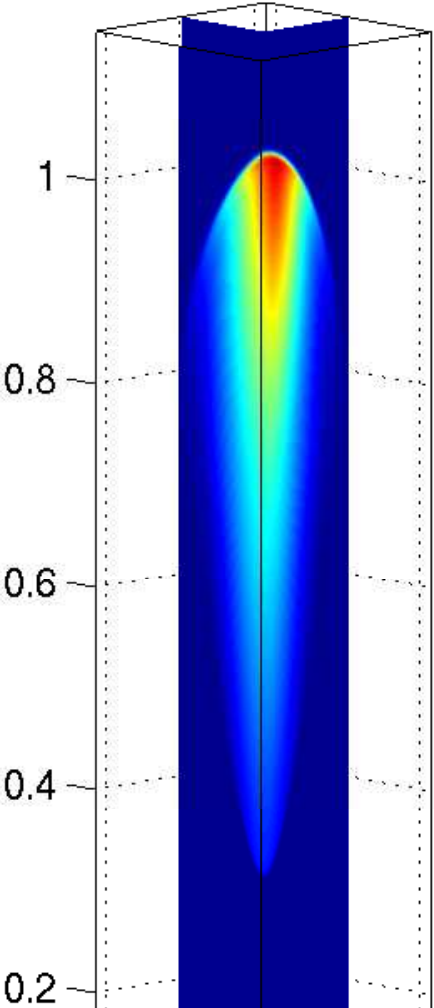}
\includegraphics[width=.04\textwidth,viewport=225 110 275 400, clip]{figures_pdf/fig1_colorbar.pdf} %White space
\\
% Third row
\includegraphics[width=.12\textwidth,viewport=25 110 150 400, clip]{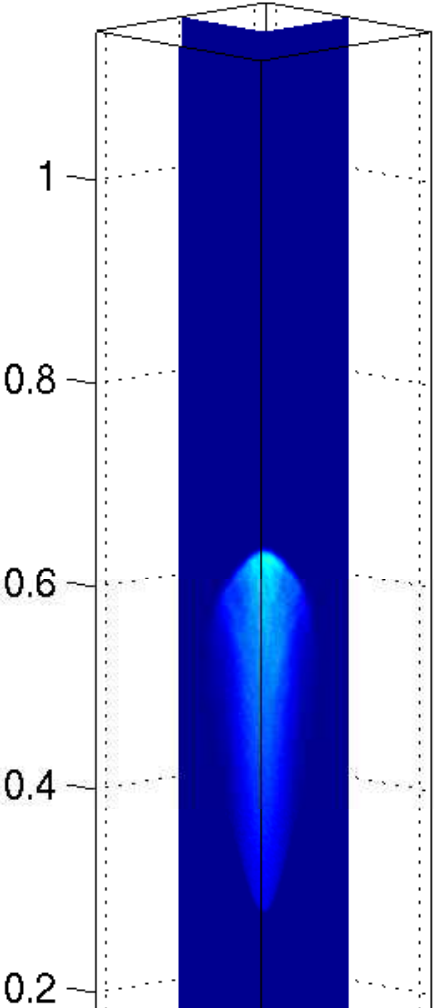}
\includegraphics[width=.12\textwidth,viewport=25 110 150 400, clip]{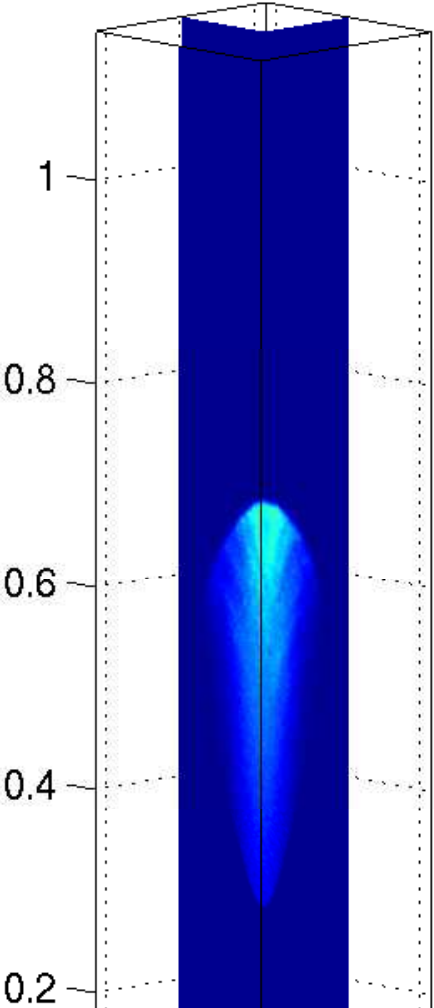}
\includegraphics[width=.12\textwidth,viewport=25 110 150 400, clip]{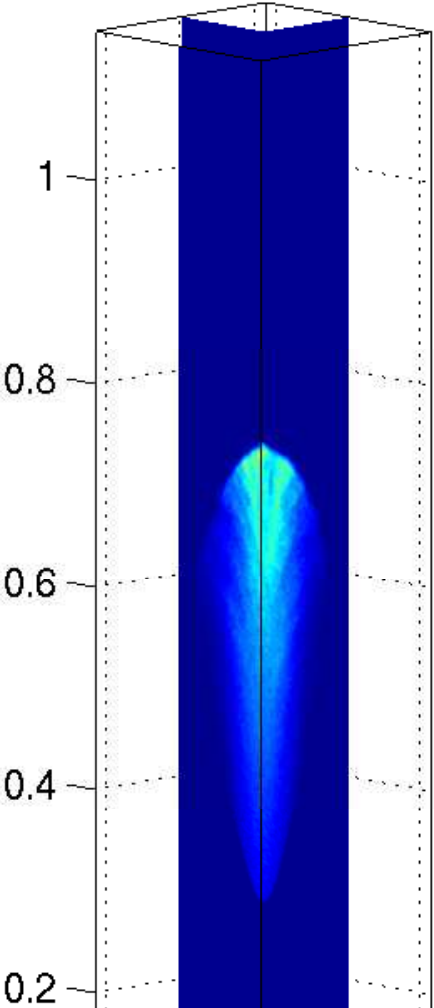}
\includegraphics[width=.12\textwidth,viewport=25 110 150 400, clip]{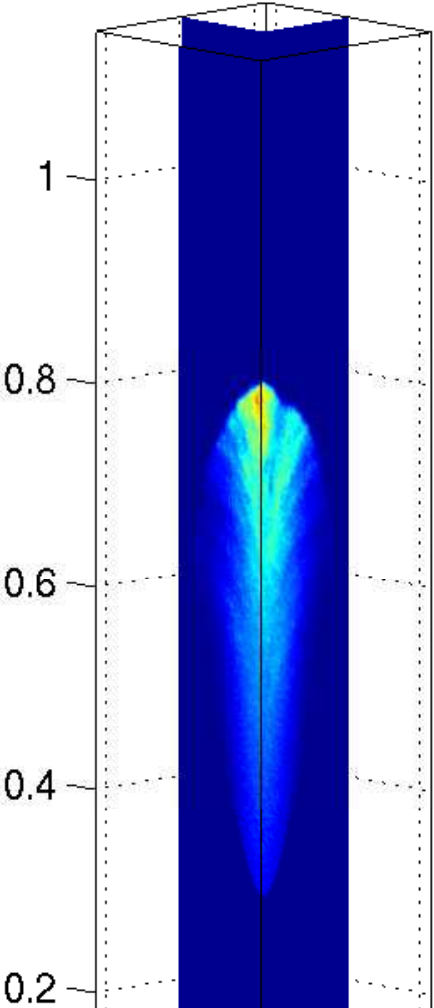}
\includegraphics[width=.12\textwidth,viewport=25 110 150 400, clip]{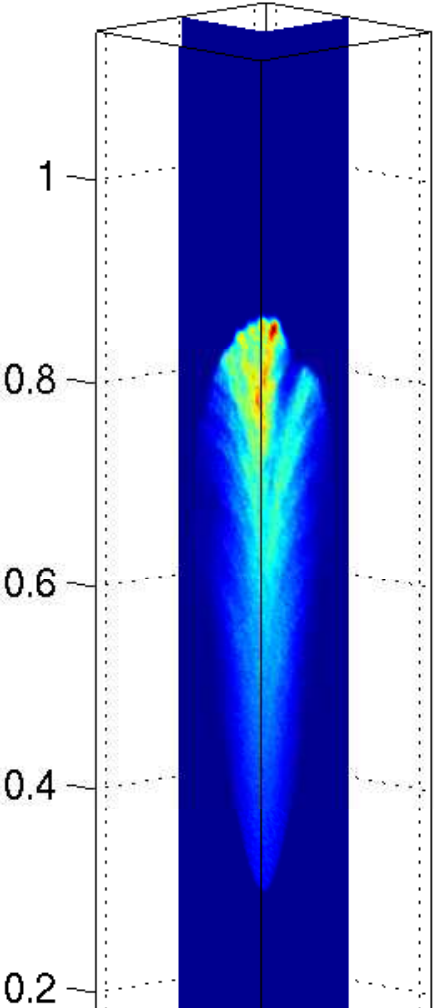}
\includegraphics[width=.12\textwidth,viewport=25 110 150 400, clip]{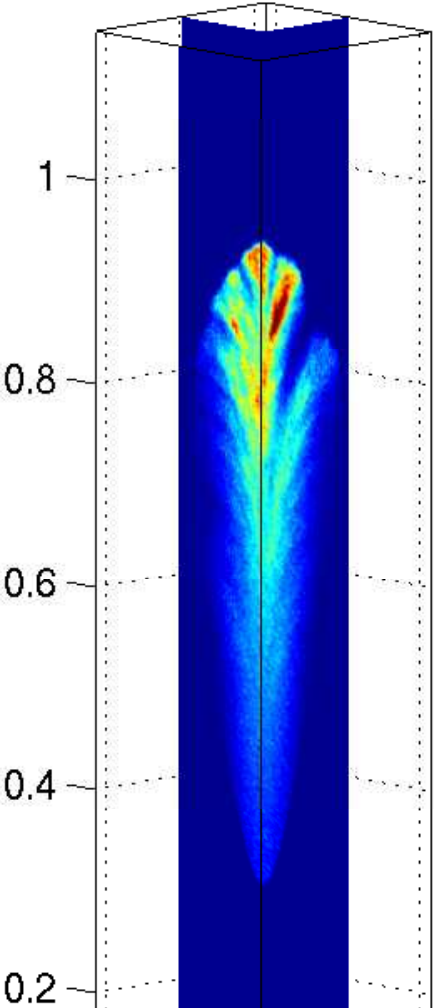}
\includegraphics[width=.12\textwidth,viewport=25 110 150 400, clip]{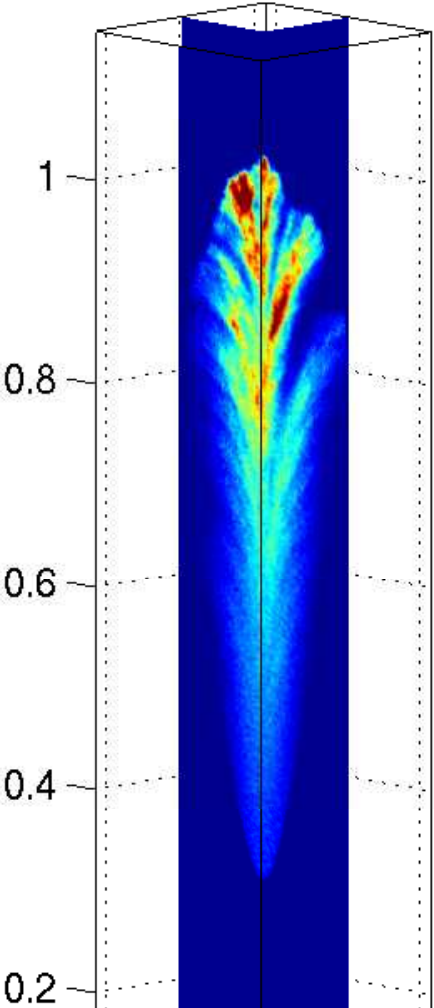}
\includegraphics[width=.04\textwidth,viewport=225 110 275 400, clip]{figures_pdf/fig1_colorbar.pdf} %White space
\\
% Fourth row
\includegraphics[width=.12\textwidth,viewport=25 30 150 400, clip]{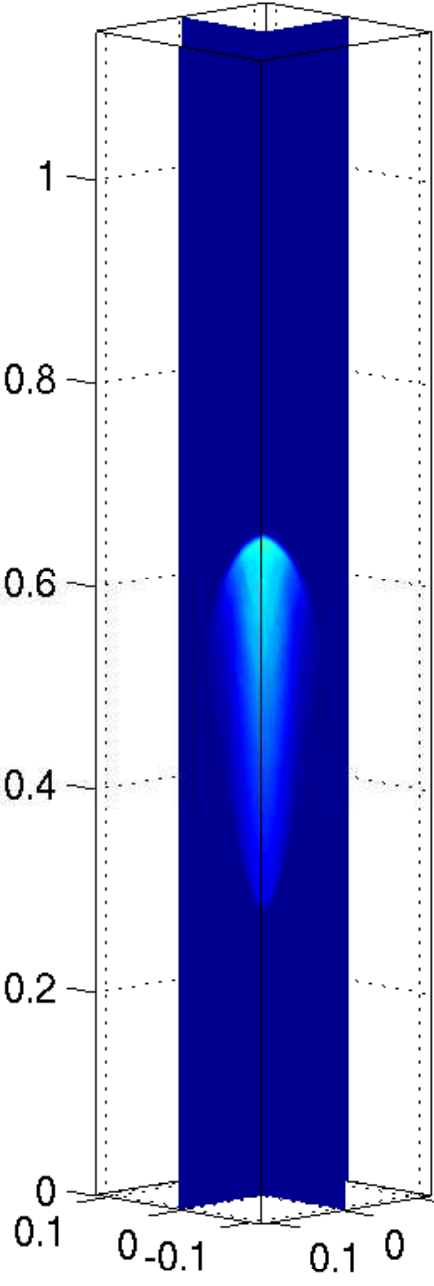}
\includegraphics[width=.12\textwidth,viewport=25 30 150 400, clip]{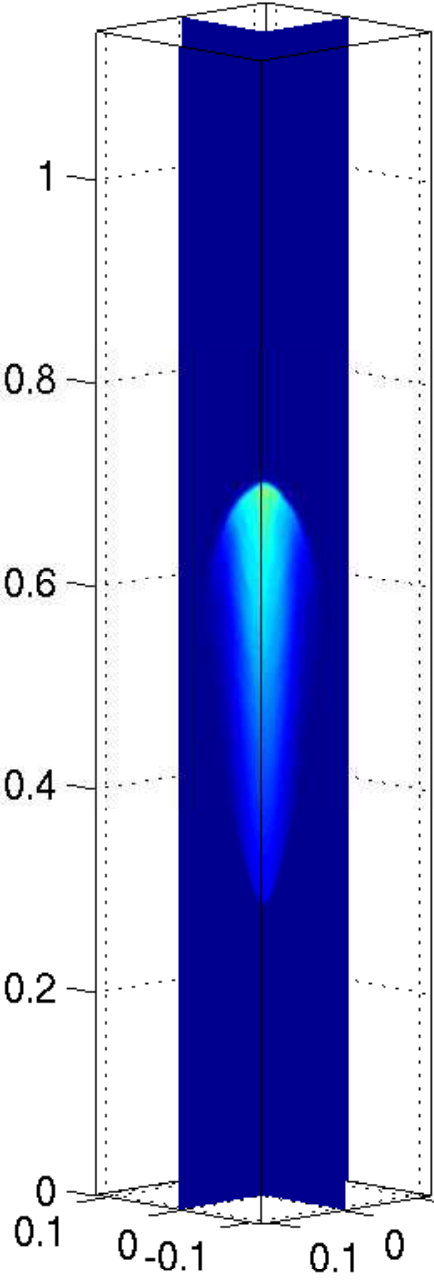}
\includegraphics[width=.12\textwidth,viewport=25 30 150 400, clip]{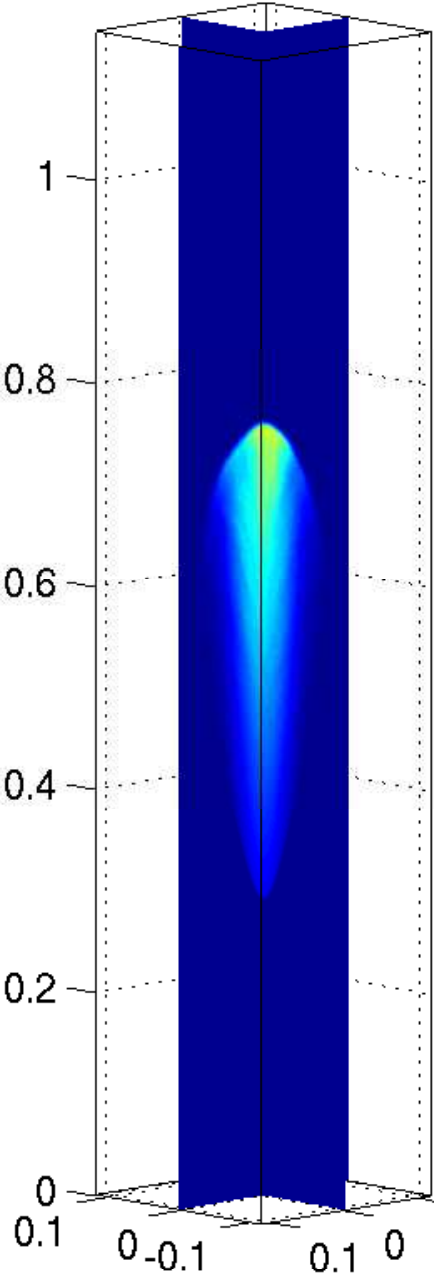}
\includegraphics[width=.12\textwidth,viewport=25 30 150 400, clip]{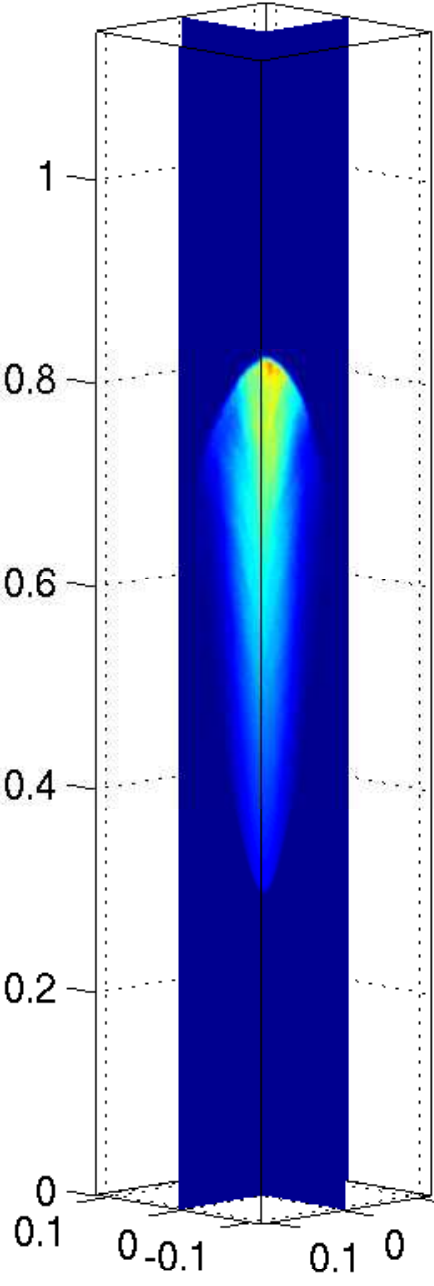}
\includegraphics[width=.12\textwidth,viewport=25 30 150 400, clip]{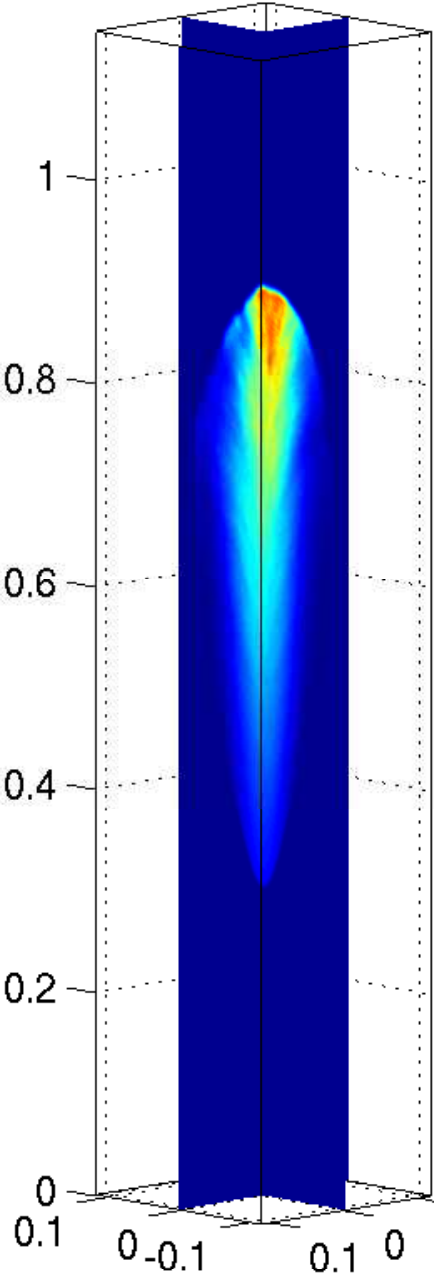}
\includegraphics[width=.12\textwidth,viewport=25 30 150 400, clip]{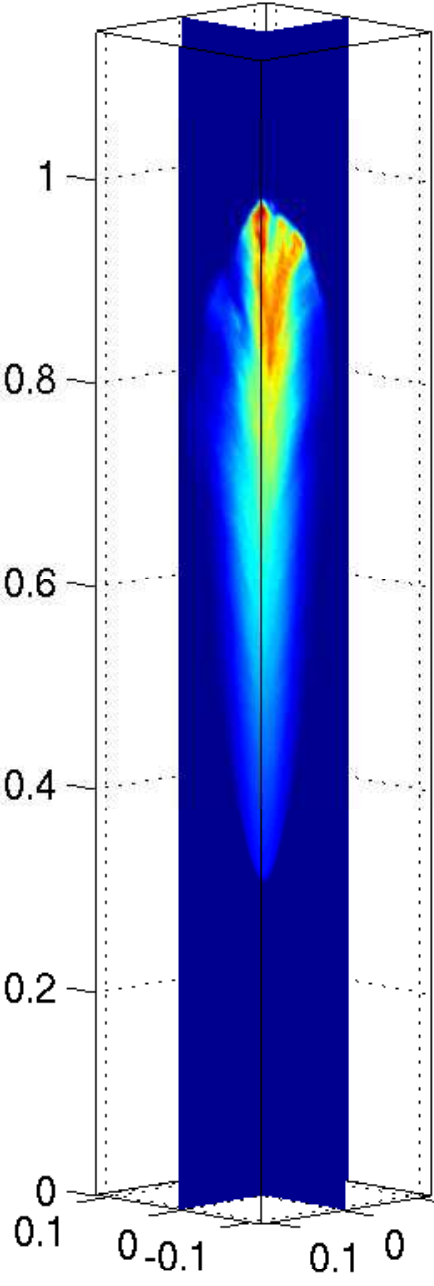}
\includegraphics[width=.12\textwidth,viewport=25 30 150 400, clip]{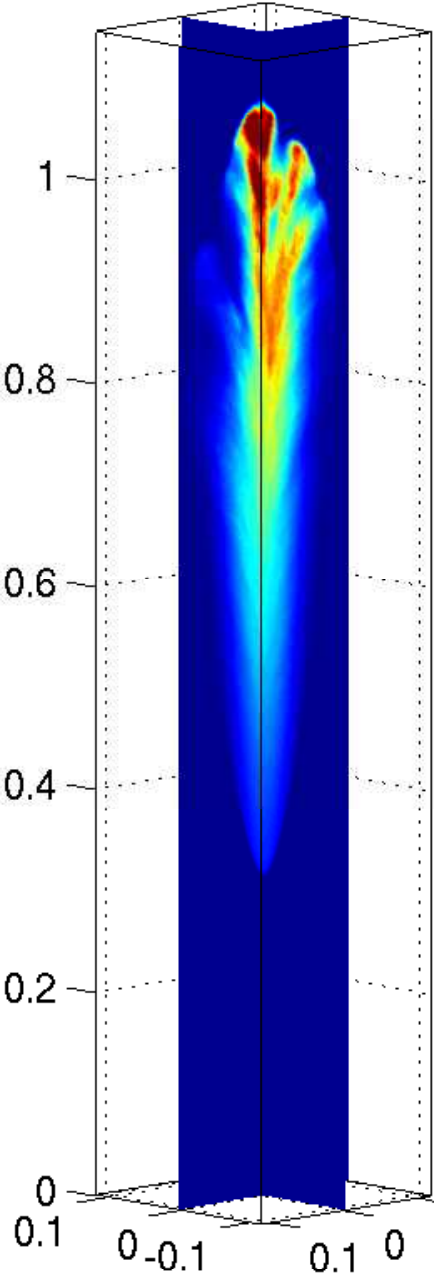}
\includegraphics[width=.04\textwidth,viewport=314 0 364 432, clip]{figures_pdf/fig1_colorbar.pdf}
\caption{
The electron density in classical fluid model (first row), extended fluid model (second row), particle model (third row) and hybrid model (fourth row). The columns show the temporal evolution from time 0.72~ns to 0.9~ns in steps of 0.03~ns. The densities are plotted on two orthogonal planes intersecting with the 3D structure. The same color coding for the densities is used in all panels, densities range from 0 (blue) to $1.4\cdot10^{15}$/cm$^3$ (red), as indicated by the color bar. The full height of 1.17~mm of the simulated system is shown in the lowest row, while the upper rows are truncated below 0.2~mm. The lateral directions are truncated from $\pm0.29$ to $\pm0.1$~mm in all panels.
}
\label{fig:elecdens1}
\end{figure}

\begin{figure}
\centering
\includegraphics[width=.120\textwidth,viewport=25 30 180 400, clip]{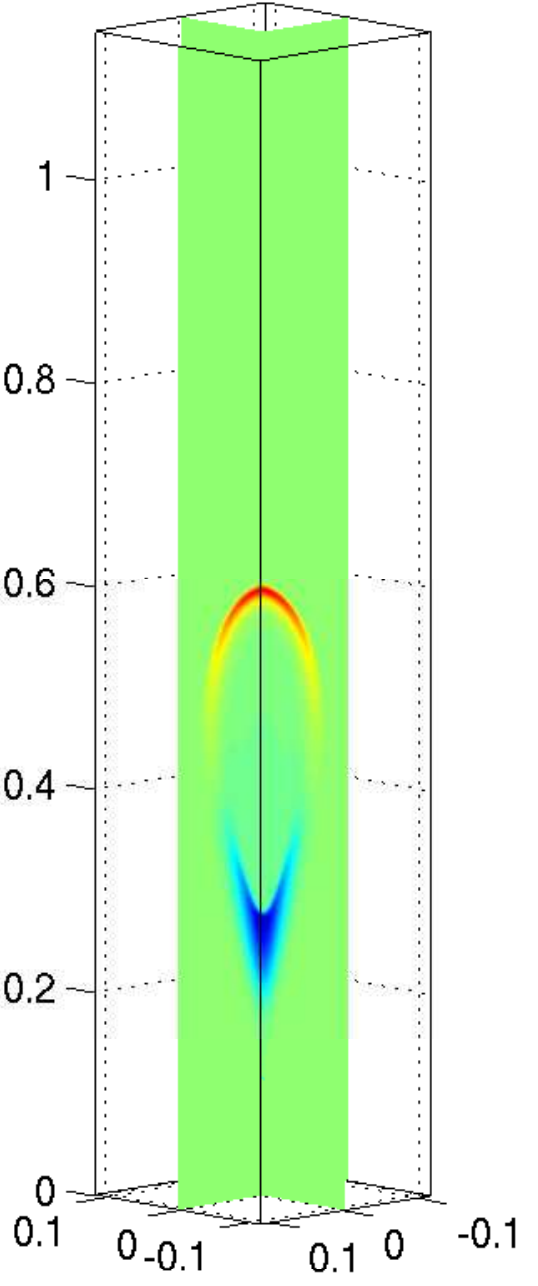}
\includegraphics[width=0.034\textwidth,viewport=314 0 364 432, clip]{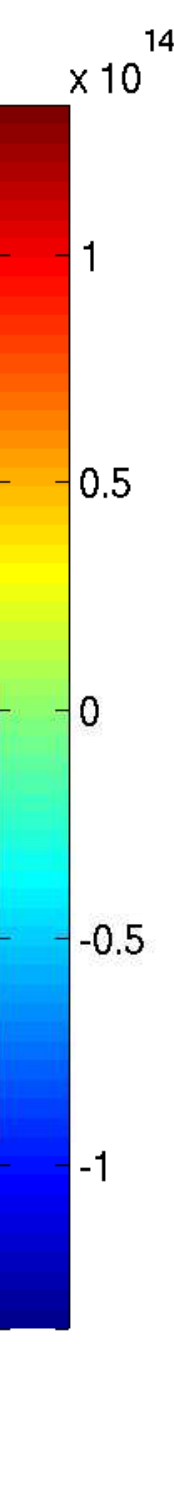}
\hspace{0.042\textwidth}
\includegraphics[width=.120\textwidth,viewport=25 30 180 400, clip]{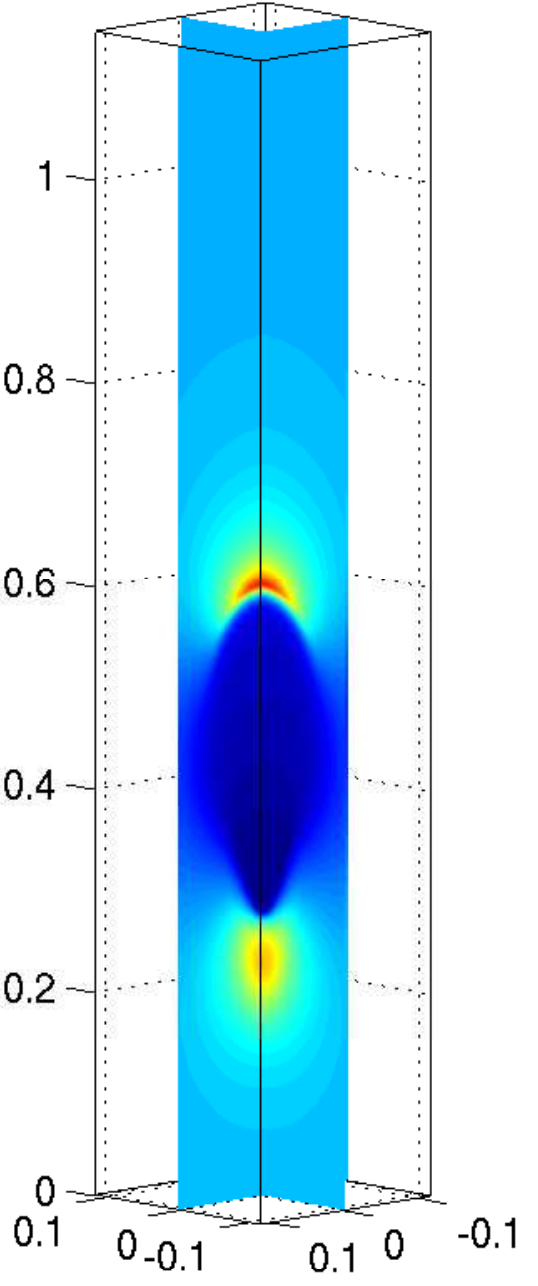}
\includegraphics[width=0.034\textwidth,viewport=314 0 364 432, clip]{figures_pdf/fig2_classfluid_24aN.pdf}
\hspace{0.042\textwidth}
\includegraphics[width=.120\textwidth,viewport=25 30 180 400, clip]{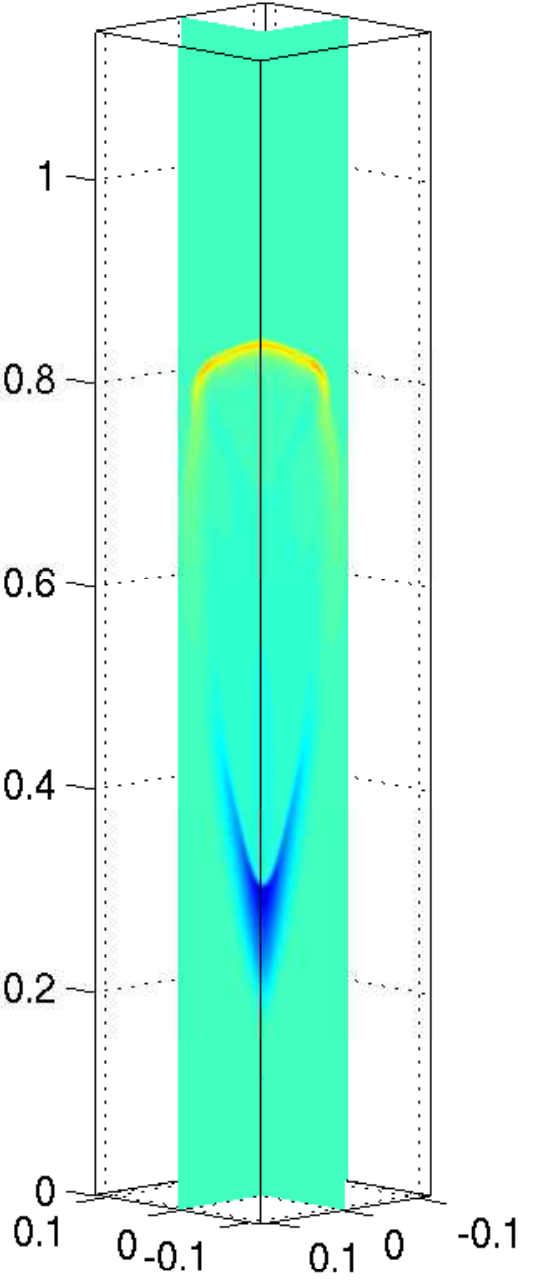}
\includegraphics[width=0.034\textwidth,viewport=314 0 364 432, clip]{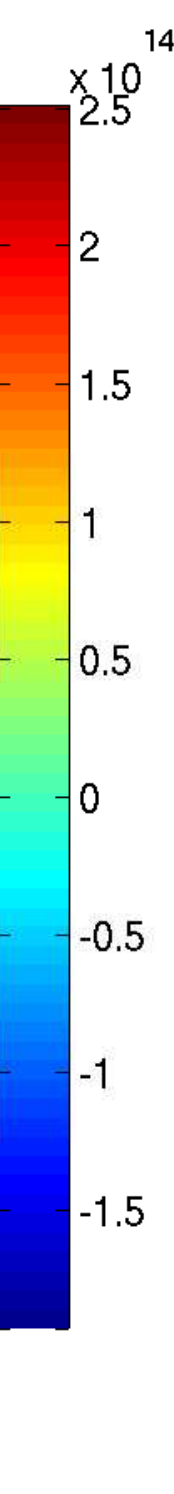}
\hspace{0.042\textwidth}
\includegraphics[width=.120\textwidth,viewport=25 30 180 400, clip]{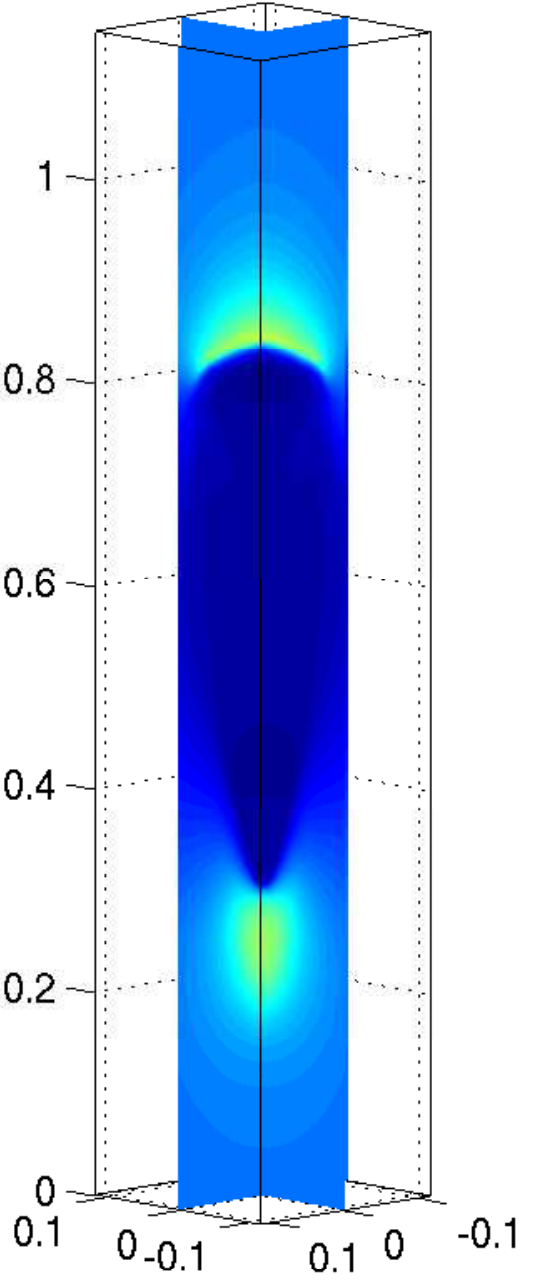}
\includegraphics[width=0.034\textwidth,viewport=314 0 364 432, clip]{figures_pdf/fig2_classfluid_30aN.pdf}

\includegraphics[width=.120\textwidth,viewport=25 30 180 400, clip]{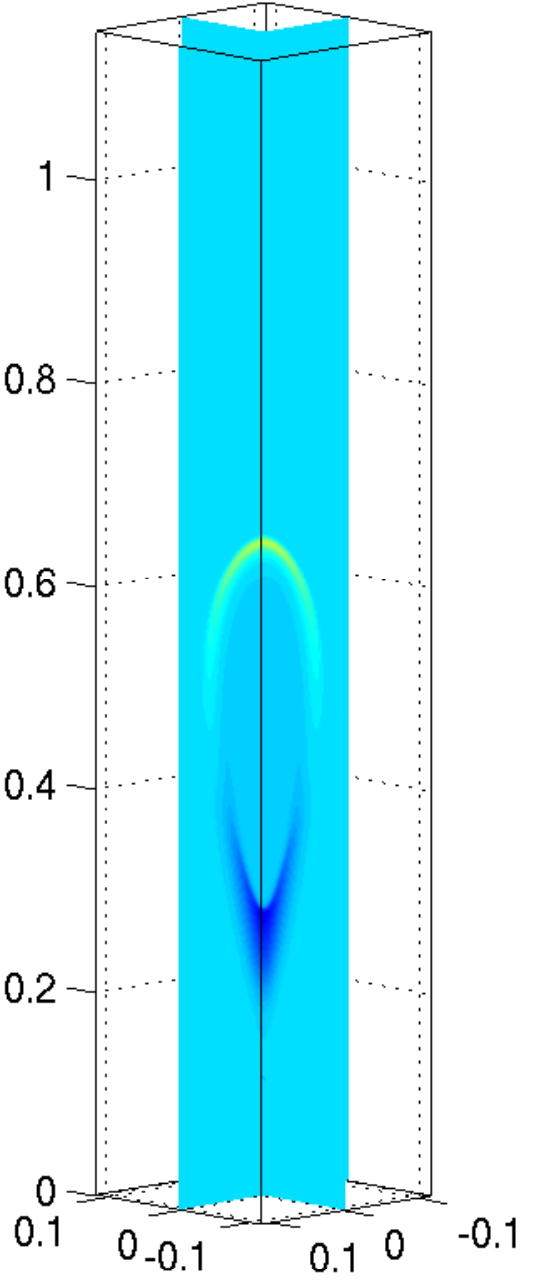}
\includegraphics[width=0.034\textwidth,viewport=314 0 364 432, clip]{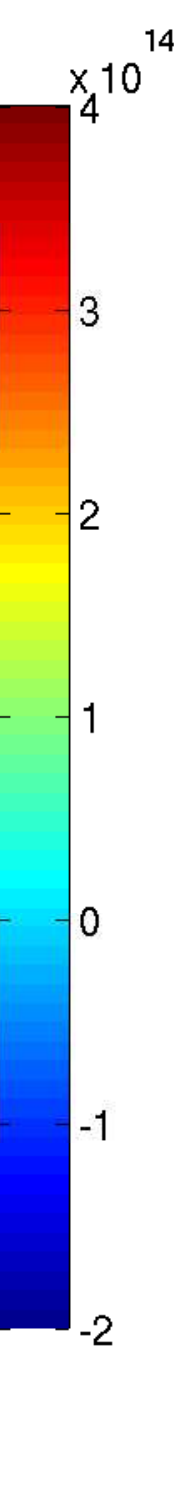}
\hspace{0.042\textwidth}
\includegraphics[width=.120\textwidth,viewport=25 30 180 400, clip]{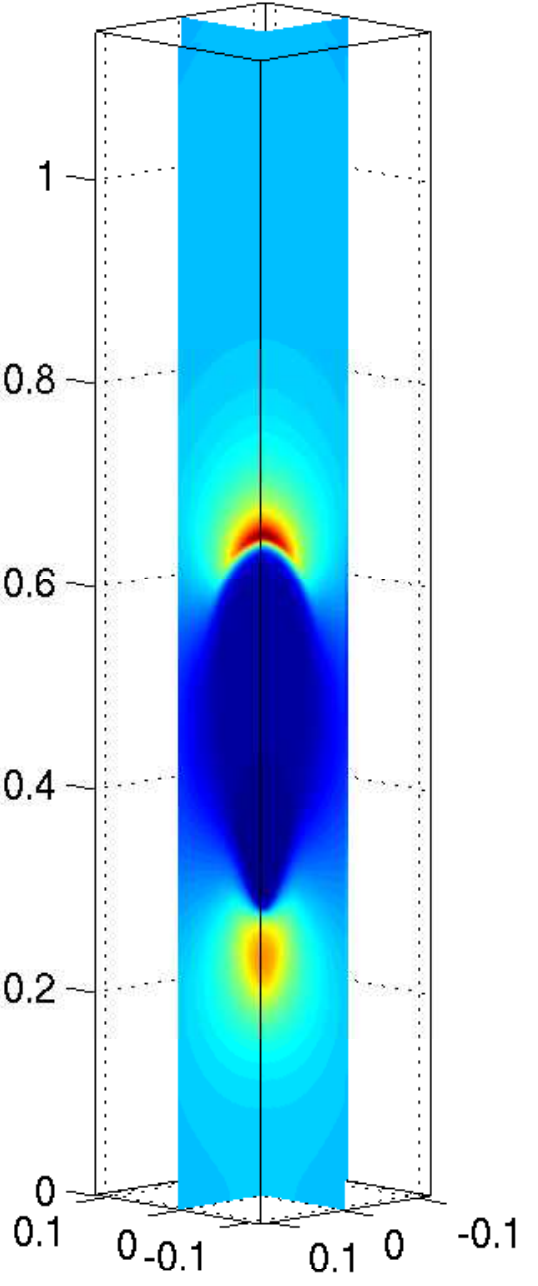}
\includegraphics[width=0.034\textwidth,viewport=314 0 364 432, clip]{figures_pdf/fig2_extfluid_24aN.pdf}
\hspace{0.042\textwidth}
\includegraphics[width=.120\textwidth,viewport=25 30 180 400, clip]{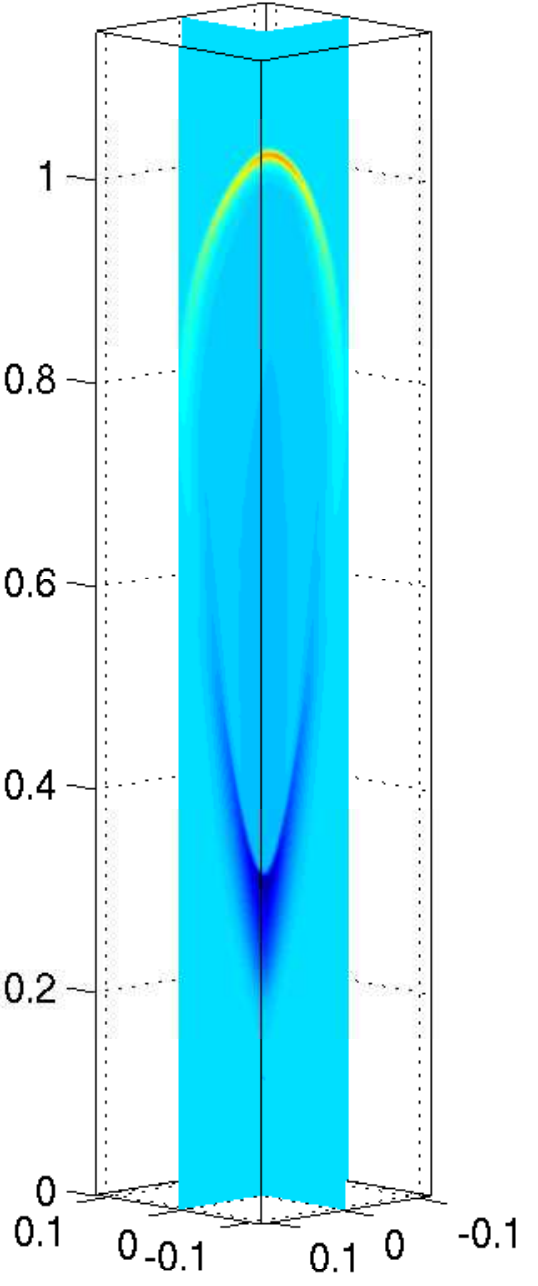}
\includegraphics[width=0.034\textwidth,viewport=314 0 364 432, clip]{figures_pdf/fig2_extfluid_colorbar30bN.pdf}
\hspace{0.042\textwidth}
\includegraphics[width=.120\textwidth,viewport=25 30 180 400, clip]{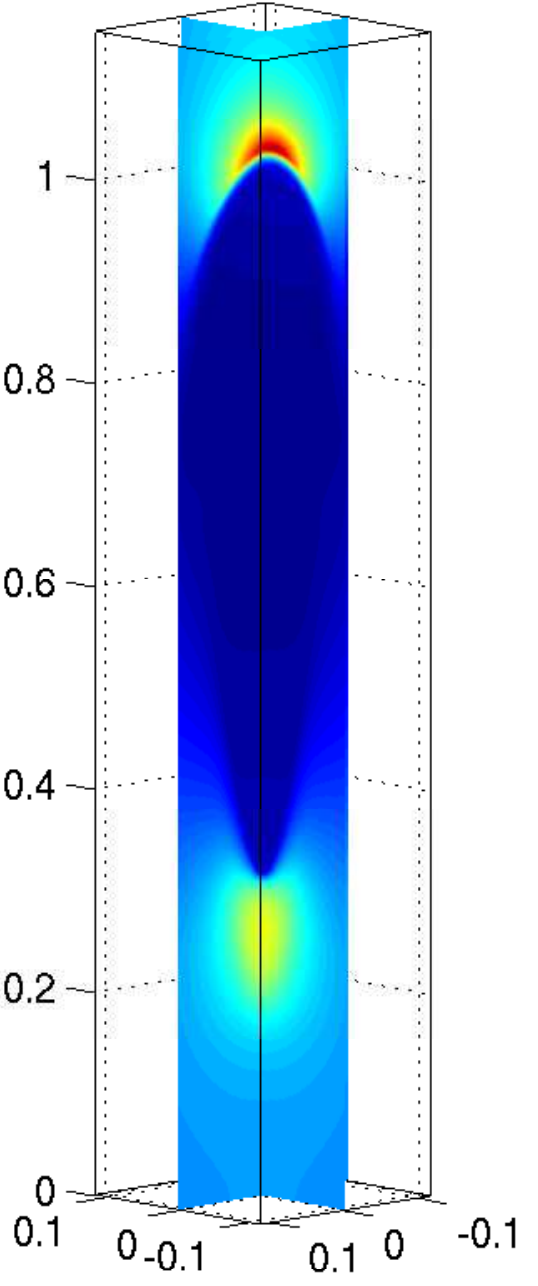}
\includegraphics[width=0.034\textwidth,viewport=314 0 364 432, clip]{figures_pdf/fig2_extfluid_30aN.pdf}

\includegraphics[width=.120\textwidth,viewport=25 30 180 400, clip]{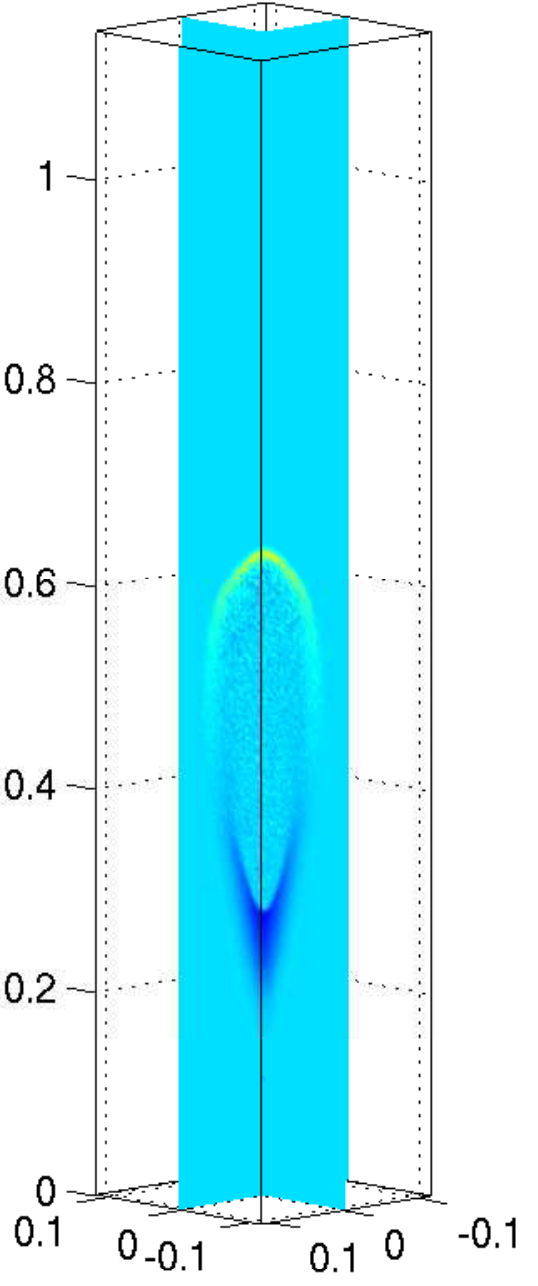}
\includegraphics[width=0.034\textwidth,viewport=314 0 364 432, clip]{figures_pdf/fig2_extfluid_colorbar30bN.pdf}
\hspace{0.042\textwidth}
\includegraphics[width=.120\textwidth,viewport=25 30 180 400, clip]{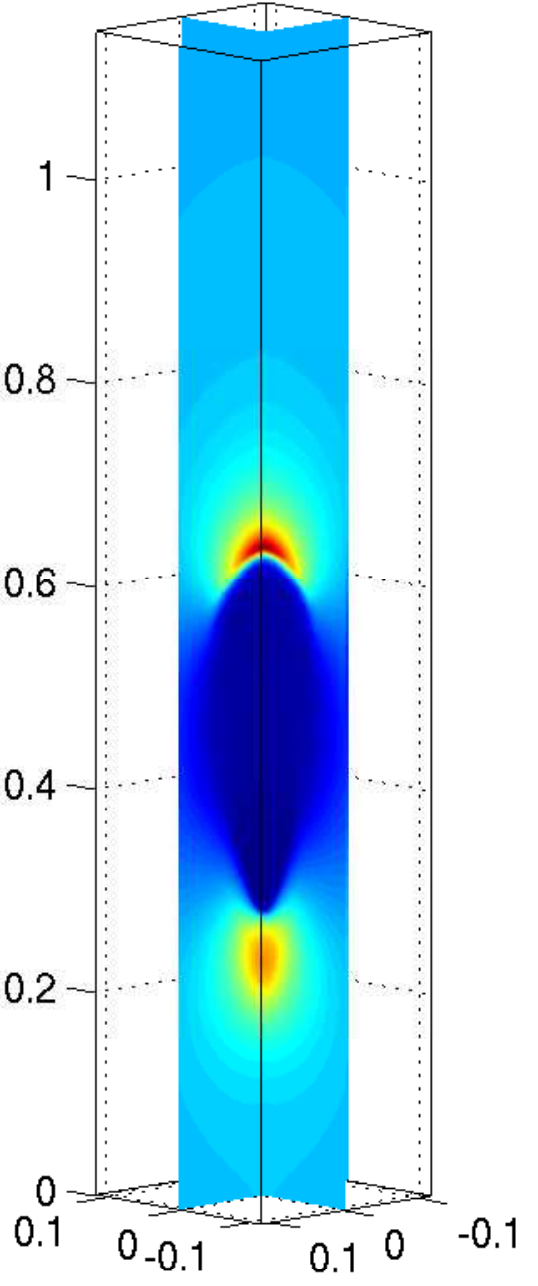}
\includegraphics[width=0.034\textwidth,viewport=314 0 364 432, clip]{figures_pdf/fig2_classfluid_24aN.pdf}
\hspace{0.042\textwidth}
\includegraphics[width=.120\textwidth,viewport=25 30 180 400, clip]{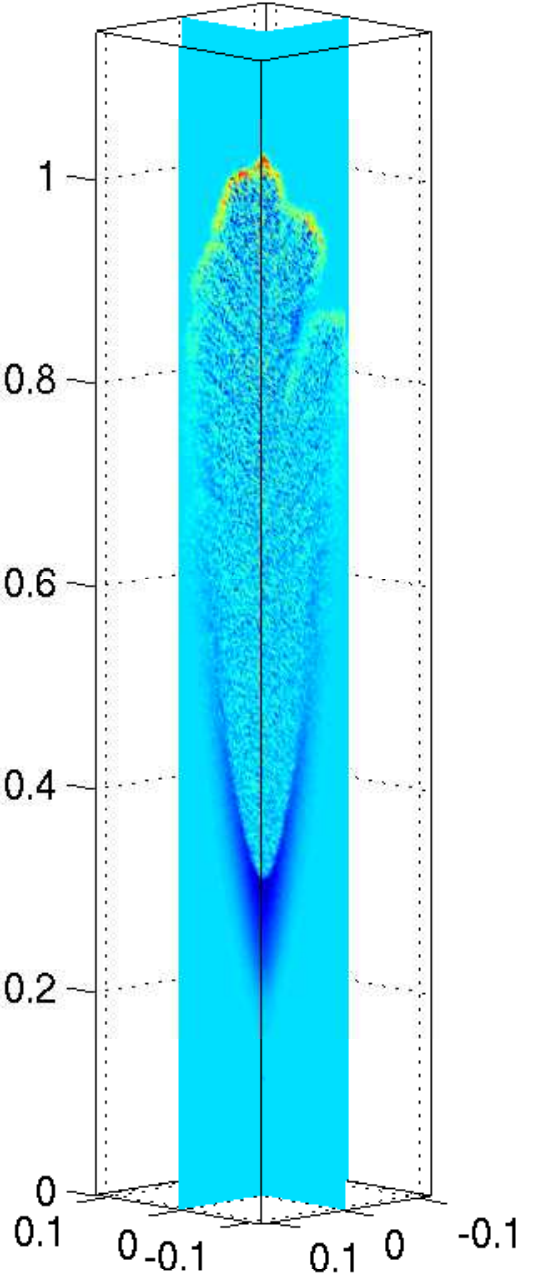}
\includegraphics[width=0.034\textwidth,viewport=314 0 364 432, clip]{figures_pdf/fig2_extfluid_colorbar30bN.pdf}
\hspace{0.042\textwidth}
\includegraphics[width=.120\textwidth,viewport=25 30 180 400, clip]{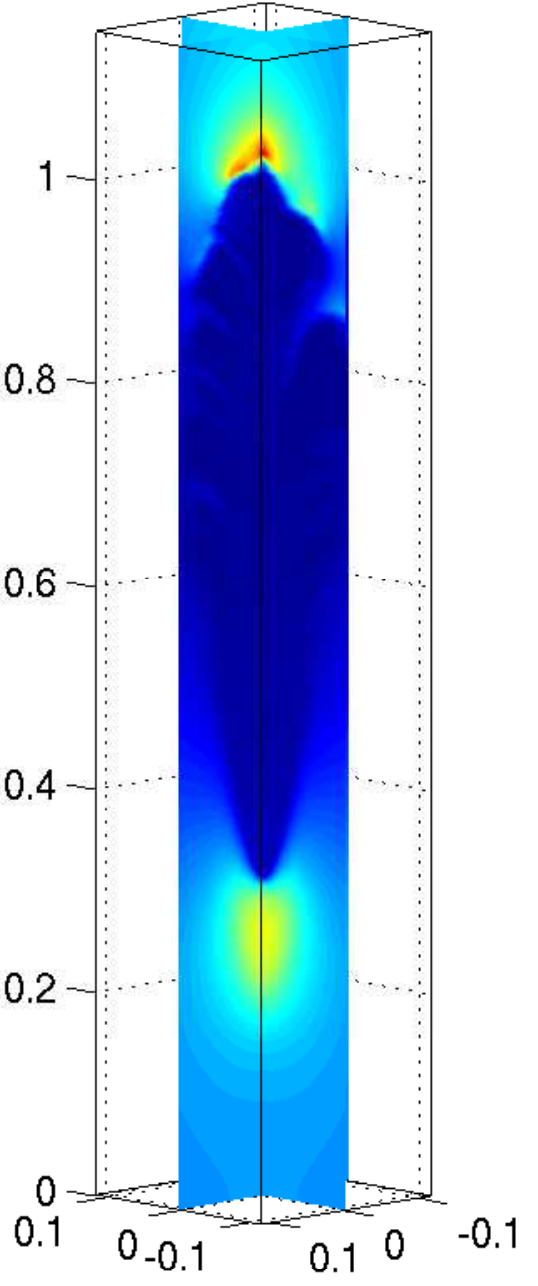}
\includegraphics[width=0.034\textwidth,viewport=314 0 364 432, clip]{figures_pdf/fig2_classfluid_30aN.pdf}

\includegraphics[width=.120\textwidth,viewport=25 30 180 400, clip]{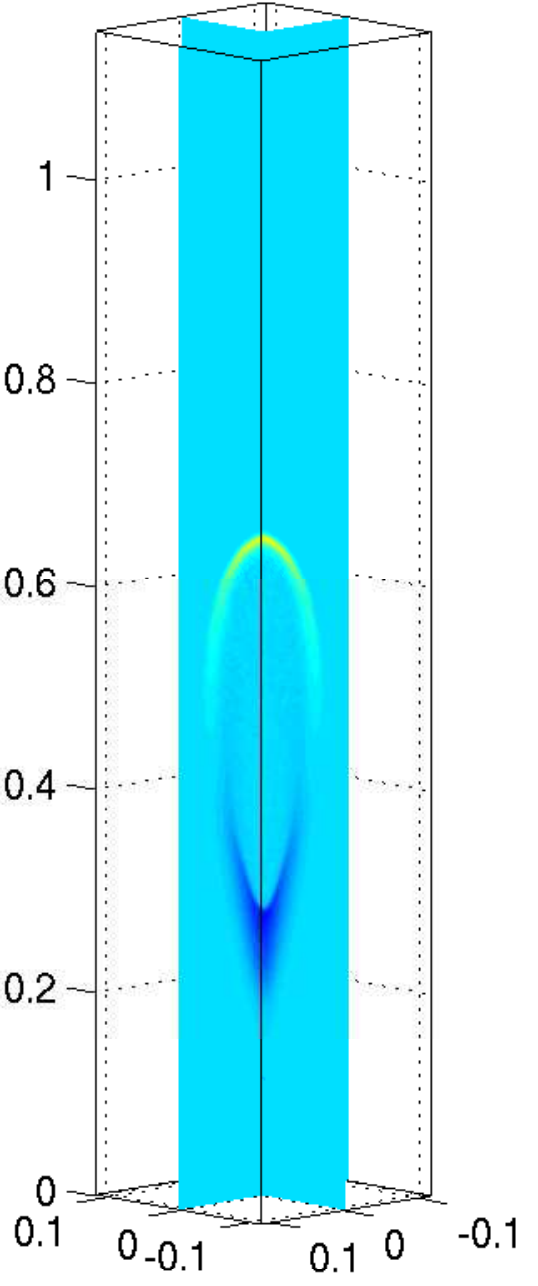}
\includegraphics[width=0.034\textwidth,viewport=314 0 364 432, clip]{figures_pdf/fig2_extfluid_colorbar30bN.pdf}
\hspace{0.042\textwidth}
\includegraphics[width=.120\textwidth,viewport=25 30 180 400, clip]{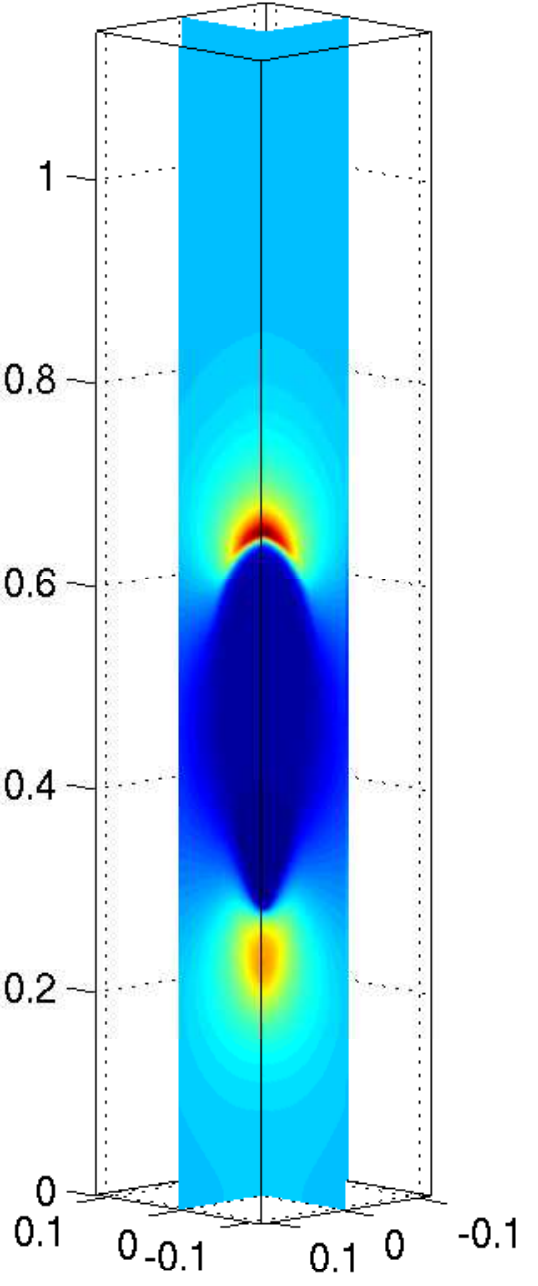}
\includegraphics[width=0.034\textwidth,viewport=314 0 364 432, clip]{figures_pdf/fig2_classfluid_24aN.pdf}
\hspace{0.042\textwidth}
\includegraphics[width=.120\textwidth,viewport=25 30 180 400, clip]{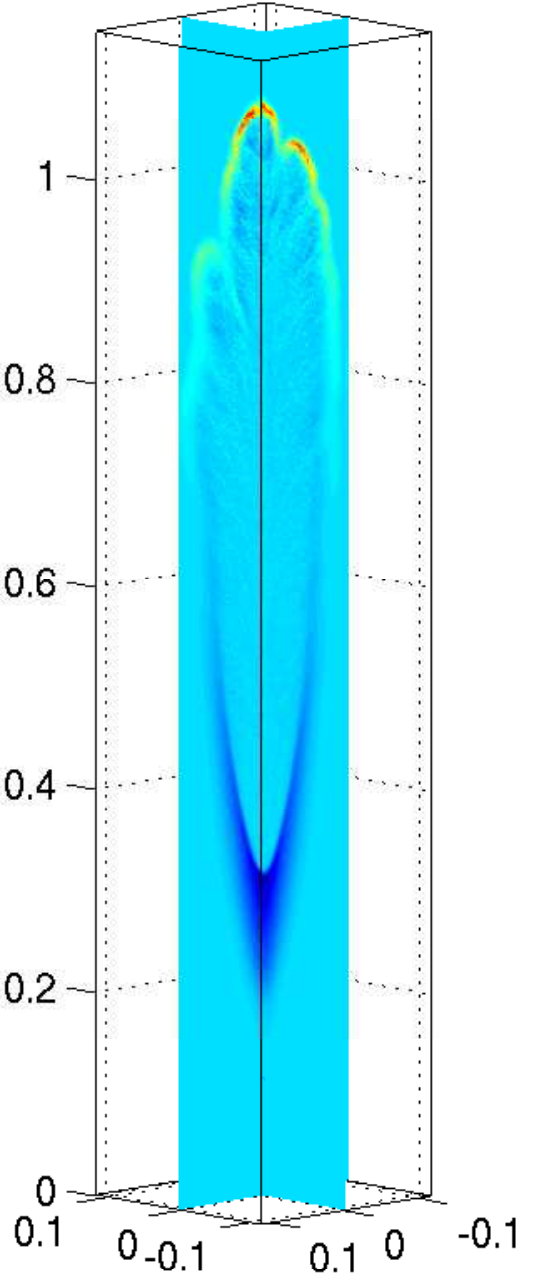}
\includegraphics[width=0.034\textwidth,viewport=314 0 364 432, clip]{figures_pdf/fig2_extfluid_colorbar30bN.pdf}
\hspace{0.042\textwidth}
\includegraphics[width=.120\textwidth,viewport=25 30 180 400, clip]{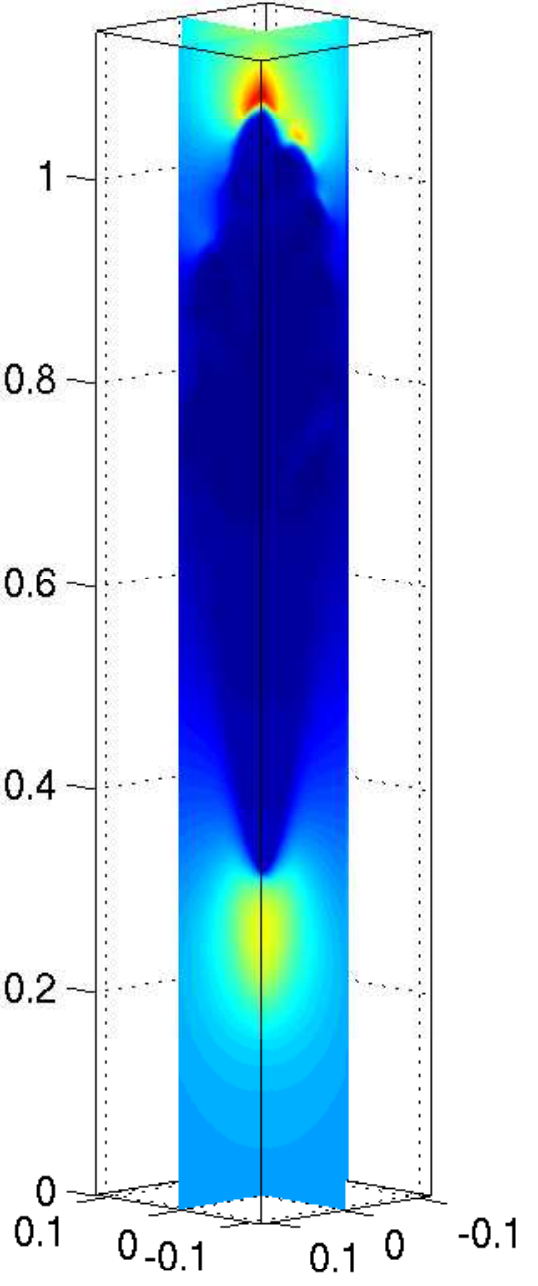}
\includegraphics[width=0.034\textwidth,viewport=314 0 364 432, clip]{figures_pdf/fig2_classfluid_30aN.pdf}

\label{fig:Fig2}
\caption{As in figure~\ref{fig:elecdens1}, the rows show from top to bottom: classical fluid model, extended fluid model, particle model and hybrid model. The columns show the negative space charge density and the electric field at the first and the last time step shown in figure~\ref{fig:elecdens1}, i.e., at times 0.72~ns and 0.9~ns. The color coding of densities and fields is the same in each column, except for the charge densities in the classical fluid model (first row). The color bar for the charge densities gives multiples of $-$e/cm$^3$ where e is the elementary charge, and the color bar for fields gives kV/cm.
}
\label{fig:elecdens2}
\end{figure}

Figure~\ref{fig:elecdens1} shows the evolution of the electron density in the streamer in seven stages, from time 0.72~ns up to 0.9~ns, with time increments of 0.03~ns, and figure~\ref{fig:elecdens2} shows the electric charge densities and the electric fields at 0.72~ns and at 0.9~ns.
The rows in figures~1 and 2 present from top to bottom: the classical fluid model, the extended fluid model, the particle model and the hybrid model.

\begin{figure}
\centering
\includegraphics[width=0.6\textwidth]{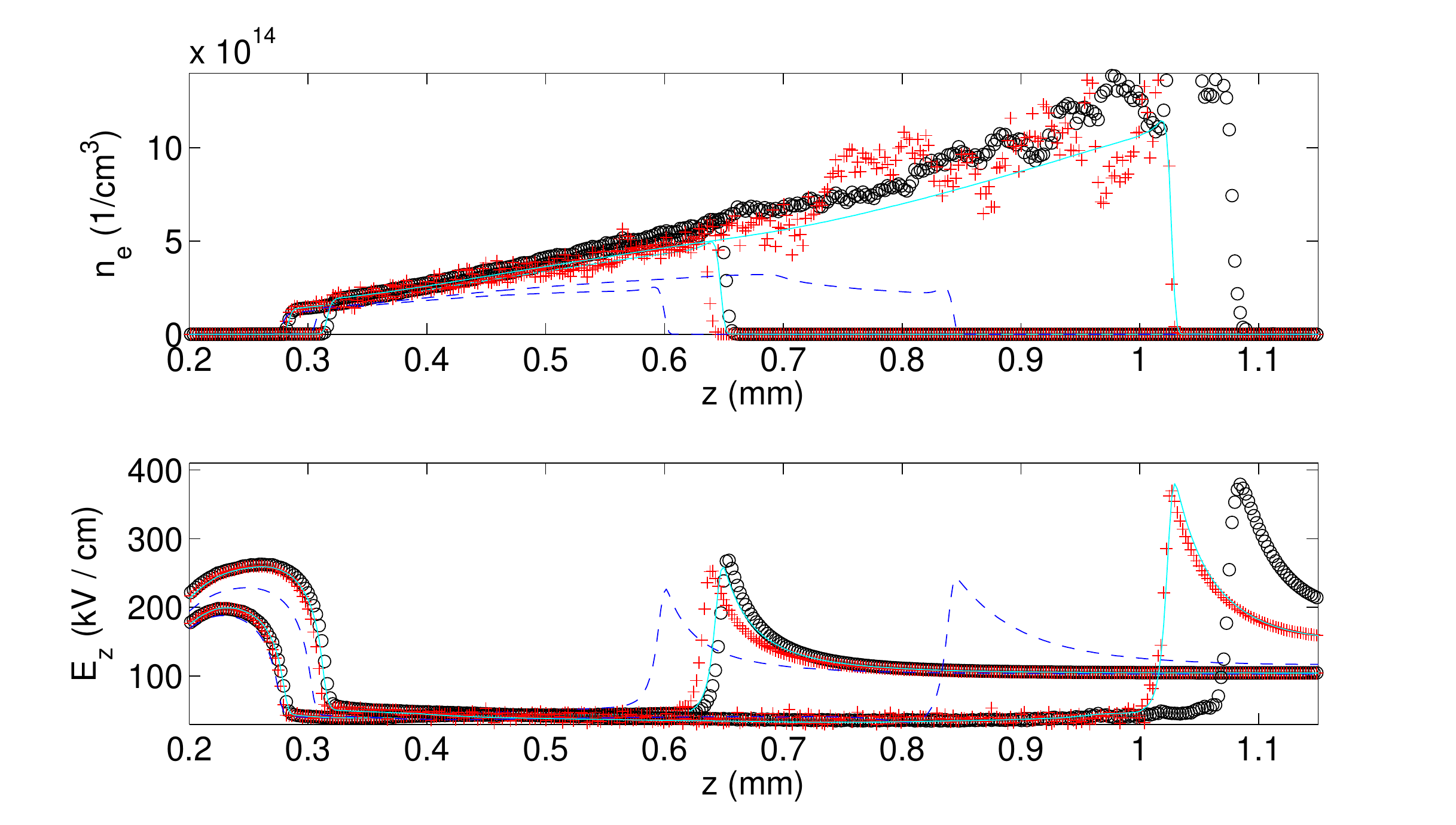}
\caption{Electron density (upper plot) and electric field strength (lower plot) on the vertical $z$ axis for the same two time steps 0.72~ns and 0.9~ns as in the previous plot. The different models are classical fluid model (dashed dark blue line), extended fluid model (solid light blue line), particle model (red crosses), and hybrid model (black circles).
}
\label{fig:1D_com}
\end{figure}

Figure~\ref{fig:1D_com} shows the electron density and the electric field on the z-axis for the four different models at 0.72~ns and 0.9~ns.

Fluctuation and destabilization effects can be seen more clearly in figures~4--6 that zoom into the propagating streamer heads.
The figures show electron density, negative space charge density and electric field for the same seven time steps as in figure~\ref{fig:elecdens1} for the extended fluid model, the particle model and the hybrid model.
The classical fluid model is not included since figures~1--3 demonstrate clearly that it does not approximate the particle dynamics well.

%The color codes for densities and fields are the same in each column, except for the classical fluid model at the second time step 0.9~ns. This is because densities and fields are so much lower than in the other models that they would not be clearly visible if plotted in the same color code.

\subsection{Streamer propagation in the four models}

At the earlier stage of 0.72 ns, figures~1 and 2 show that a streamer has emerged and grown to about the same length in all models, and that it has approximately the same radius and field enhancement at the tip.
The streamer in the classical fluid model (upper row) has stayed a bit behind, and electron and charge density are lower, though the field enhancement is still similar.
The propagation differences can be seen more clearly in figure~\ref{fig:1D_com}, which shows the electron density and electric field on the z-axis at 0.72~ns and 0.9~ns.
At 0.72~ns, the profiles of extended fluid, particle and hybrid model are about the same, but the classical fluid model has a lower field enhancement, lower electron density and shorter propagation length.
That the streamer in the classical fluid model grows more slowly both in space and in electron density, while the other models have comparable results, is also reflected in the total number of electrons: it is $(5.3\pm0.2)\times 10^8$ in particle, hybrid and
extended fluid model (more precisely 5.1, 5.6 and 5.3$\times10^8$), while it is only $2.9\times10^8$ in the classical fluid simulation.
We will discuss the dynamics of the classical fluid approximation in more detail in section~\ref{sec:classicalfluid}.

At this stage, the particle model uses $1.36\times10^7$ super-particles with a weight of 32 real electrons, while the hybrid model follows $2\times10^7$ real electrons and leaves the rest to the fluid region.
The super-particles in the particle model already create visible fluctuations of the space charge density, as discussed earlier in~\cite{Li2008:2}.
The fluctuations of the local field create numerical heating, and this effect increases as time evolves.
Such numerical artifacts can be somewhat suppressed if the super-particles are formed adaptively using particle coalescence techniques~\cite{Cha2010, Lapenta2002, Bagdonat2002, Welch2007}.
But as the number of particles increases, the increased fluctuations will affect the simulations, especially for negative streamers where perturbations in the electron density can grow as they move outwards.
The hybrid model does not suffer from such artifacts.

\subsection{Front destabilization in the four models}

During the further evolution up to time 0.9~ns, the streamer ionization front destabilizes in three of the four models, but in characteristically different manners.

The streamer in the classical fluid model is destabilizing into off-axis branches, which are quite symmetric (as we expect in the deterministic fluid model, and as we have seen previously in the simulations of Montijn {\it et al.}~\cite{Caro2006}).
The actual branching can also be seen in the plot for time 0.9~ns in figure~\ref{fig:1D_com}: the electron density on the z-axis in the classical fluid model starts to decrease for $z>0.69$~mm.
As the ionization density is determined by the electric field at the front at the moment when it passed that particular position, the electric field at the front increases until the position 0.69~mm is reached, and decreases thereafter on the axis as the lateral protrusions grow and screen the electric field on the axis.

The streamer in the extended fluid model propagates in a stable manner until the last time step 0.9~ns.
The off-axis branching of the classical fluid model is suppressed by the higher ionization rates on the axis in the extended fluid model, that are due to the gradient correction in the ionization term.
However, it cannot be excluded a priori that the streamer later destabilizes along a different mode.

The streamers in the particle and in the hybrid model both clearly show density fluctuation effects, both at the front and in the interior charge density.
The fluctuations in the particle model are unphysical due to super-particle artifacts while the fluctuations in the hybrid model stay physical.
The fluctuations at the front destabilize the streamer into several branches in both models, but the streamer tips close to the axis keep the strongest field enhancement and screen new branches up to the end of the simulation at 0.9~ns.
It should be mentioned here that the hybrid model operates with real particles up to 0.69~ns, but has introduced super-particles of weight 8 at 0.9 ns.
The weight of the super-particles in the particle model at 0.9~ns is 16 times higher, namely 128.

The extended fluid model and the particle model agree very well in propagation velocity, field enhancement and ionization density, up to the large super-particle fluctuations in the particle model.
The hybrid model has similar a ionization density and electric field profile, but is ahead of the other models.
At 0.9~ns, there are 3.9, 4.6 and 3.5$\times 10^9$ electrons in particle, hybrid and extended fluid model, and only 1.3$\times 10^9$ in the classical fluid model.

\subsection{Computing times}

To obtain the results presented here, the computing time for the fluid simulations was about 1 week, for the hybrid simulation about 1.5 weeks, and for the particle simulation about 2 weeks (all on an Intel Q6600 2.4~GHz quadcore processor).
The cost of solving the Poisson equation at every timestep dominated the total computational cost, therefore computing times are similar for the different models.
All the simulations ran sequentially on a single core.

\section{Discussion of the results}
\label{sec:discussion}

\subsection{Classical fluid model\label{sec:classicalfluid}}

Our results show that the streamers in the classical fluid model develop lower velocities, field enhancement and ionization densities than in the other models; the model clearly approximates the microscopic dynamics quite badly, as also found previously in~\cite{Li2007}.
This is the case even though the transport and reaction coefficients were derived from swarm simulations in the particle model for consistency, and though bulk coefficients were used.
As the fields do not exceed 250~kV/cm in the simulations with the classical model, the fluid coefficients were only used in the parameter range in which they were actually derived in~\cite{Li2011:JCP}.
By construction, the classical fluid model with bulk coefficients models electron swarms in a constant electric field well, and earlier numerical studies as well analytical arguments have shown that also the velocity of a planar front in a fixed electric field is well approximated~\cite{Li2007,Li2010:1}.
However, the ionization density behind a planar
front in a fixed field is too low in the classical fluid model with bulk coefficients when the maximal field exceeds 50~kV/cm~\cite{Li2007}; this is always the case in the present calculations.

In the 3D simulations, the deviation from the other models is larger than in 1D~\cite{Li2007}.
We argue that this is because in our 1D front simulations, the electric field ahead of the front is fixed, while in 3D the field falls off with distance and varies in time.
As the same field in the front creates a lower ionization density in the classical fluid model, also the conductivity in the streamer channel and the consecutive electric screening are lower than in the other models.
Therefore the field enhancement is less in 3D and leaves an even lower ionization density behind as the ionization level depends on the field at the front.
The lower field enhancement also explains why the streamer is slower than in the other models.

Choosing flux rather than bulk coefficients had not resolved the discrepancy with other models either, as already discussed in section~2.2.
The model with flux coefficients does not reproduce the swarm results in a constant electric field.
Furthermore, with flux coefficients the electron mobility had been considerably lower (cf.~figure~3 in~\cite{Li2011:JCP}); therefore the front had been even slower than with bulk coefficients.

The streamers within the classical fluid model destabilize and branch at about the same time as in hybrid and particle model, but in a more symmetric manner, as the destabilization is not supported by electron density fluctuations.
This deterministic branching in a fluid model is well approximated by moving boundary models as studied in~\cite{Kao} and reviewed in~\cite{Ebert2011/Nonlinearity}.

We remark that the front destabilization in the present fully three-dimensional simulations of the classical fluid model for negative streamers without photo-ionization occurs in a very similar manner as in previous high accuracy calculations under the constraint of cylindrical symmetry~\cite{Caro2006}.
Earlier simulations~\cite{PRL2002} demonstrated the mechanism but suffered from lower numerical resolution.
In~\cite{Caro2006}, streamers in the same background field were studied, but with a more ionized initial condition and attached to a planar electrode.
In those simulations, the branching instability occurred after 1.09~ns, while here it occurs after 0.81~ns.
The somewhat longer evolution time until branching in~\cite{Caro2006} could be due to the different initial and boundary conditions, or to the less accurate transport and reaction coefficients in~\cite{Caro2006} or to the symmetry constraint; this is subject of future research.
In any case, the results support the argument given in~\cite{ReplyPRL2002} that the branching time under the symmetry constraint is an approximation and upper bound of the branching time in the fully 3D calculation.

\subsection{Extended fluid model, particle model and hybrid model}

The extended fluid model was constructed to cure the deficiencies of the classical model.
It was shown already in~\cite{Li2007,Li2010:1} that with the extension in the reaction term and with flux coefficients, it approximates the growth and propagation of particle swarms and of planar streamer fronts well, including the ionization density behind an ionization front.
It should be noted though that the reaction and transport coefficients are used here for up to 400~kV/cm while they were derived only for up to 250~kV/cm in~\cite{Li2011:JCP} and extrapolated to higher field values.
The hybrid model uses the fluid coefficients only in the range in which they were derived.

We already discussed above that destabilization into off-axis branches in the extended fluid model is less likely than in the classical fluid model, but we have currently no explanation why branching does not occur at all; possibly this is a mere coincidence and branching does occur at some time after the end of the present simulations at 0.9~ns.
The front destabilization in particle and hybrid model occurs at a similar time as in the classical fluid model, but in a different manner: the fastest propagating branch stays close to the axis and screens the other branches that therefore keep staying behind.

The figures show that the extended fluid model approximates particle and hybrid model well up to the moment of destabilization, after this moment there are characteristic differences due to the density fluctuation effects caused by the discreteness of the electrons.
These fluctuations have an unphysical distribution when the particle model needs to use super-particles, and therefore the hybrid simulations should be closer to the true dynamics.
The stronger destabilization in the particle model seems to be compensated by the increased noise in the space charge distribution (see figure~\ref{fig:chargedens}), so that in the end the front moves with essentially the same velocity as in the extended fluid model.

\subsection{The front velocity in the different models compared to an analytical result}

The front positions (see figure~\ref{fig:frontposition}a) and the density and field profiles on the axis (see figure~\ref{fig:1D_com}) agree very well between the extended fluid model and the particle model even at the latest stages, when the particle model shows strong fluctuation effects; the streamer in the hybrid model is a bit faster at the latest stages.
This could be due to two different reasons:
%The question rises why the streamer in the hybrid model is a bit faster than in particle and extended fluid model. Two explanations are possible.
\begin{itemize}
\item The single particle fluctuations resolved in the hybrid model cover rare electron run-away effects better.
This could create more new avalanches ahead of the front, so that eventually the front jumps forward when the electron density within the avalanches has increased sufficiently.
In this case the front would be moving faster than expected from reaction, drift and diffusion in the local electric field.

\item Or the single particle fluctuations create more branching and thinner streamers with more field enhancement.
The front would then propagate faster because the local field is higher, and not because electrons run away.
\end{itemize}

The maximal electric field as a function of time is plotted in figure~\ref{fig:frontposition}b.
Comparison with the front position in figure~\ref{fig:frontposition}a already points to the second statement: where the hybrid model is ahead of the other models, the maximal field enhancement is higher as well.
The question is further analyzed in figure~\ref{fig:headposition}.
For planar fronts in a slowly varying electric field $E$ and with a sufficiently rapidly decay of the electron density ahead, the front velocity in the classical fluid model is given by~\cite{Ebert1997}
\begin{equation} \label{eq1}
v^*=\mu_e|E|+2\sqrt{D_e \mu_e|E|\alpha},
\end{equation}
where $\mu_e$ is the electron mobility, $D_e$ the electron diffusion constant and $\alpha$ the effective ionization coefficient.
We now use this equation for all models, not only for the classical fluid model.
We insert the maximal electric fields $E(t)$ on the z-axis of the respective models into this equation and evaluate it with our flux coefficients for $\mu_e(E)$, $D_e(E)$ and $\alpha(E)$.
As the front velocity and the maximal field in the hybrid and particle model fluctuate heavily during the late stages, we do not compare velocities, but the resulting front displacements of the models in figure~\ref{fig:headposition}.
Here the lines indicate the simulation results and the extended symbols the front displacement as predicted by Eq.~(\ref{eq1}).

Up to time 0.6~ns and a maximal field of about 200~kV/cm, the models agree very well with the analytical approximation of~(\ref{eq1}).
During the further evolution, the classical fluid model shows almost no deviation from~(\ref{eq1}).
(We should remark that this is somewhat accidental, as the classical fluid model uses bulk coefficients and the approximation flux coefficients.) For extended fluid model, particle model and hybrid model, the deviations follow a similar trend at later times: the simulation models are always a bit ahead of the analytical approximations, and in both the hybrid model is ahead of particle and extended fluid model.

One can conclude (i) that the hybrid model is ahead of the others because the field enhancement is higher, and (ii) that equation (\ref{eq1}) is a reasonable approximation of the front velocity in all models, but somewhat too low at higher fields.
This might be related to the fact that the transport and reaction coefficients were extrapolated from 250~kV/cm to higher fields.

Some high energy electrons can ``run away'' from the front in the hybrid and in the particle model.
The first electrons with energies above a typical run-away threshold of 200 eV appear at time 0.585~ns in the hybrid model, when the maximal electric field reaches 190 kV/cm, in agreement with the results in~\cite{Li2011:JCP}.
In the particle model the first electrons pass the 200~eV threshold a bit later, at time 0.65~ns.
At the end of the simulation the energies of some electrons exceed 1~keV both in hybrid and in particle model.
The role of electron run away for the front speed will be investigated further in future work, but it does not seem to have a significant influence on the results presented here, according to our velocity analysis above.

Finally, it should be noted that we used the maximal electric field on the axis in our analysis, while after front destabilization at ~0.75~ns, the true maximum fluctuates over some near-axis positions.

\section{Summary and outlook}

\subsection{Summary}

We have tested four 3D models for negative streamers in air without photo-ionization in overvolted gaps, and we have found a clear advantage for the hybrid model.
It offers a fast and accurate method to model streamers in cases when rare events are significant, like electron run-away or like ionization avalanches created by single electrons.
Simulations with the hybrid model run faster than with the particle model, with computing times approaching those of the fluid models, and they do not suffer from super-particle artifacts.
The extended fluid model is a good approximation up to the moment of front destabilization, but lacks realistic fluctuations.
In the classical fluid model propagation speeds and ionization densities and field enhancement are too low.

We have studied short negative streamers in high fields, because this allows us to run physically meaningful 3D simulations with all models, without the need to introduce (adaptive) grid refinement; at this moment, adaptive grid refinement is only available in our fluid model~\cite{Luque2010/JCPprep}, but introducing it across the model boundary in the hybrid model still poses a major challenge.
This forced us to exclude photo-ionization from the model for reasons discussed below in the outlook, and in that sense our present results directly apply to discharges in gases like high purity nitrogen~\cite{Nijdam2010/JPD,Wormeester2010/JPD} or like the atmosphere of Venus~\cite{Daria} where photo-ionization is very weak.

Our study lays a basis for future quantitative investigations of electron run-away from streamers and streamer branching, based on the methodology of analysis and numerical models presented here.

\subsection{Outlook on physical implications}

As the presented investigations do not include photo-ionization, % and apply to high purity nitrogen or to the Venusian atmosphere when modified to the somewhat different cross sections for these gases.
the question rises how they change when the nonlocal photo-ionization mechanism is included for realistic models in atmospheric air.
Basically this depends on whether the background field is below or above the breakdown field, i.e., whether the gap is undervolted or overvolted.

Now our choice of system parameters was constraint by the fact that we had to resolve both the space charge structure of the ionization front and the complete system with developed streamer in 3D without grid refinement.
Therefore we chose a high background field to initiate the streamer sufficiently rapidly.
This field is well above the break-down value where impact ionization balances electron attachment, i.e., the gap is overvolted.
A consequence is that results will change greatly, if photo-ionization is included: whereever the nonlocal photo-ionization mechanism creates a new electron ion pair, a new local ionization avalanche will appear.
The emerging structure was demonstrated with the 3D hybrid model in Fig.~1 of ~\cite{Li2011/IEEE} and in Fig.~14 of~\cite{Li2011:JCP}, and with Luque's density fluctuation model in~Fig.~2 of~\cite{Luque2011}.
The figures show how eventually the whole space is filled with new avalanches at stochastic locations, suppressing the field enhancement at the streamer tip.
Depending on initial conditions, eventually the whole region above the breakdown field will be filled with plasma.
This prediction is completely in agreement with experiments where the high field region around a needle electrode fills up with an ionization cloud that breaks up into streamers only beyond some critical radius~\cite{Nijdam11,Nijdam12}.
(Of course, this behavior also depends on the initial electron density distribution while streamer development is rather insensitive to this distribution in undervolted gaps~\cite{Luque2008:3}.)

%Particle and hybrid model contain accurate distributions and fluctuations of electron energies and densities, up to super-particle effects. They are the appropriate starting point for investigating electron run-away and streamer branching.

The onset of branching of positive streamers in air with photo-ionization in an undervolted gap was recently investigated in~\cite{Luque2011} with a model accounting specifically for density fluctuations.
There it was found that the density fluctuations that are due to the discrete nature of electrons, accelerate streamer branching.
Similarly, the present results show that the fluctuations of electron densities and energies destabilize the ionization front earlier than without fluctuations, but this does not need to create permanent branching in negative streamers.
These observations open up many questions: Are negative streamers more self-stabilizing than positive ones as various experiments also seem to indicate? Which differences are caused by over- or undervolted gaps or by the presence or absence of photo-ionization, next to the polarity differences? Which role is played by run-away electrons?
Our study lays a methodological basis for future studies of these questions.

\ack{} C.L. acknowledges support by STW-project 10118, and J.T.\ by STW-project 10755, STW being part of the Netherlands' Organization for Scientific Research NWO.

%%\begin{thebibliography}{00}
%\bibliography{lc_thesis}
%%\bibliography{Everything}
%%\bibliographystyle{elsarticle-num-names}
%%\end{thebibliography}

\newpage

% FIG. 1

% FIG. 2

% figure 3

% FIG. 4

\begin{figure}
\centering
\includegraphics[width=.12\textwidth,viewport=115 20 240 400, clip]{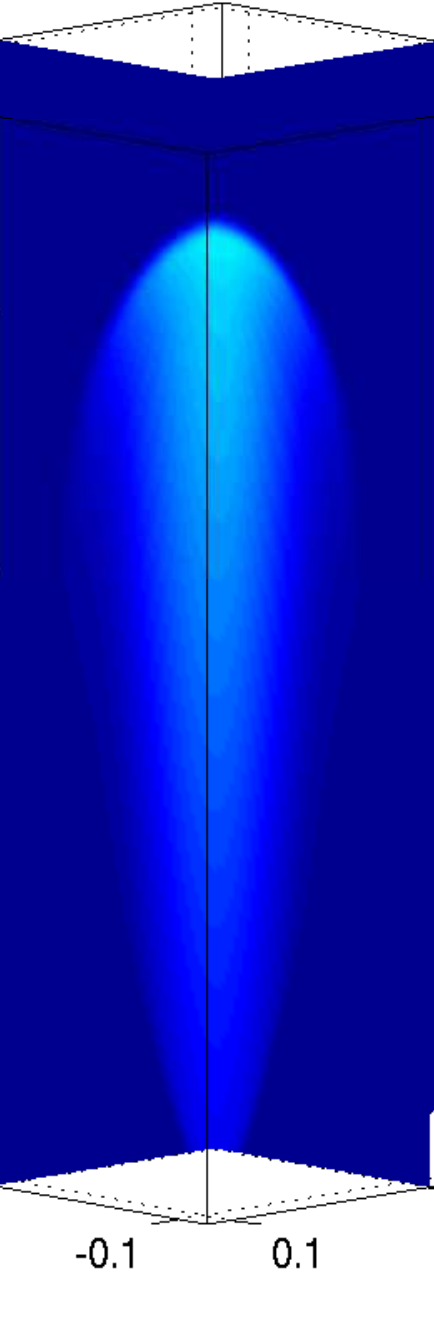}
\includegraphics[width=.12\textwidth,viewport=115 20 240 400, clip]{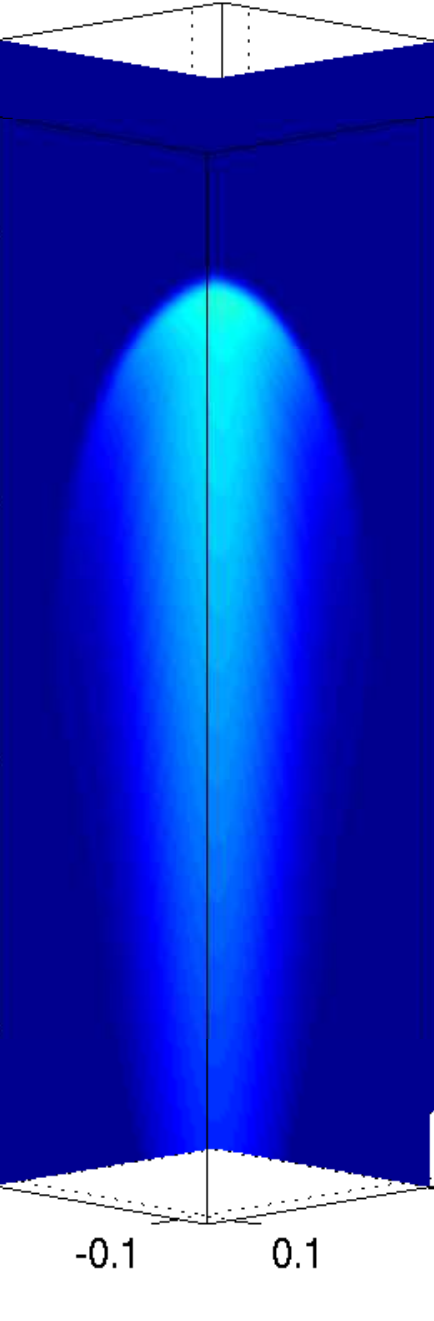}
\includegraphics[width=.12\textwidth,viewport=115 20 240 400, clip]{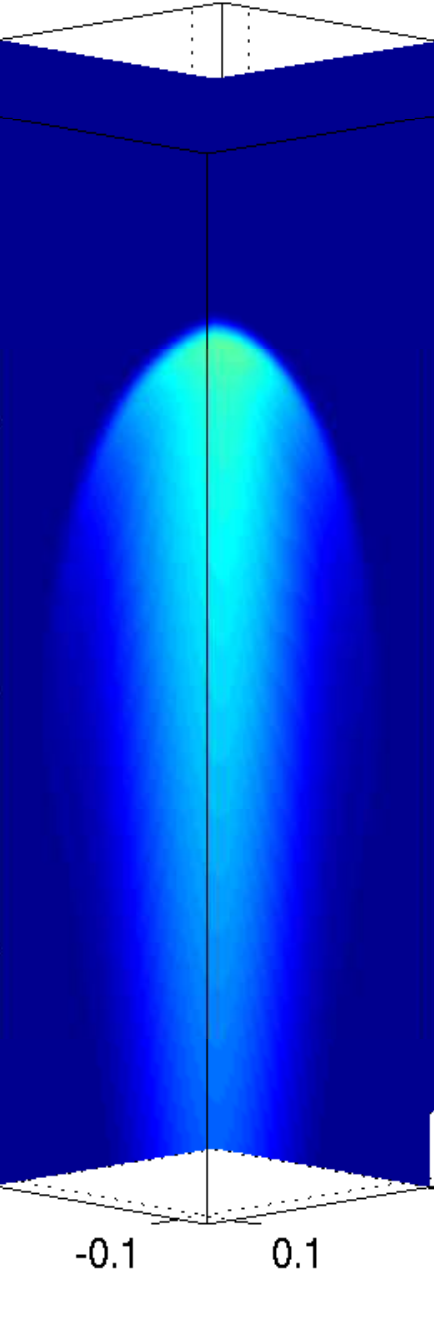}
\includegraphics[width=.12\textwidth,viewport=115 20 240 400, clip]{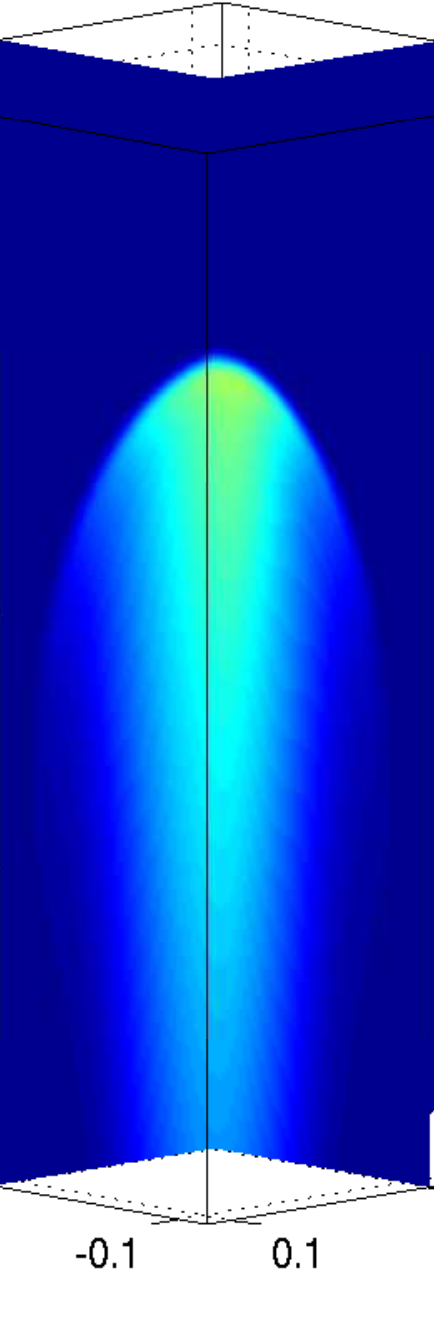}
\includegraphics[width=.12\textwidth,viewport=115 20 240 400, clip]{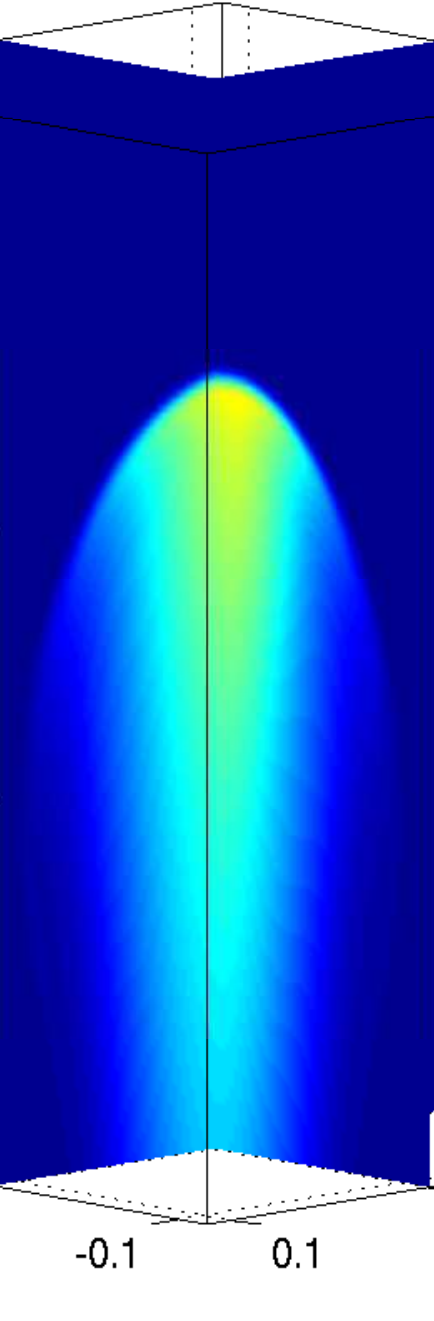}
\includegraphics[width=.12\textwidth,viewport=115 20 240 400, clip]{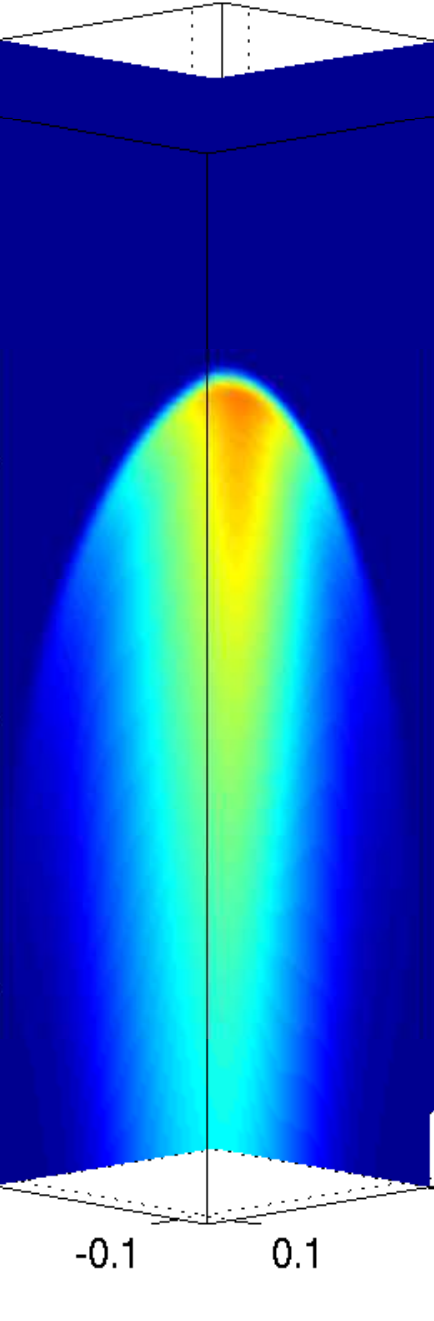}
\includegraphics[width=.12\textwidth,viewport=115 20 240 400, clip]{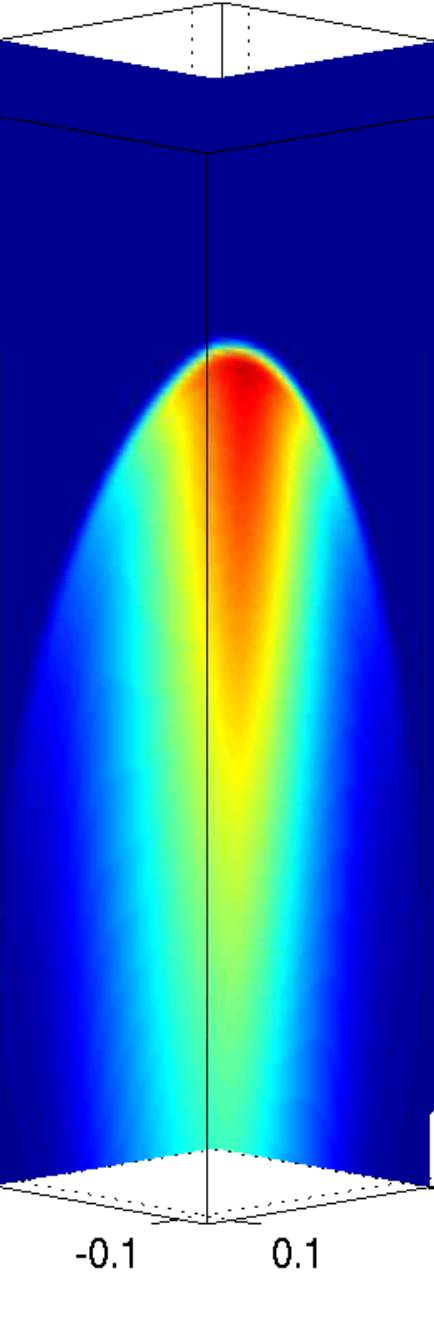}
\includegraphics[width=.04\textwidth,viewport=175 20 300 400, clip]{figures_pdf/fig1_colorbar.pdf} %White space
\\
\includegraphics[width=.12\textwidth,viewport=115 20 240 400, clip]{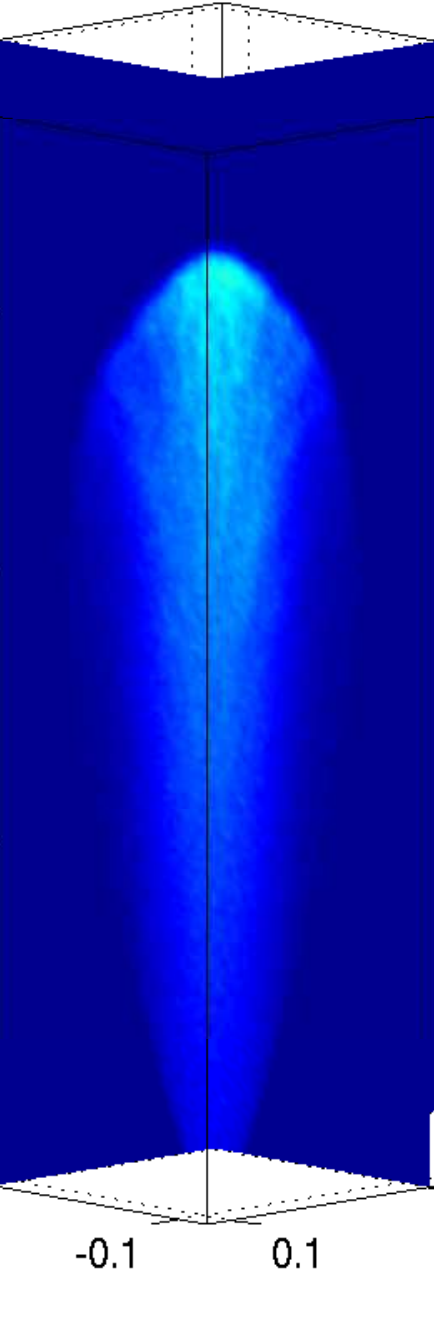}
\includegraphics[width=.12\textwidth,viewport=115 20 240 400, clip]{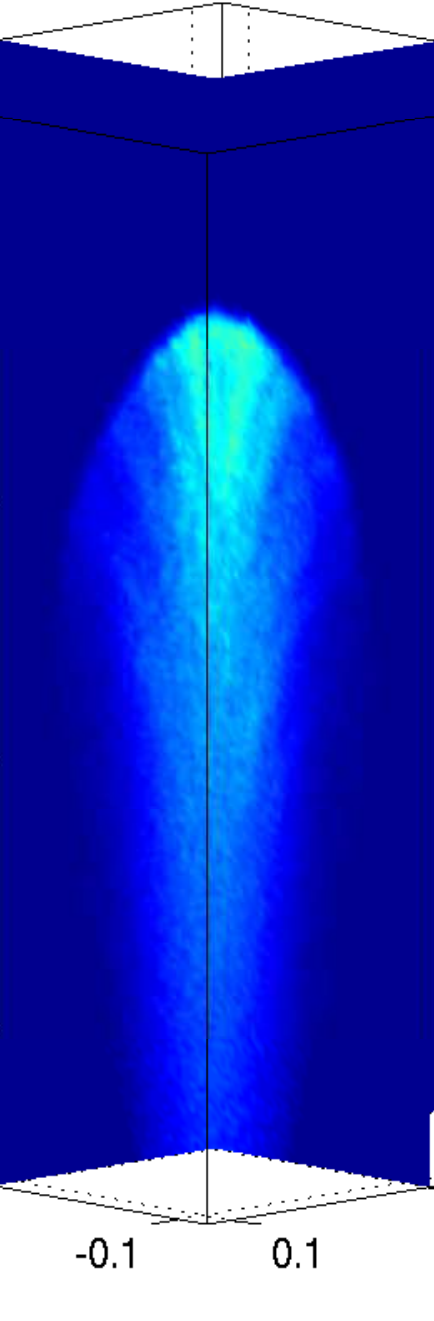}
\includegraphics[width=.12\textwidth,viewport=115 20 240 400, clip]{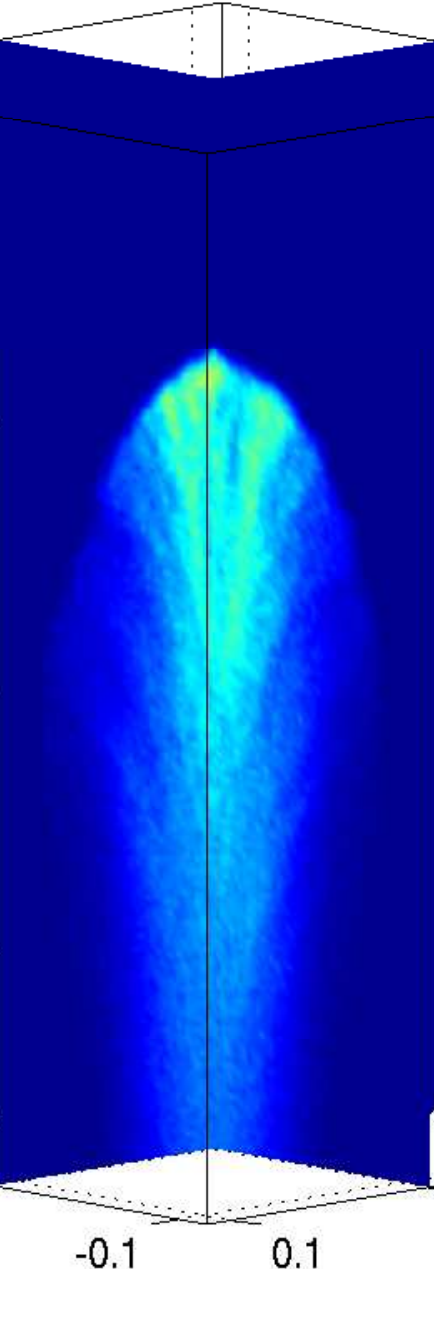}
\includegraphics[width=.12\textwidth,viewport=115 20 240 400, clip]{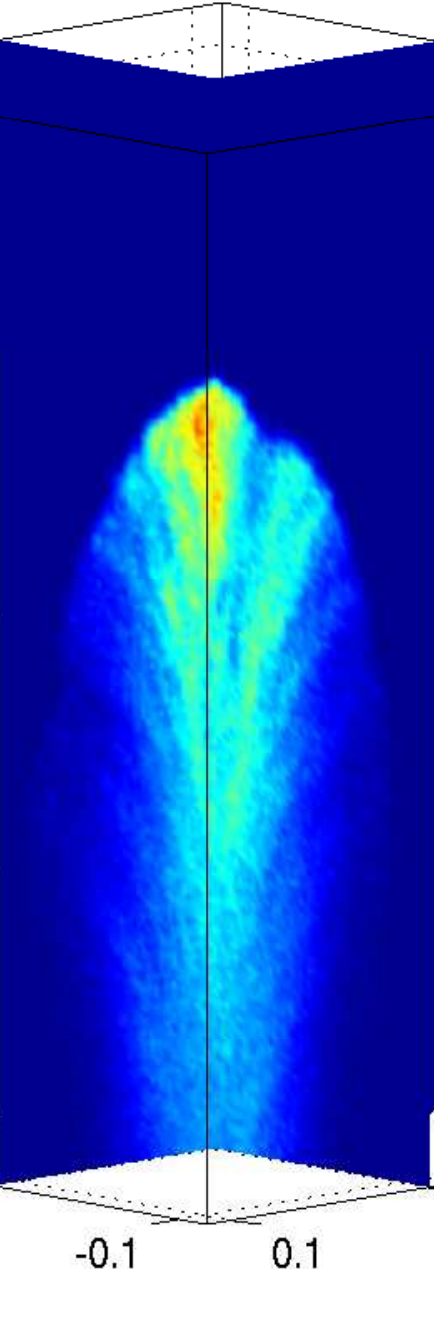}
\includegraphics[width=.12\textwidth,viewport=115 20 240 400, clip]{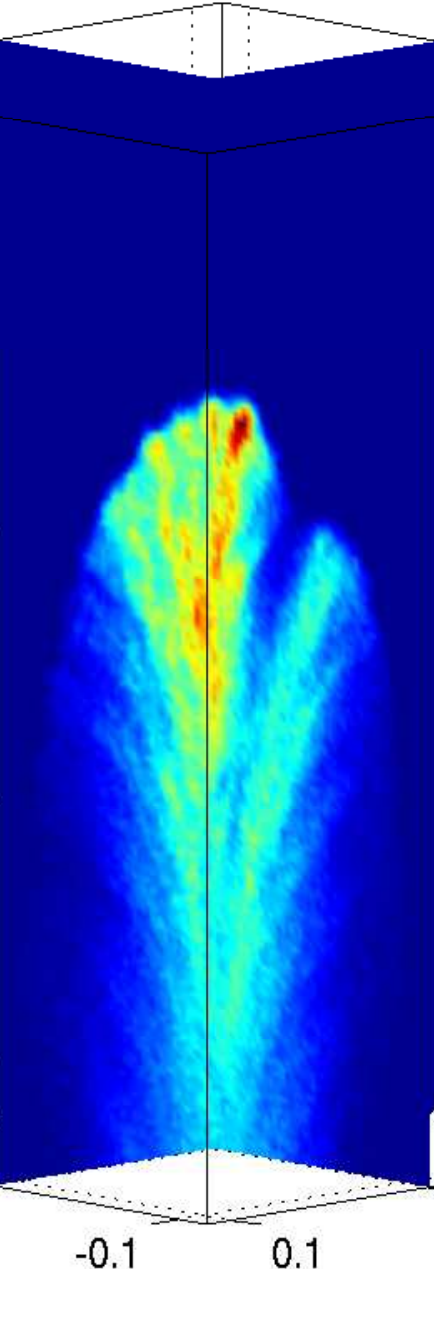}
\includegraphics[width=.12\textwidth,viewport=115 20 240 400, clip]{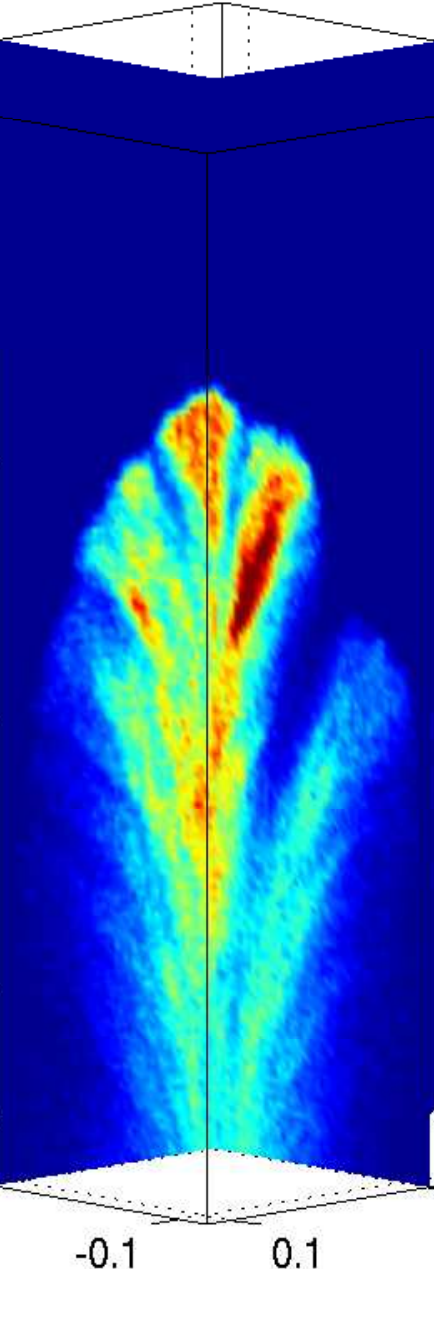}
\includegraphics[width=.12\textwidth,viewport=115 20 240 400, clip]{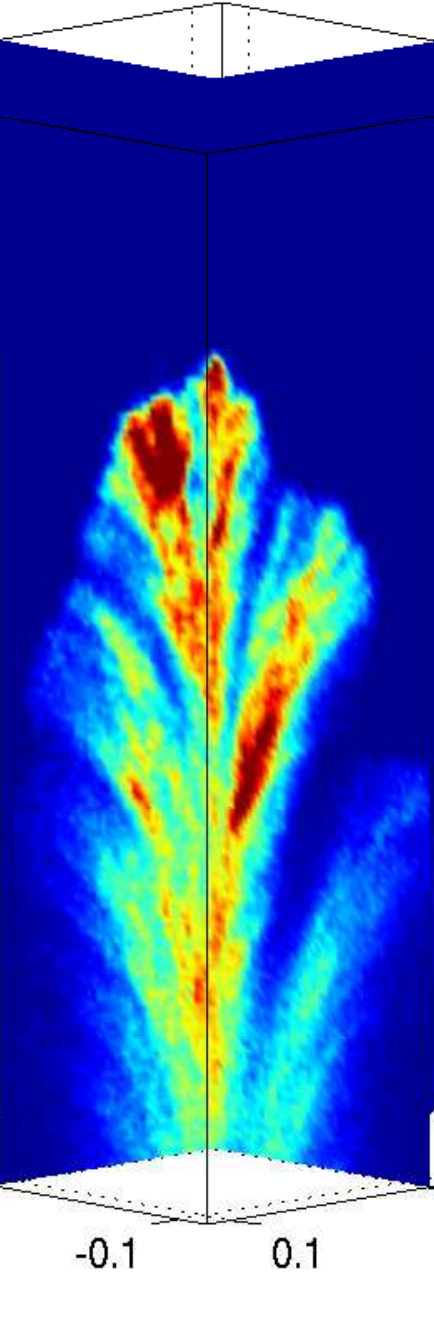}
\includegraphics[width=.04\textwidth,viewport=175 20 300 400, clip]{figures_pdf/fig1_colorbar.pdf} %White space
\\
\includegraphics[width=.12\textwidth,viewport=115 20 240 400, clip]{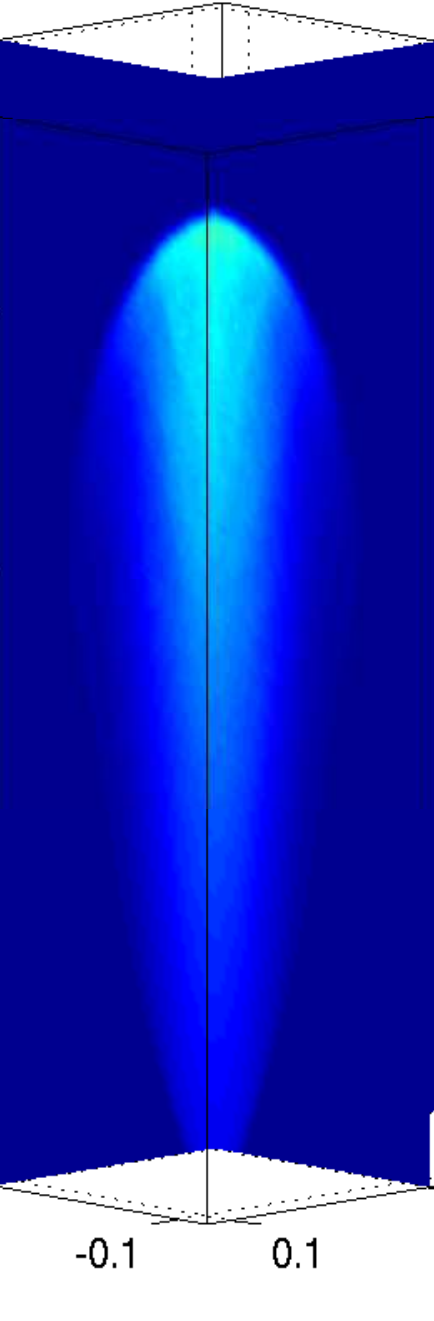}
\includegraphics[width=.12\textwidth,viewport=115 20 240 400, clip]{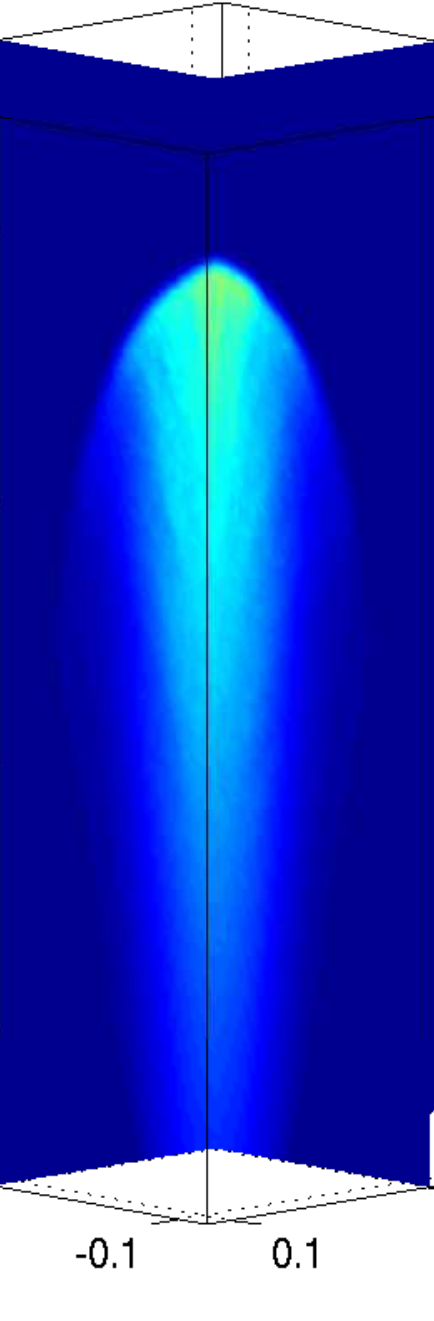}
\includegraphics[width=.12\textwidth,viewport=115 20 240 400, clip]{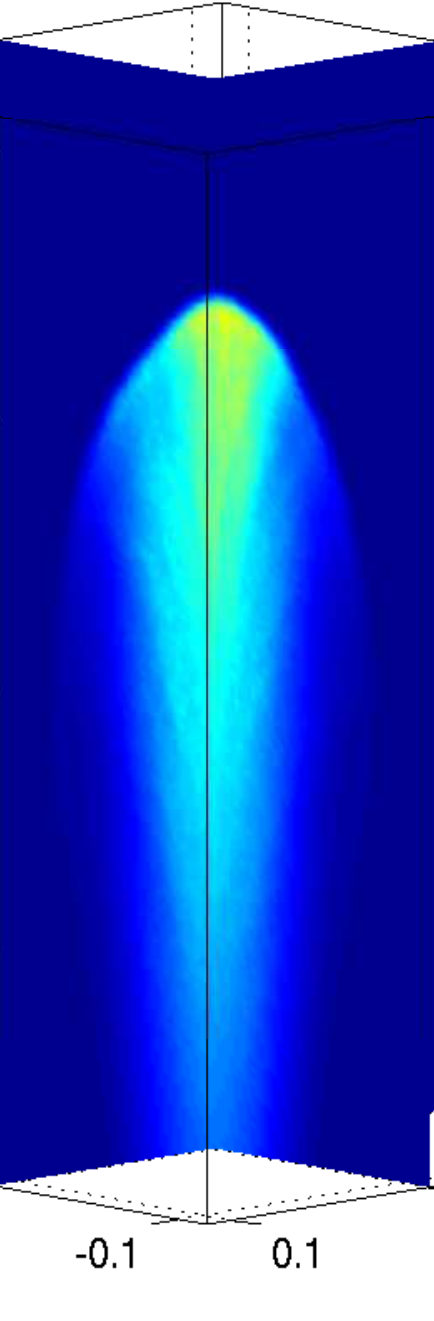}
\includegraphics[width=.12\textwidth,viewport=115 20 240 400, clip]{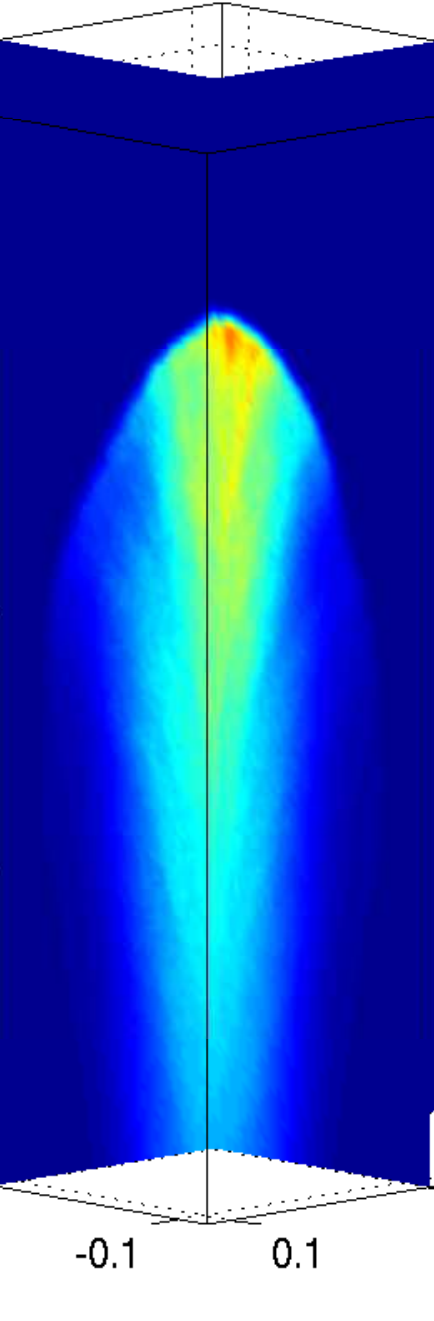}
\includegraphics[width=.12\textwidth,viewport=115 20 240 400, clip]{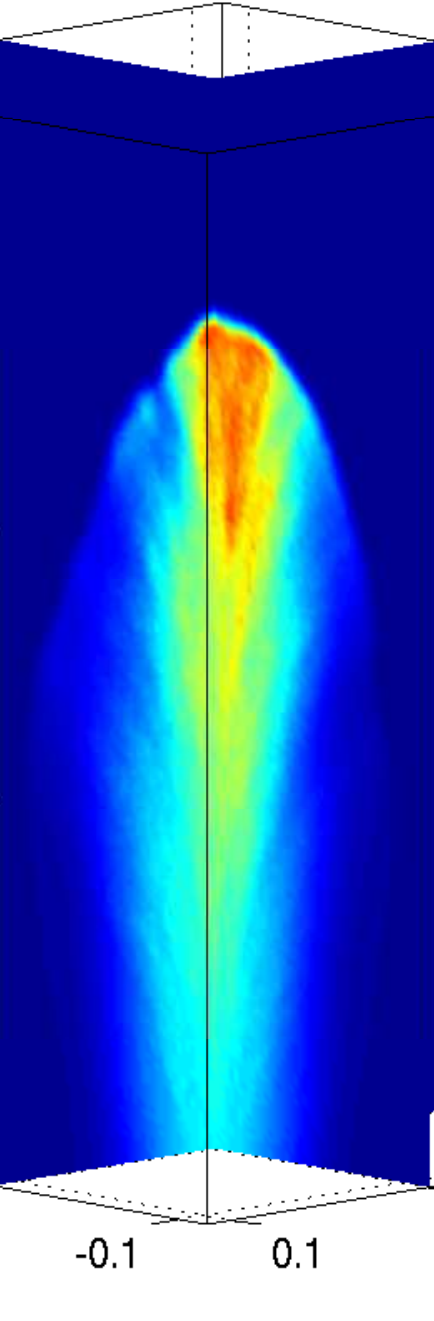}
\includegraphics[width=.12\textwidth,viewport=115 20 240 400, clip]{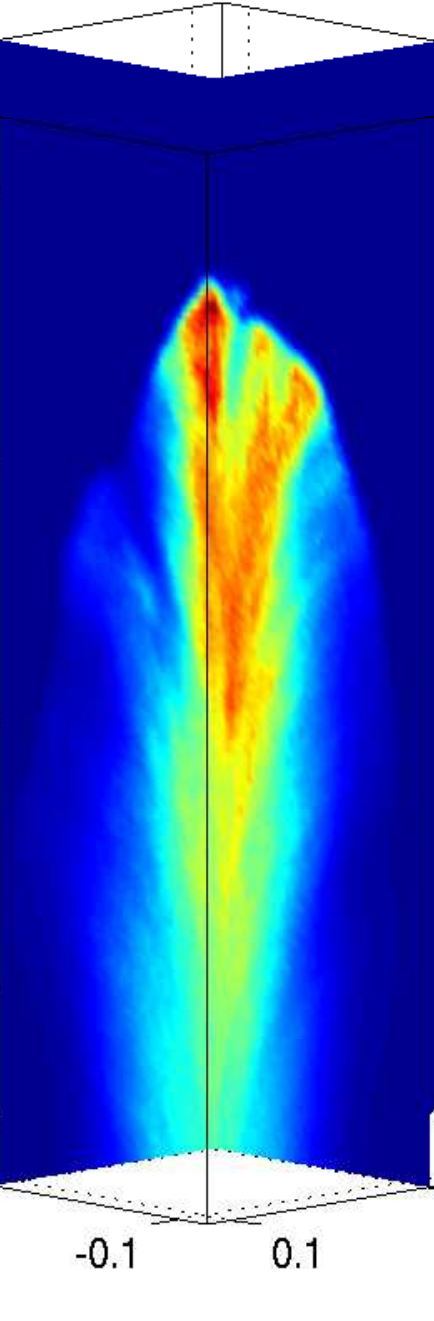}
\includegraphics[width=.12\textwidth,viewport=115 20 240 400, clip]{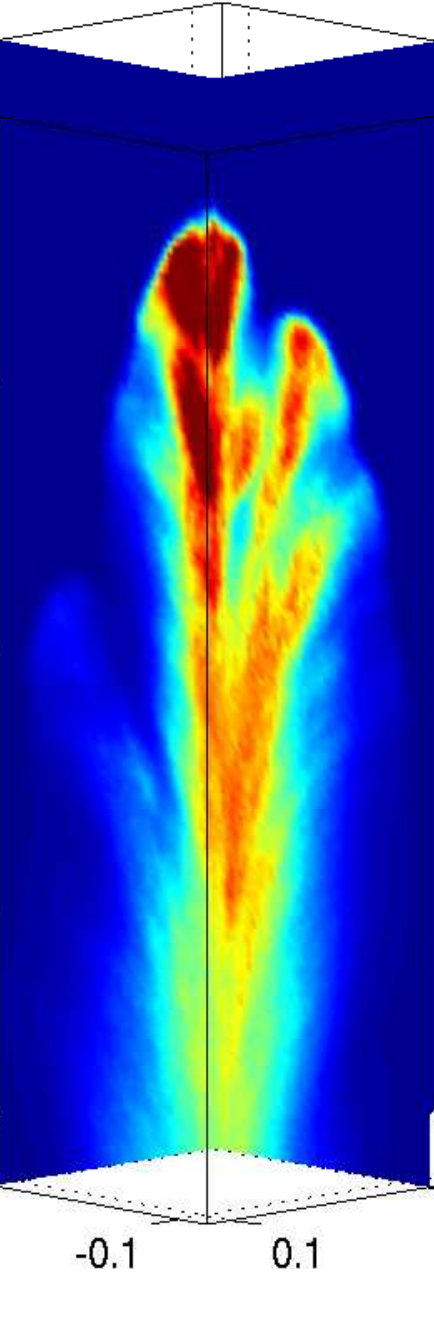}
\includegraphics[width=.04\textwidth,viewport=390 0 440 432, clip]{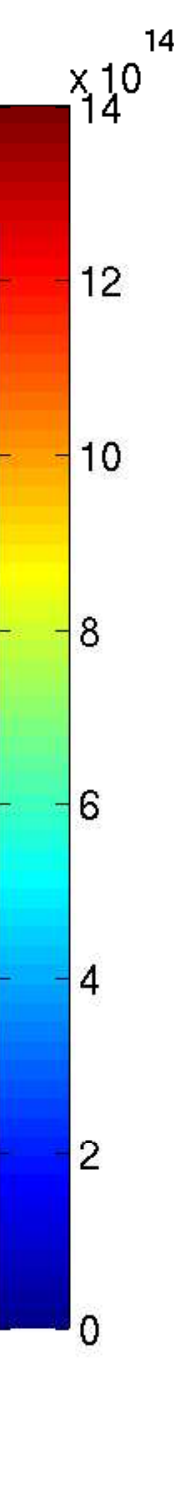}

\caption{The same electron density as in figure~\ref{fig:elecdens1}, but now zoomed into the region where the streamer grows. The classical fluid model is not shown, and the rows show extended fluid (first row), particle model (second row) and hybrid model (third row). The time steps are the same as in figure~\ref{fig:elecdens1}, but now at the first time step of 0.72~ns, only the interval from 0.3 to 0.7~mm on the vertical axis is shown. The spatial interval shifts upwards with 0.07~mm per time step of 0.03~ns, until it reaches the interval of 0.72 to 1.12~mm at the last time step of 0.9~ns.
}
\label{fig:elecdenszoom}
\end{figure}

% FIG. 5

\begin{figure}
\centering
\includegraphics[width=.12\textwidth,viewport=115 20 240 400, clip]{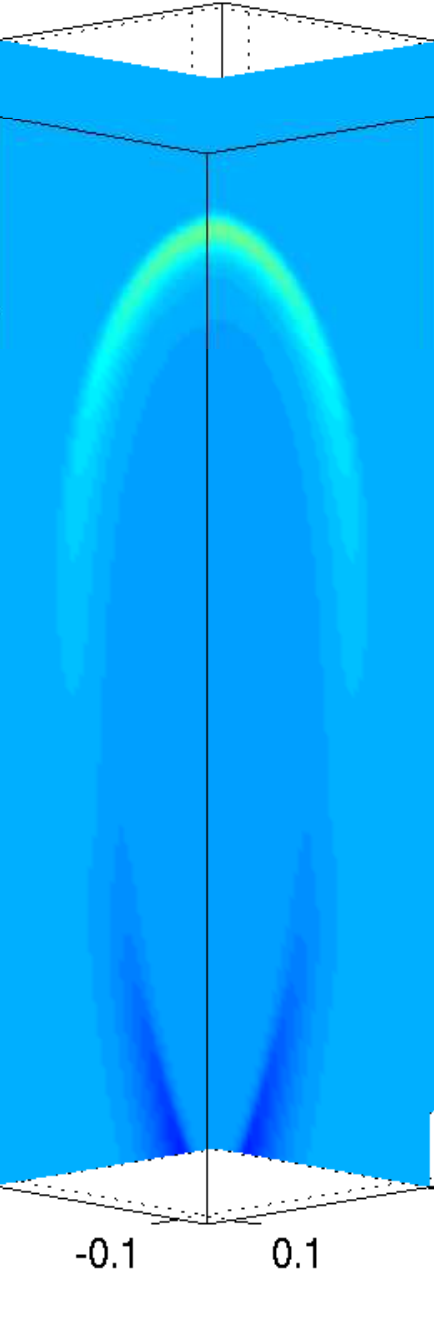}
\includegraphics[width=.12\textwidth,viewport=115 20 240 400, clip]{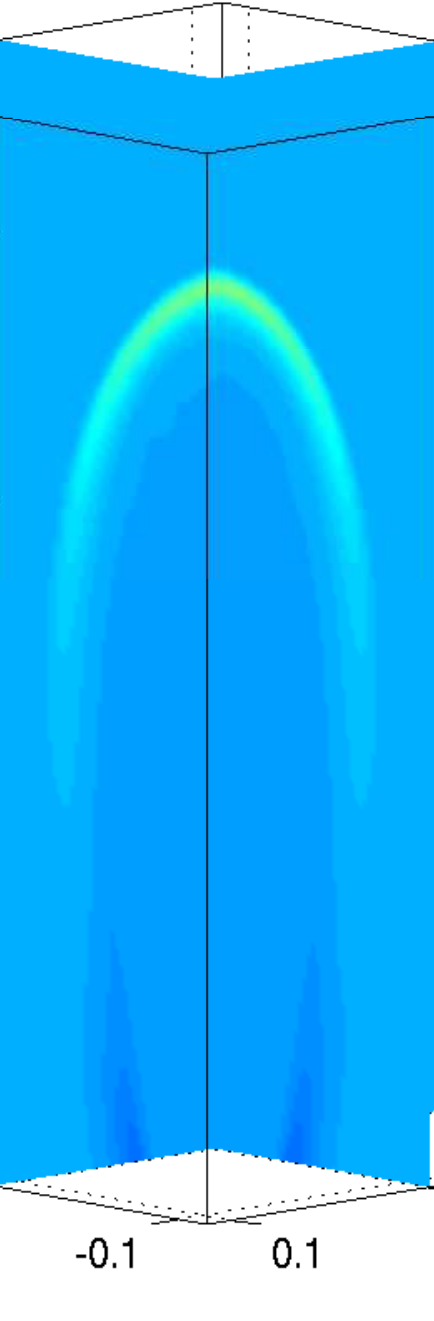}
\includegraphics[width=.12\textwidth,viewport=115 20 240 400, clip]{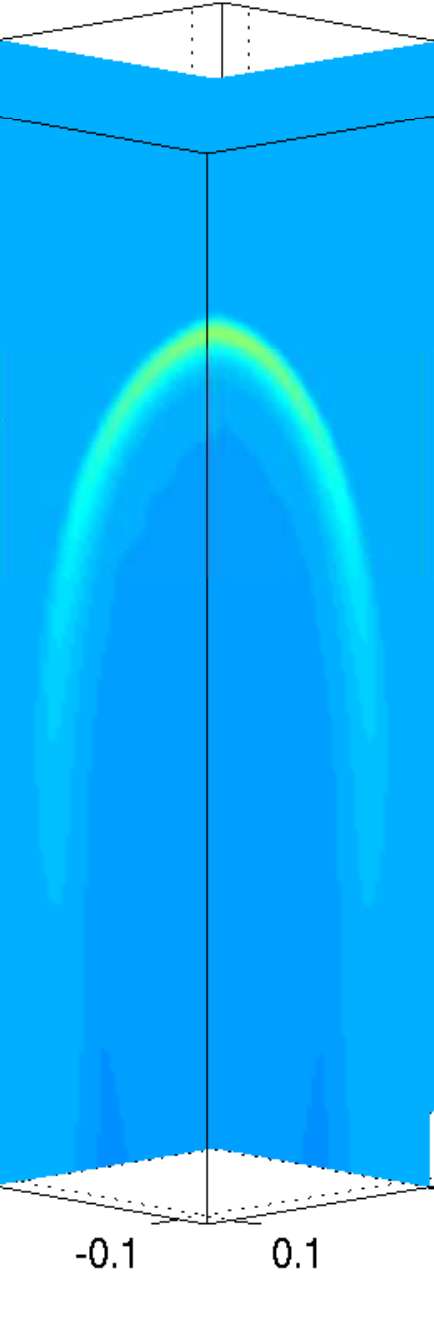}
\includegraphics[width=.12\textwidth,viewport=115 20 240 400, clip]{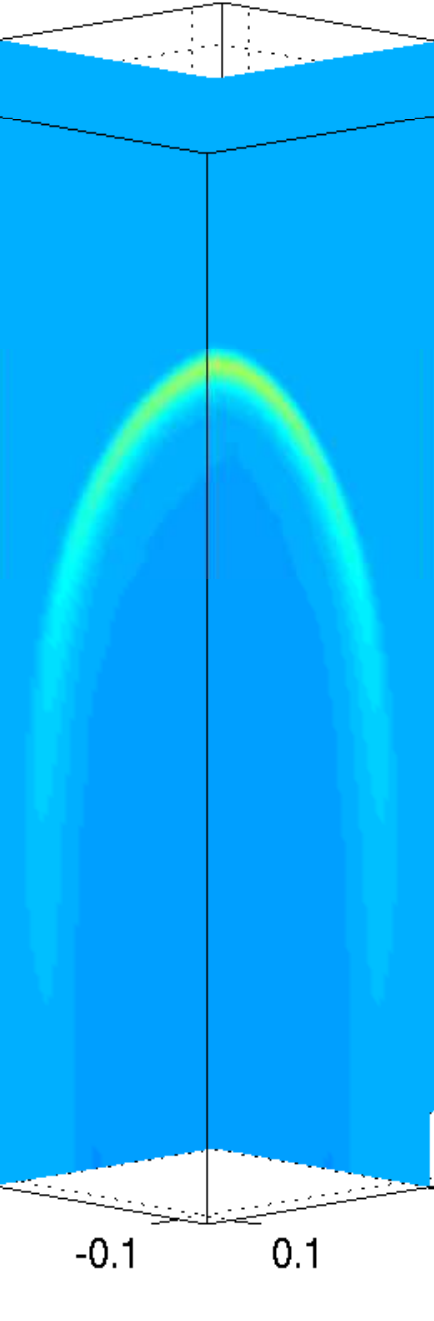}
\includegraphics[width=.12\textwidth,viewport=115 20 240 400, clip]{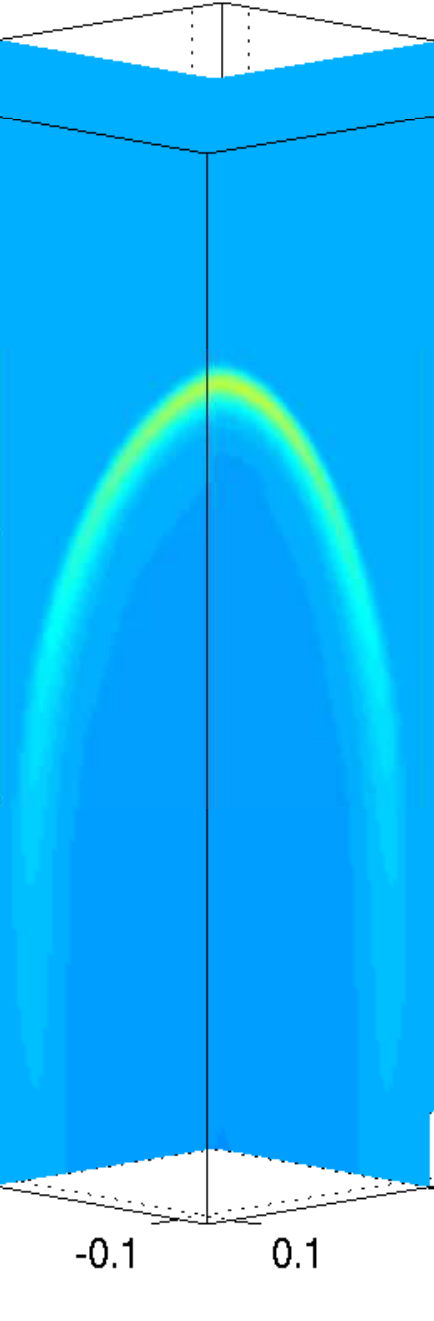}
\includegraphics[width=.12\textwidth,viewport=115 20 240 400, clip]{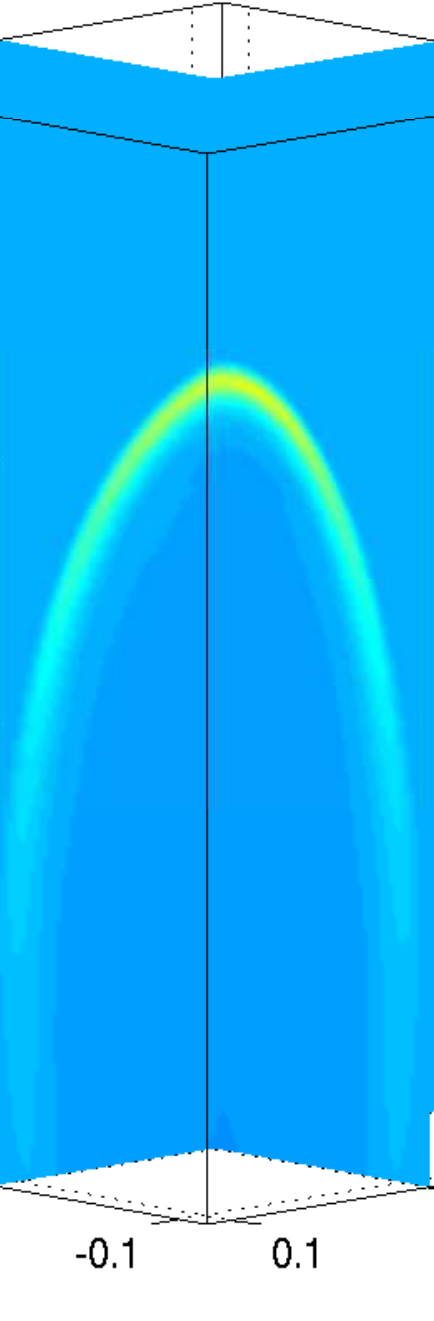}
\includegraphics[width=.12\textwidth,viewport=115 20 240 400, clip]{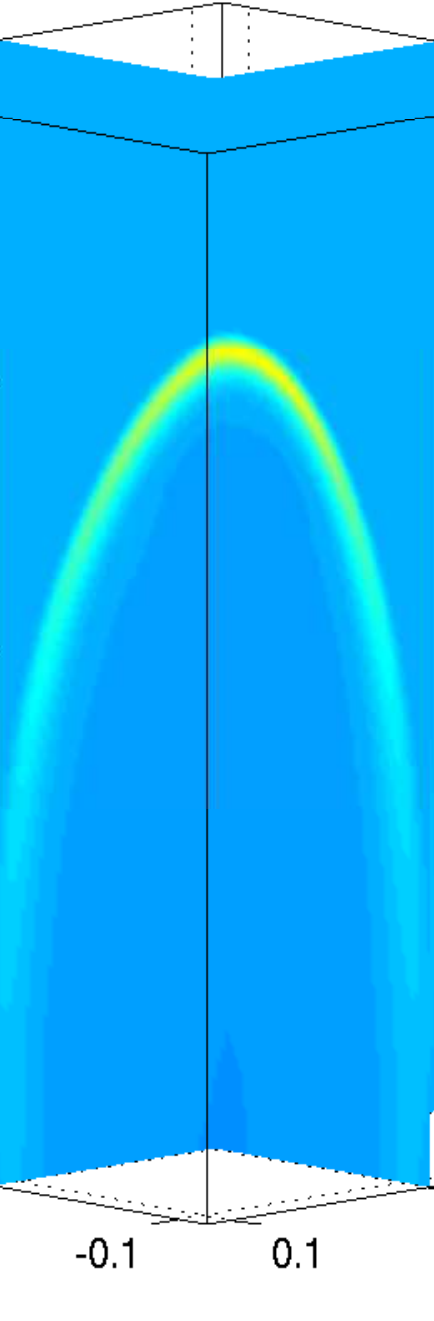}
\includegraphics[width=.04\textwidth,viewport=175 20 300 400, clip]{figures_pdf/fig1_colorbar.pdf} %White space
\\
\includegraphics[width=.12\textwidth,viewport=115 20 240 400, clip]{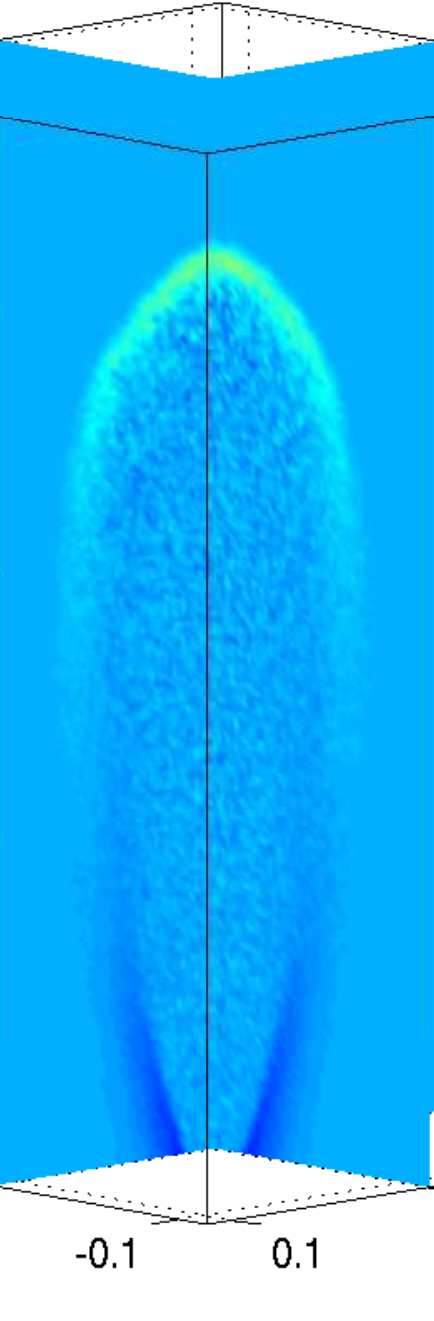}
\includegraphics[width=.12\textwidth,viewport=115 20 240 400, clip]{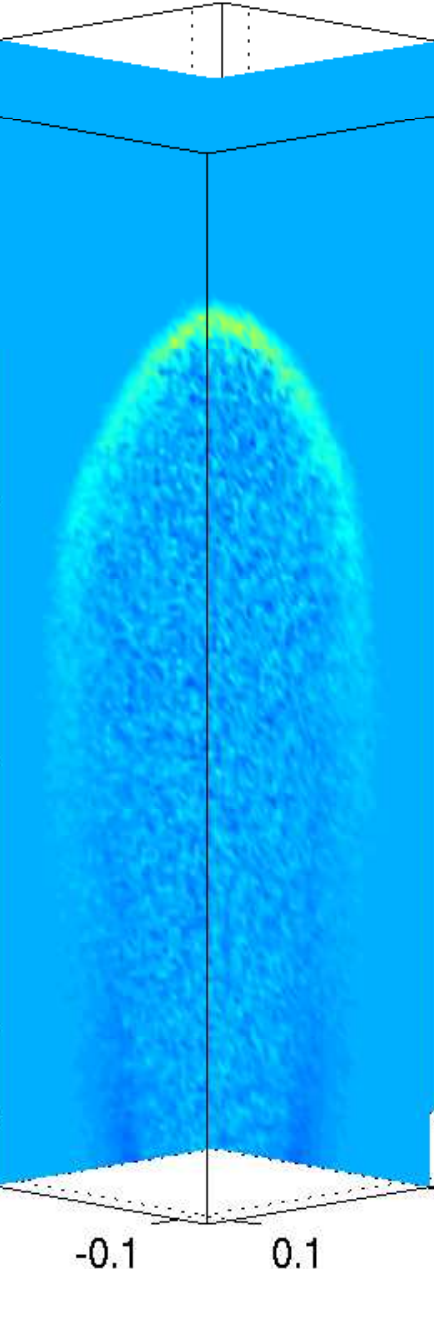}
\includegraphics[width=.12\textwidth,viewport=115 20 240 400, clip]{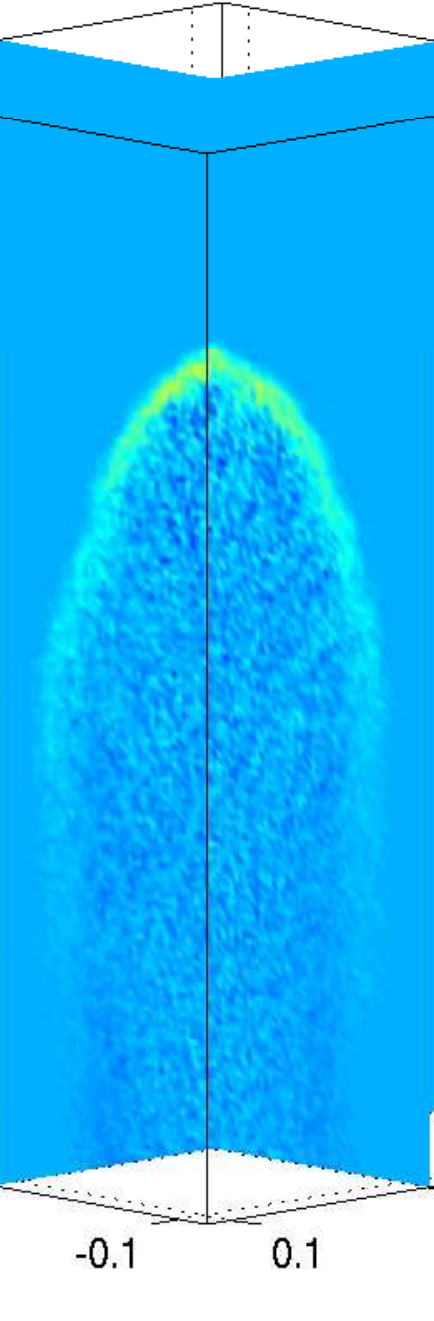}
\includegraphics[width=.12\textwidth,viewport=115 20 240 400, clip]{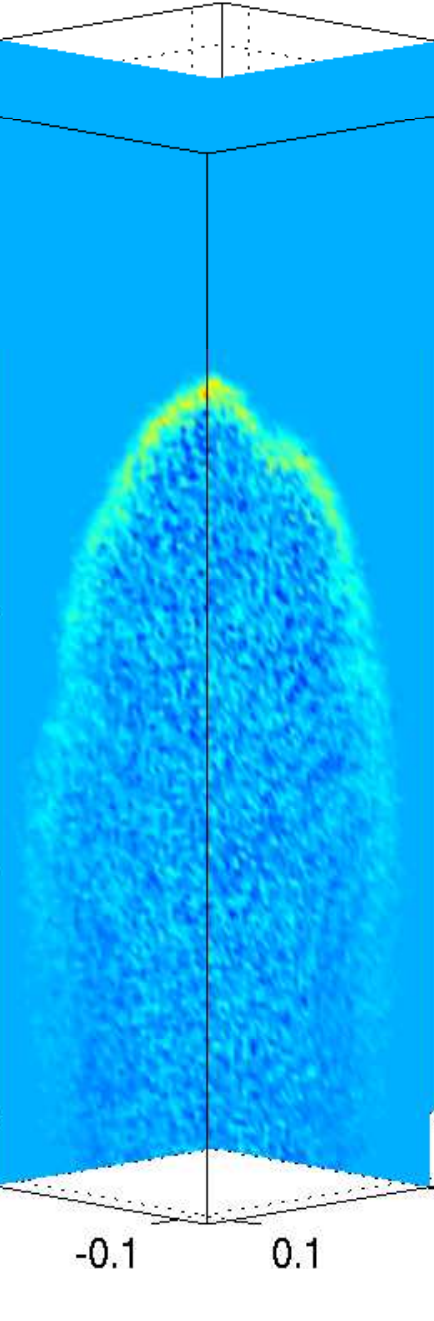}
\includegraphics[width=.12\textwidth,viewport=115 20 240 400, clip]{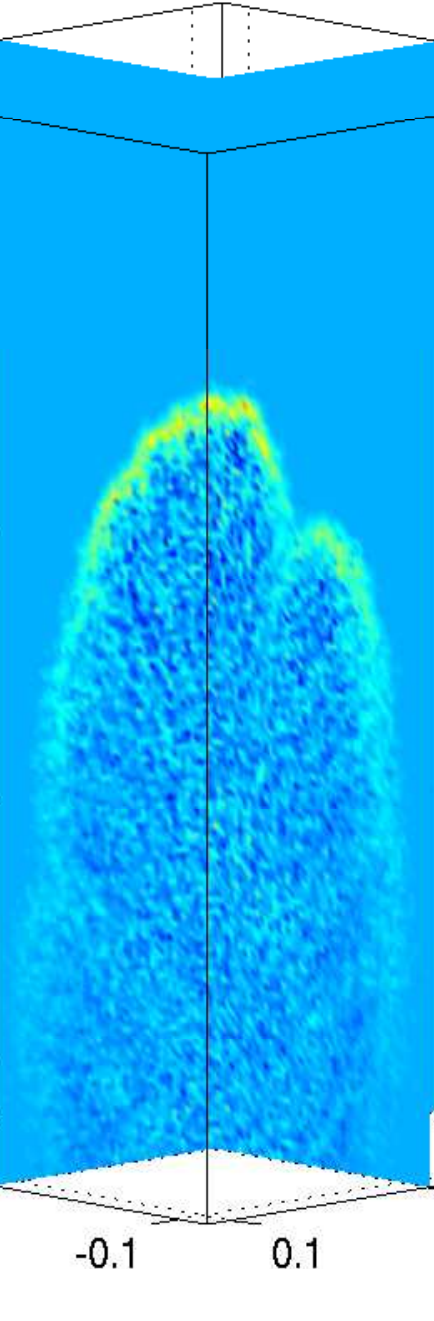}
\includegraphics[width=.12\textwidth,viewport=115 20 240 400, clip]{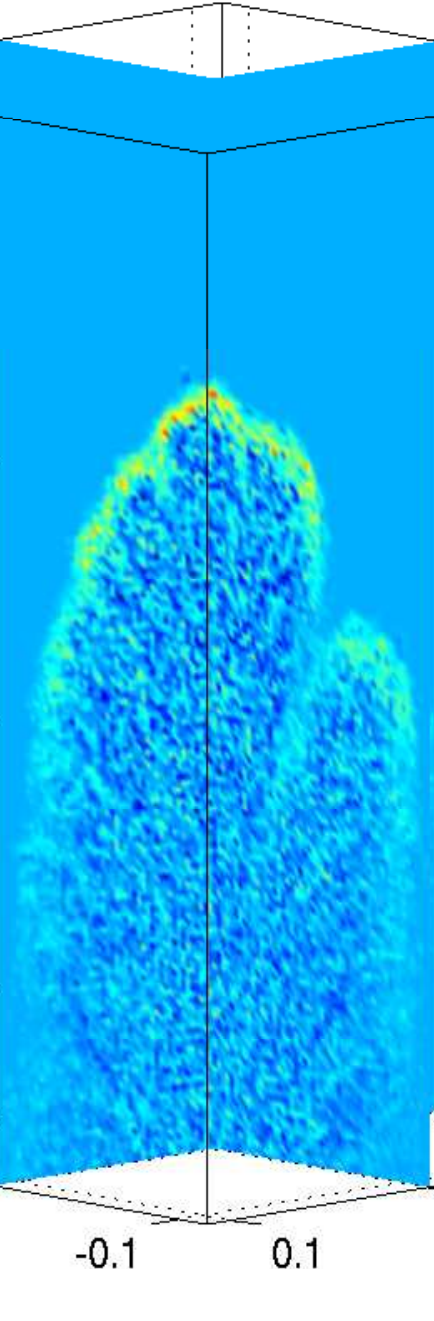}
\includegraphics[width=.12\textwidth,viewport=115 20 240 400, clip]{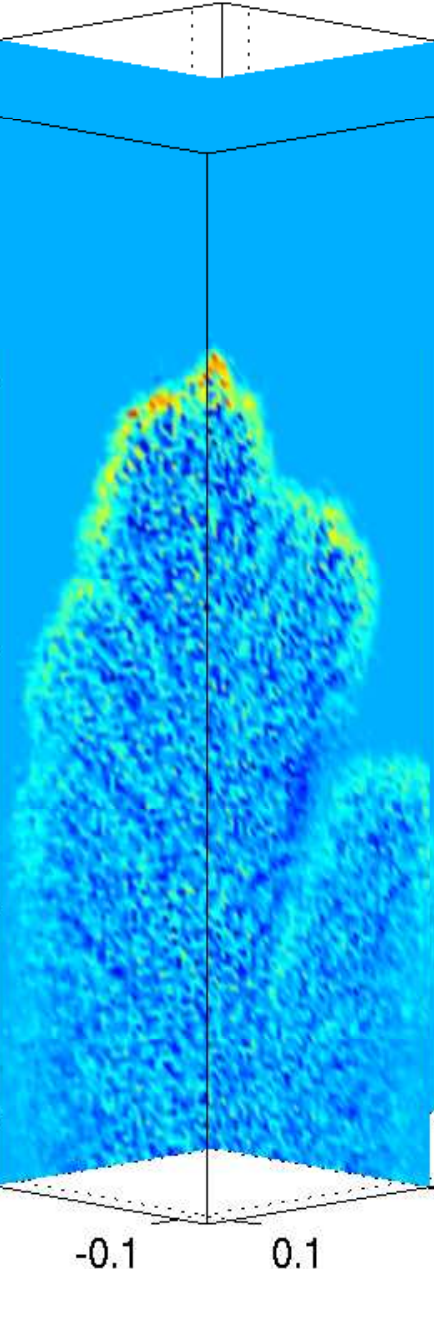}
\includegraphics[width=.04\textwidth,viewport=175 20 300 400, clip]{figures_pdf/fig1_colorbar.pdf} %White space
\\
\includegraphics[width=.12\textwidth,viewport=115 20 240 400, clip]{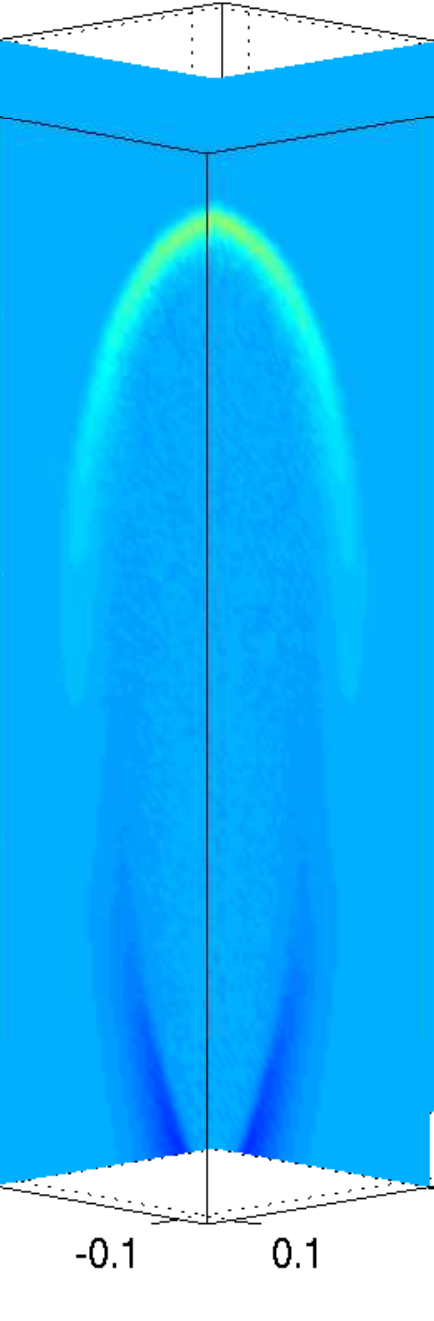}
\includegraphics[width=.12\textwidth,viewport=115 20 240 400, clip]{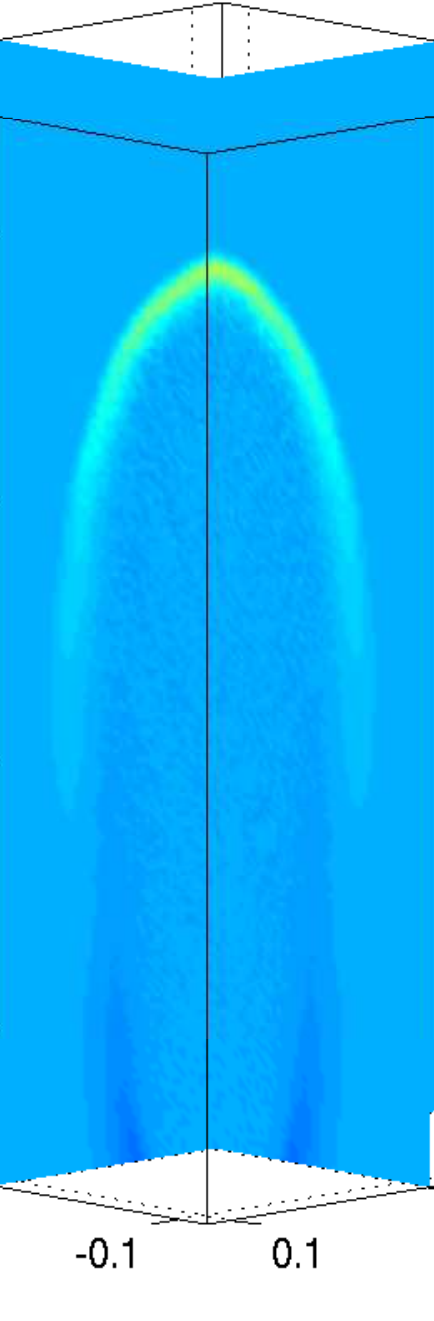}
\includegraphics[width=.12\textwidth,viewport=115 20 240 400, clip]{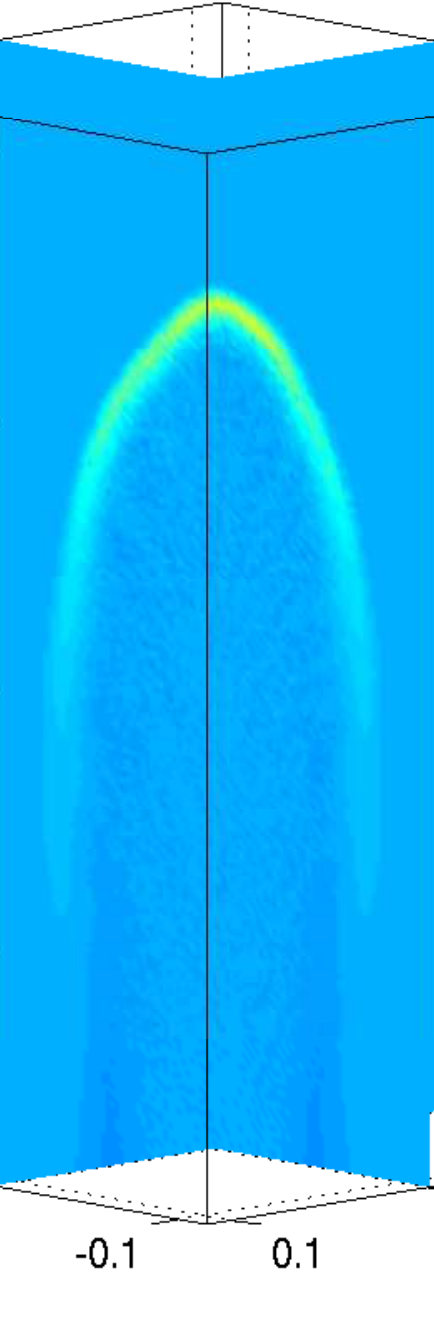}
\includegraphics[width=.12\textwidth,viewport=115 20 240 400, clip]{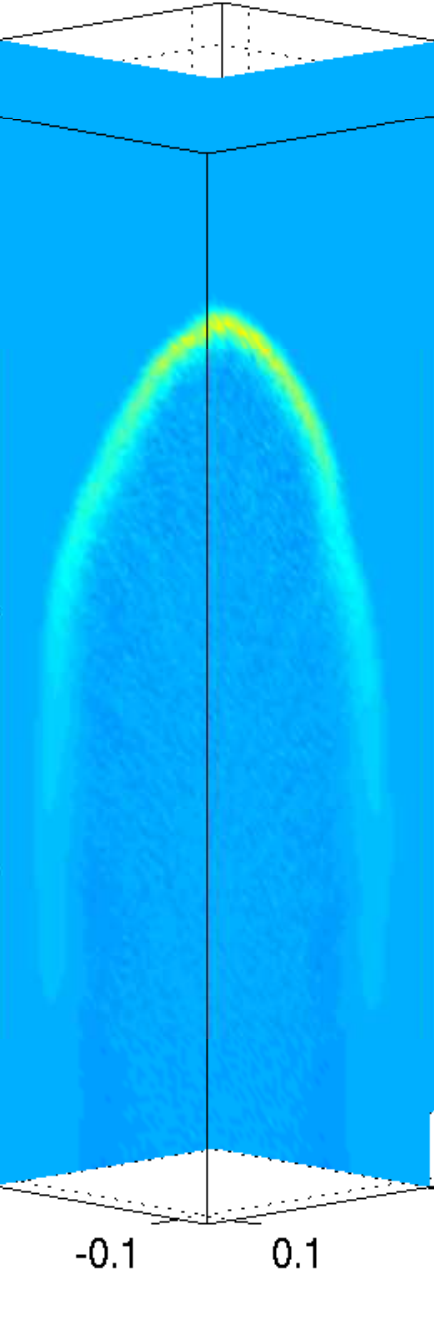}
\includegraphics[width=.12\textwidth,viewport=115 20 240 400, clip]{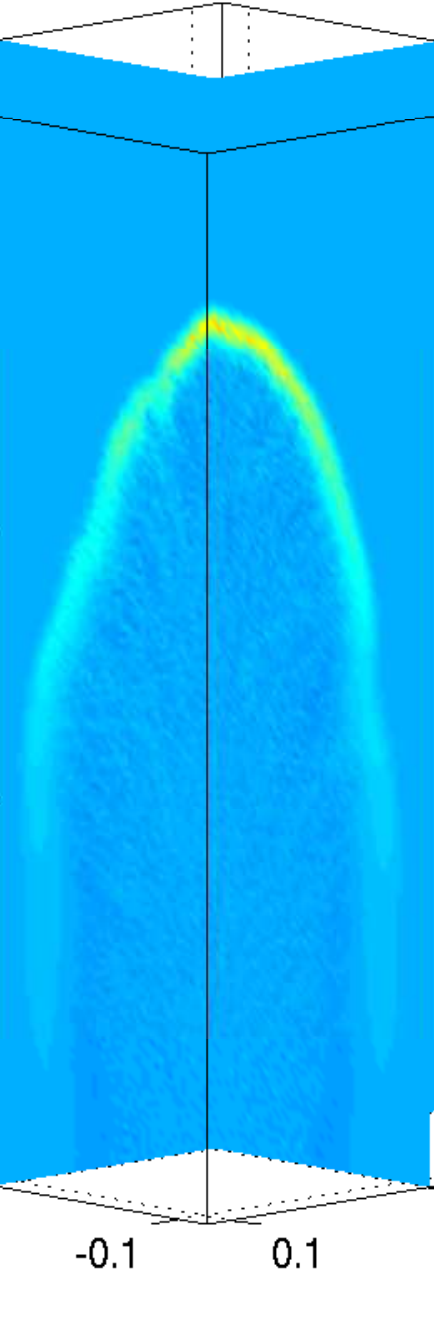}
\includegraphics[width=.12\textwidth,viewport=115 20 240 400, clip]{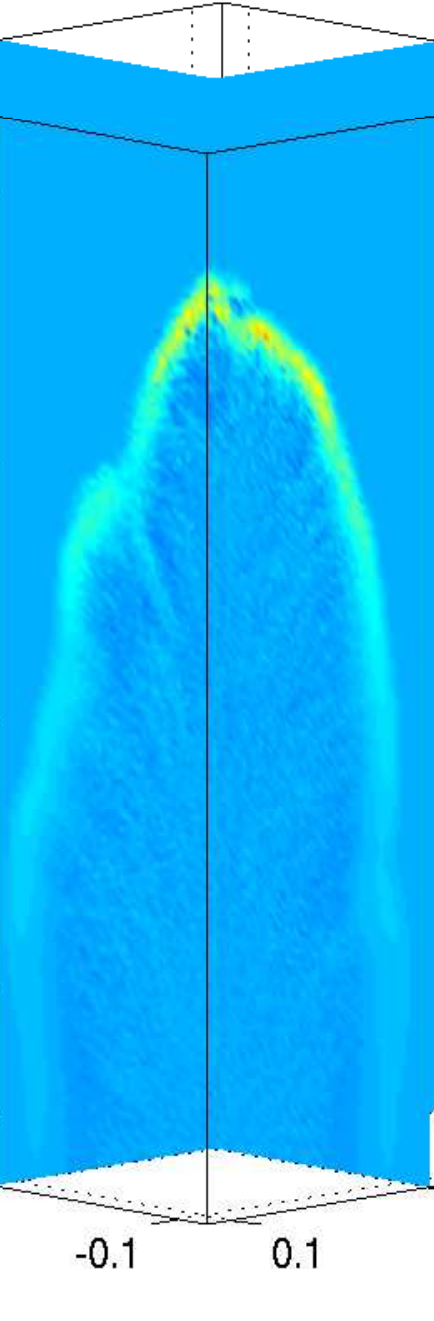}
\includegraphics[width=.12\textwidth,viewport=115 20 240 400, clip]{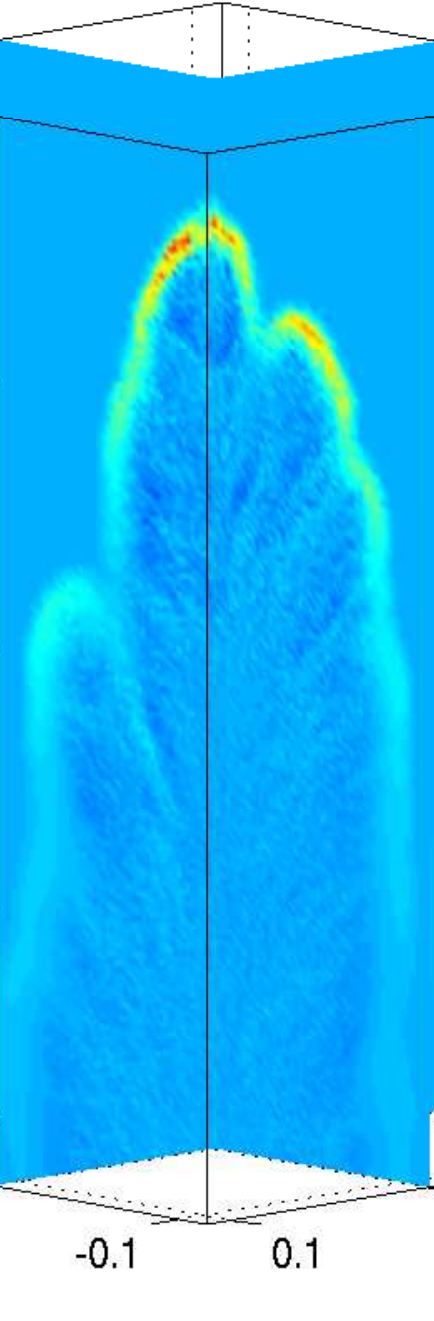}
\includegraphics[width=0.042\textwidth,viewport=390 0 440 432, clip]{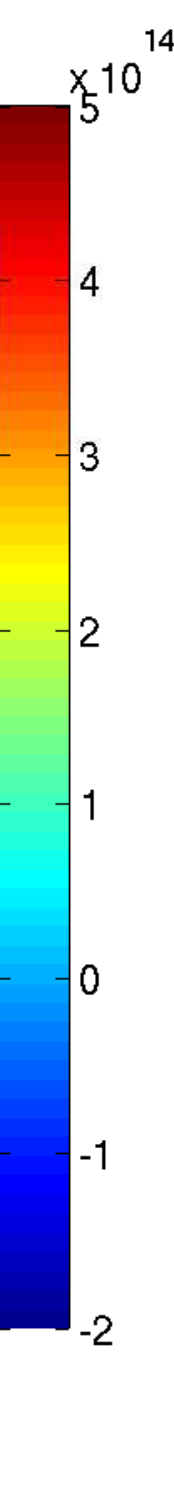}
\caption{The negative charge density in the same models, at the same time steps and with the same spatial zoom as the electron density on the previous figure~\ref{fig:elecdenszoom}. The colours indicate charge densities from $2\cdot10^{14}$ e/cm$^3$ (dark blue) to $-5\cdot10^{14}$ e/cm$^3$ (dark red) where e is elementary charge.
}
\label{fig:chargedens}
\end{figure}

% FIG. 6

\begin{figure}
\centering
\includegraphics[width=.12\textwidth,viewport=115 20 240 400, clip]{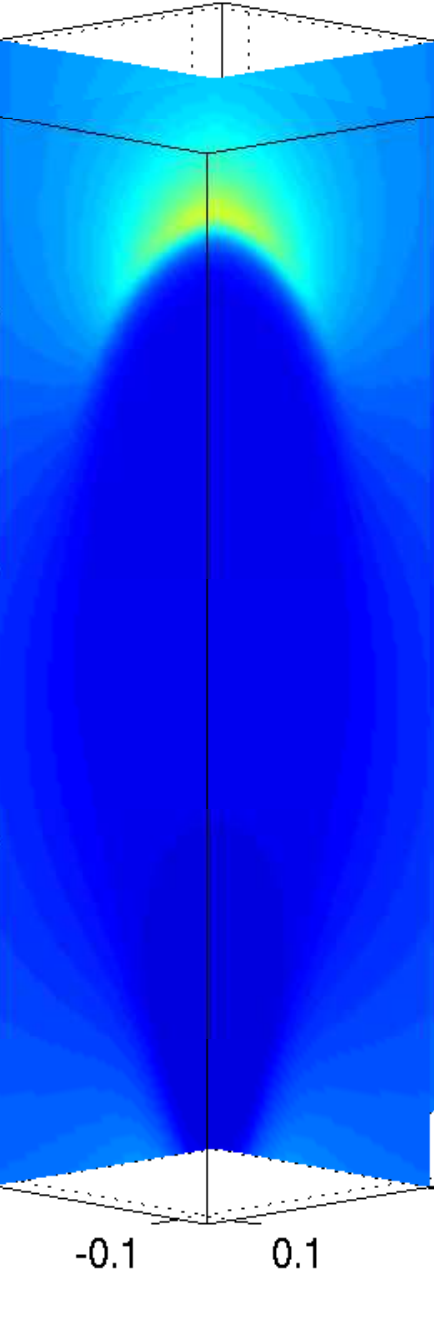}
\includegraphics[width=.12\textwidth,viewport=115 20 240 400, clip]{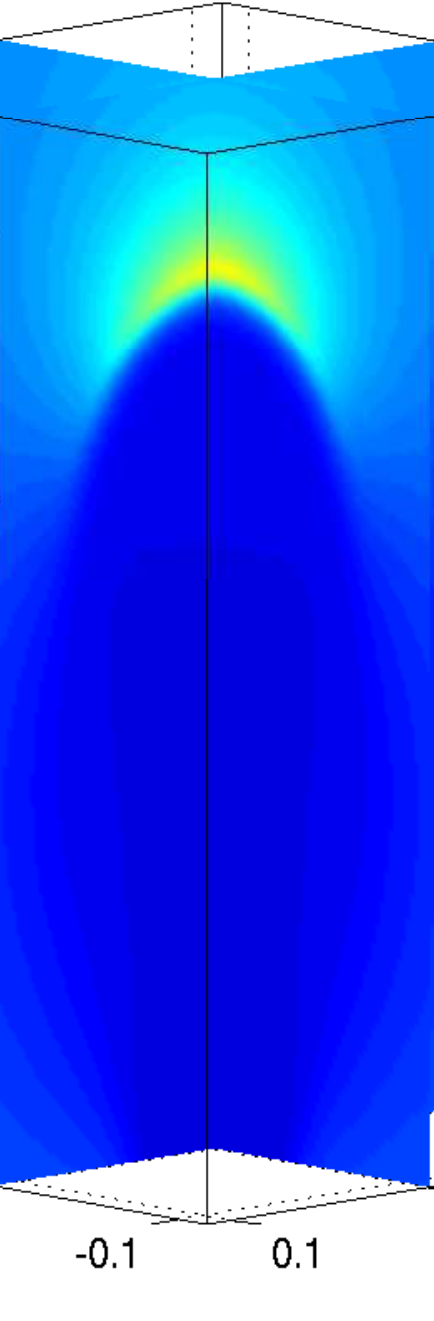}
\includegraphics[width=.12\textwidth,viewport=115 20 240 400, clip]{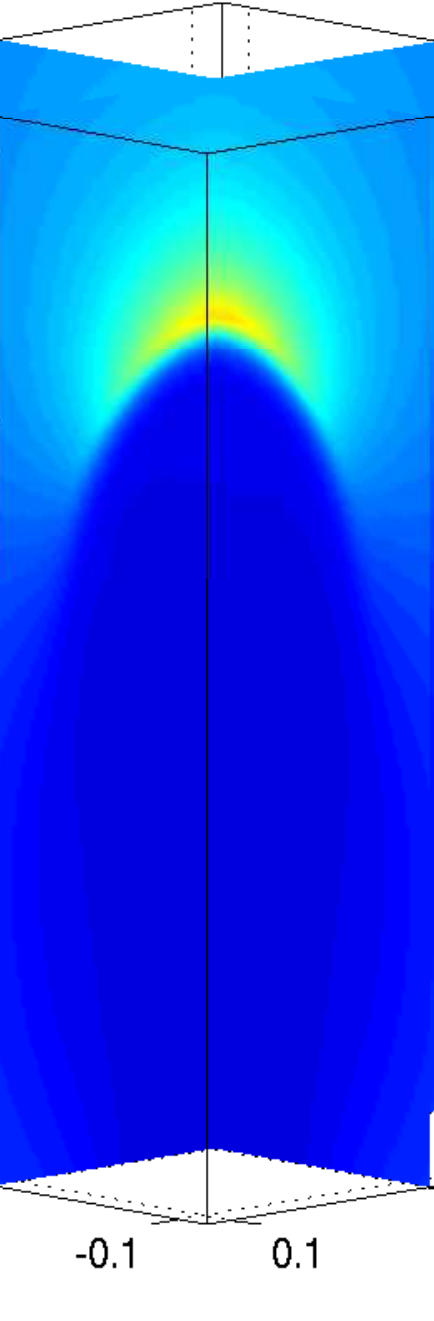}
\includegraphics[width=.12\textwidth,viewport=115 20 240 400, clip]{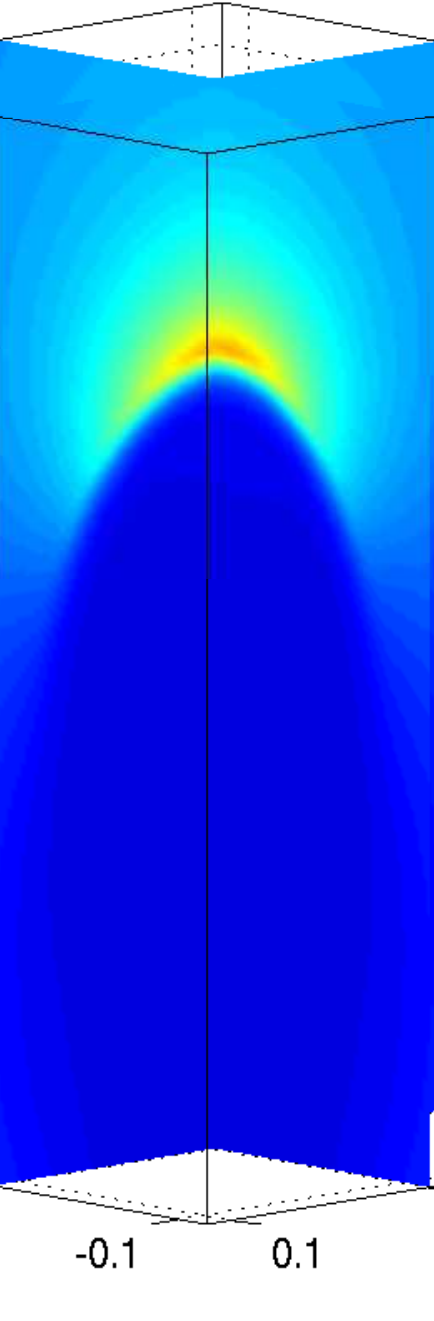}
\includegraphics[width=.12\textwidth,viewport=115 20 240 400, clip]{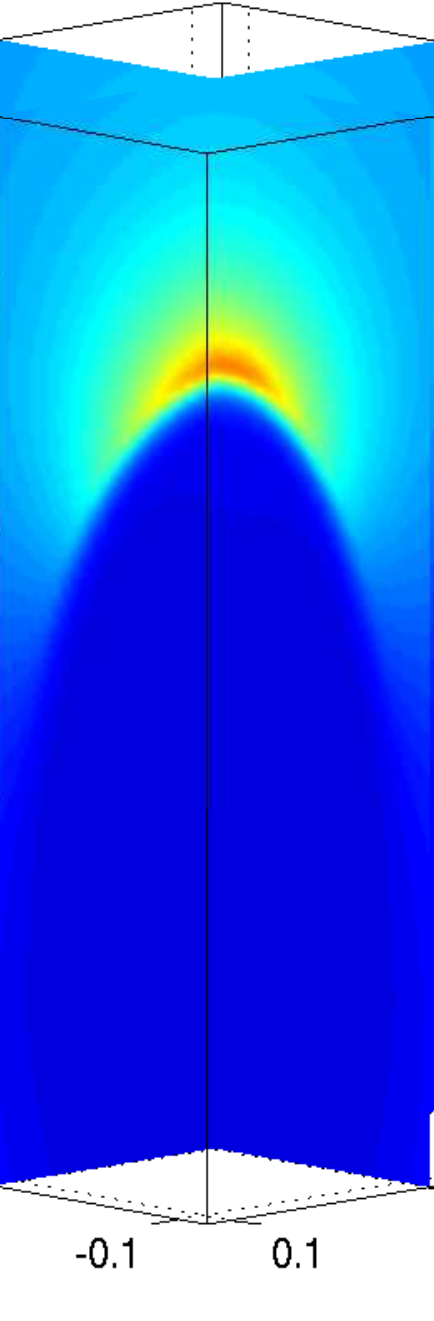}
\includegraphics[width=.12\textwidth,viewport=115 20 240 400, clip]{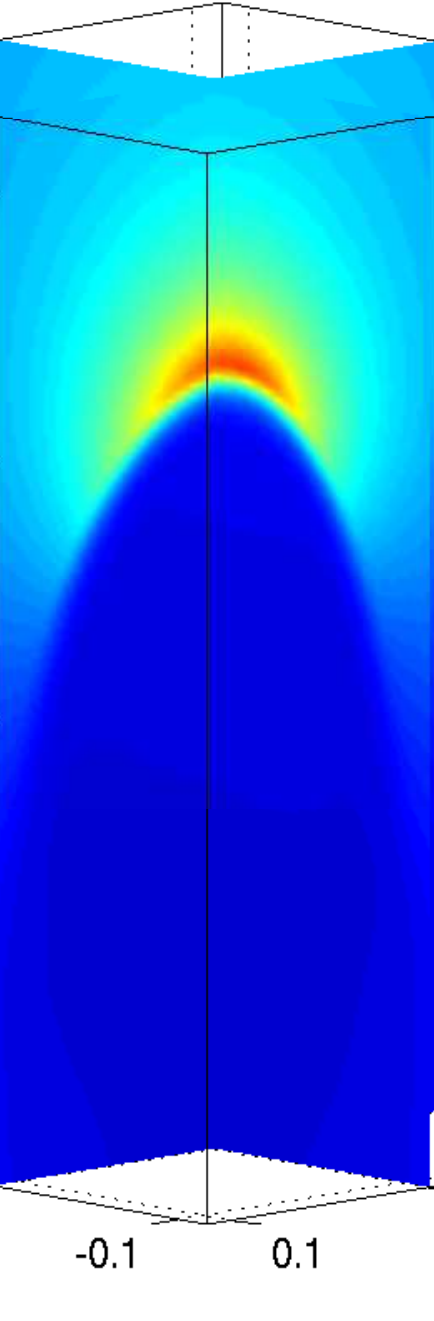}
\includegraphics[width=.12\textwidth,viewport=115 20 240 400, clip]{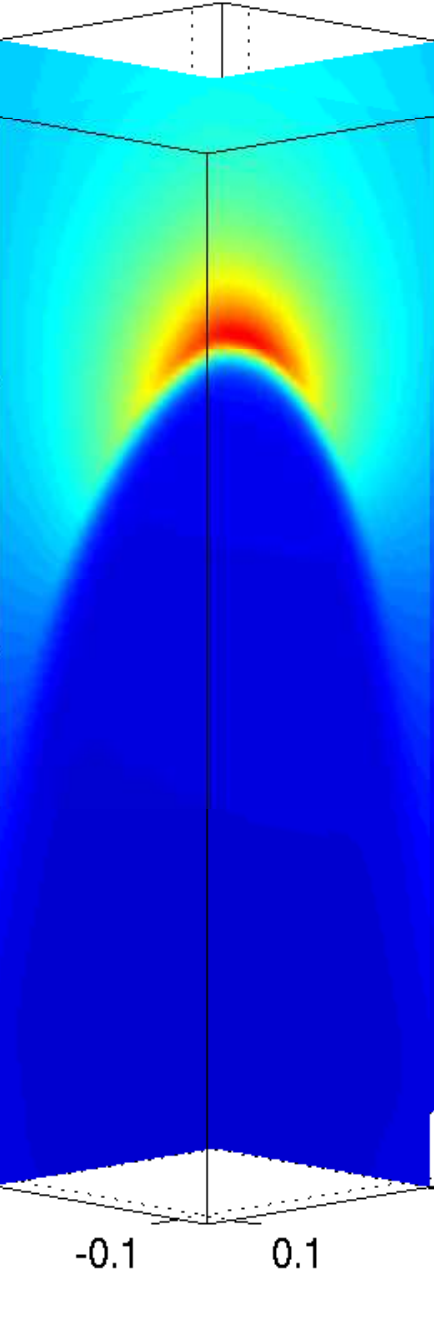}
\includegraphics[width=.04\textwidth,viewport=175 20 300 400, clip]{figures_pdf/fig1_colorbar.pdf} %White space
\\
\includegraphics[width=.12\textwidth,viewport=115 20 240 400, clip]{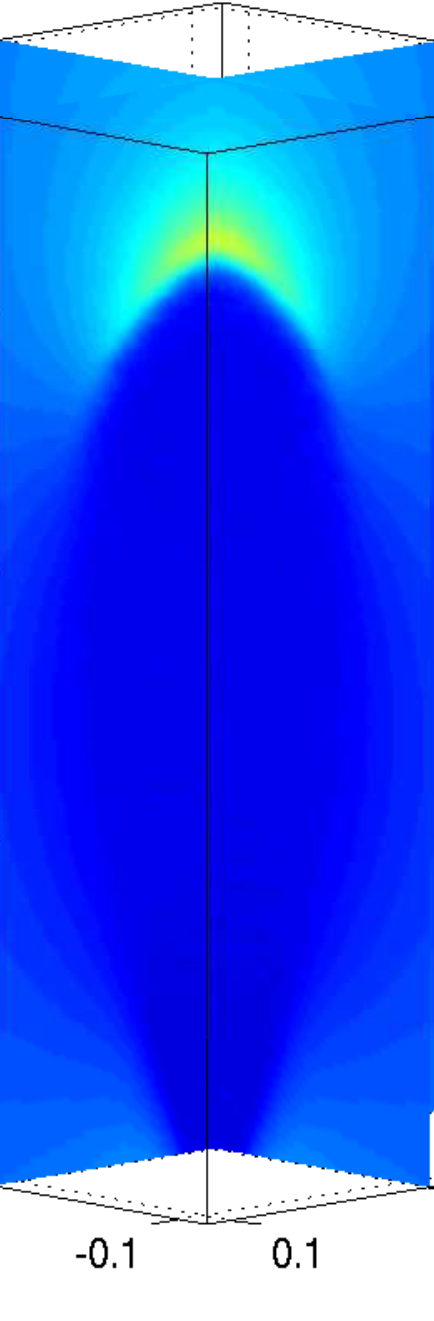}
\includegraphics[width=.12\textwidth,viewport=115 20 240 400, clip]{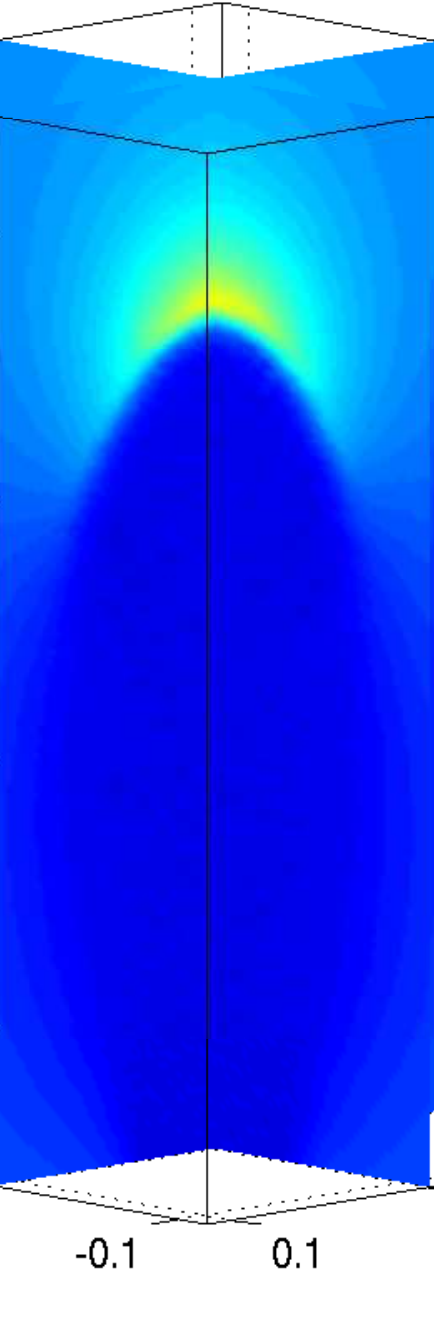}
\includegraphics[width=.12\textwidth,viewport=115 20 240 400, clip]{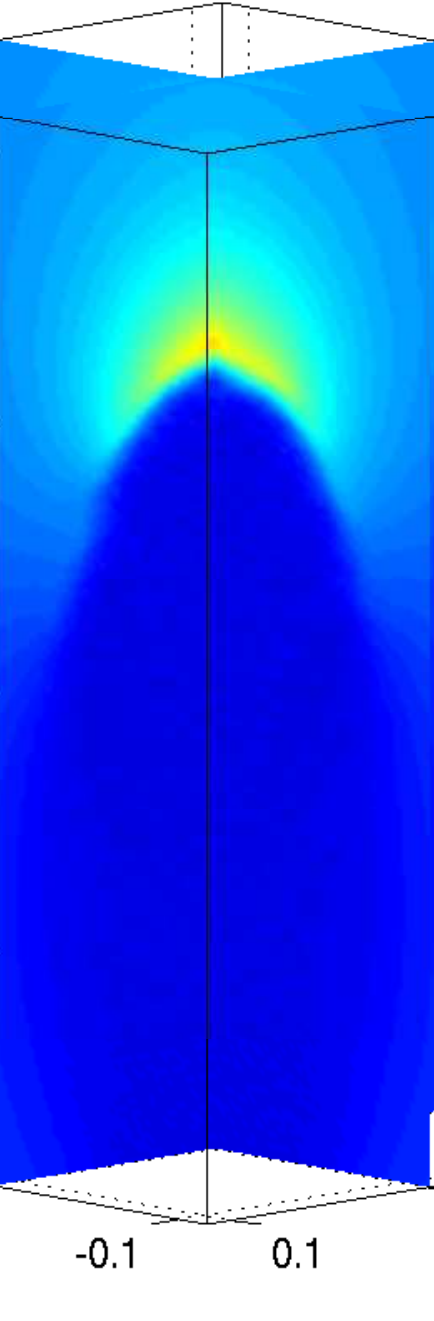}
\includegraphics[width=.12\textwidth,viewport=115 20 240 400, clip]{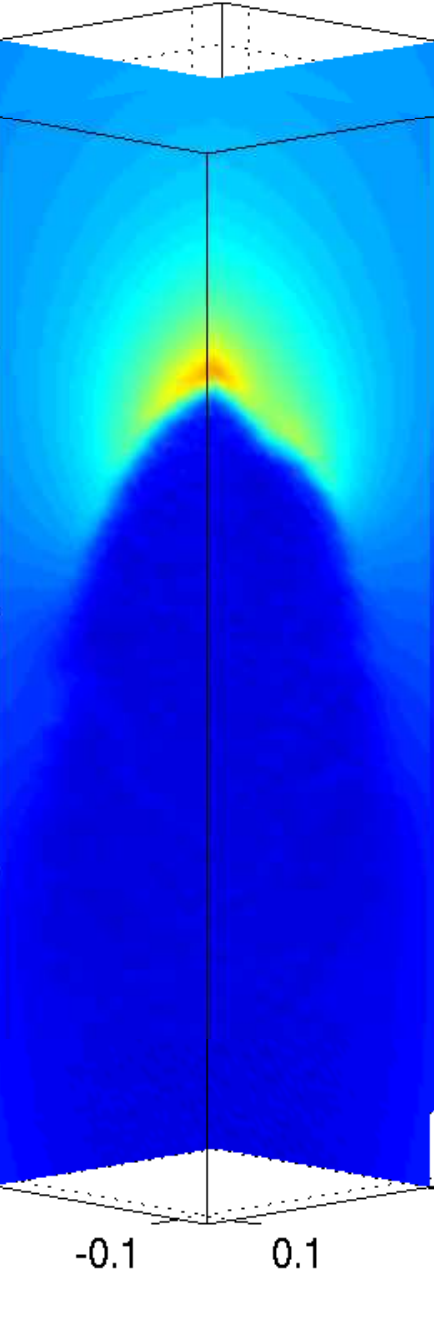}
\includegraphics[width=.12\textwidth,viewport=115 20 240 400, clip]{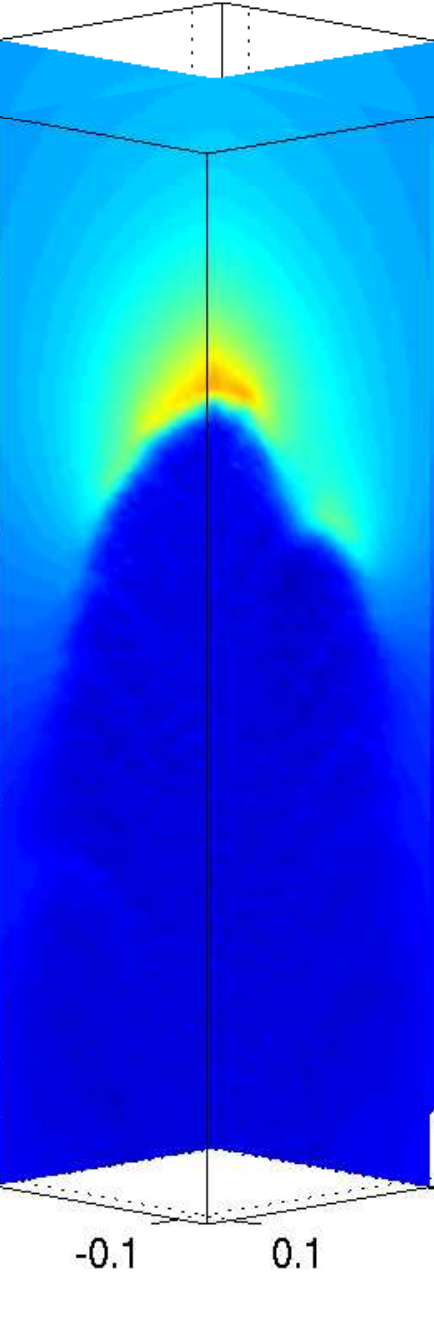}
\includegraphics[width=.12\textwidth,viewport=115 20 240 400, clip]{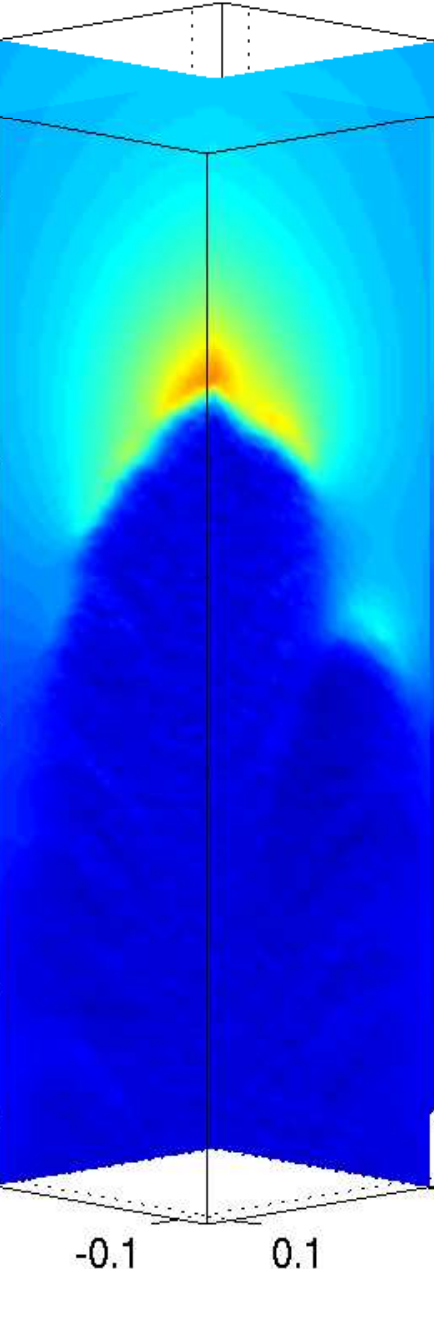}
\includegraphics[width=.12\textwidth,viewport=115 20 240 400, clip]{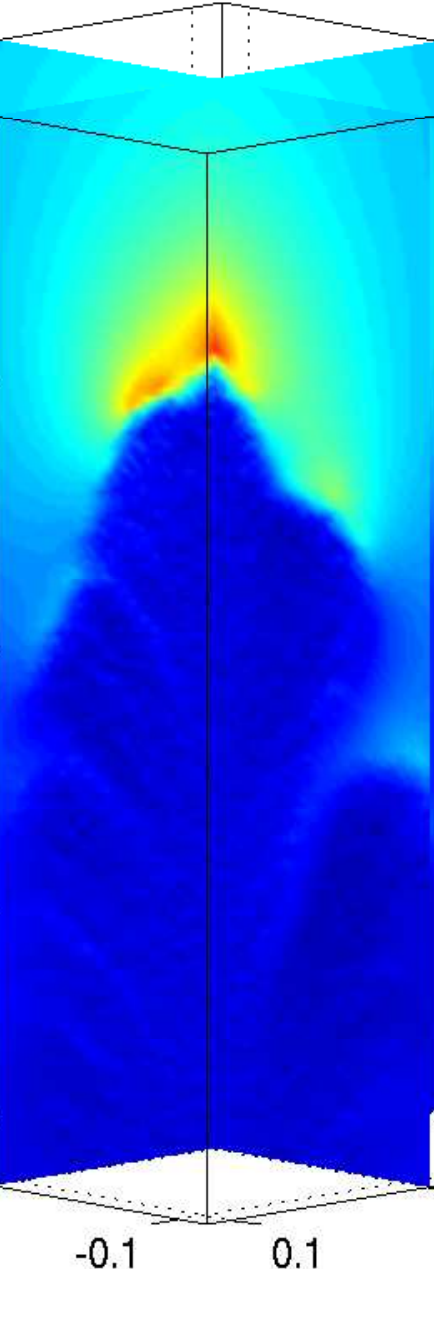}
\includegraphics[width=.04\textwidth,viewport=175 20 300 400, clip]{figures_pdf/fig1_colorbar.pdf} %White space
\\
\includegraphics[width=.12\textwidth,viewport=115 20 240 400, clip]{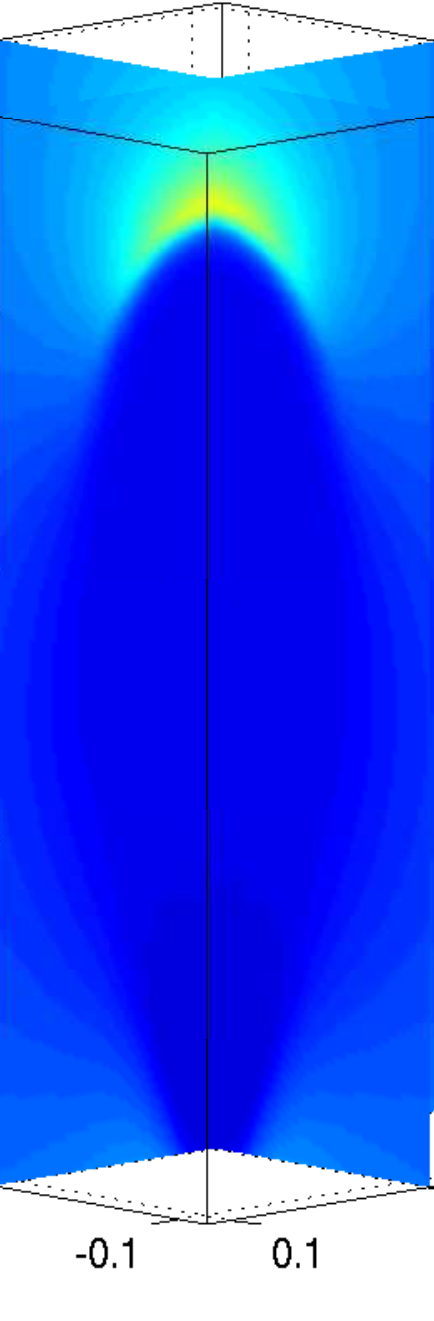}
\includegraphics[width=.12\textwidth,viewport=115 20 240 400, clip]{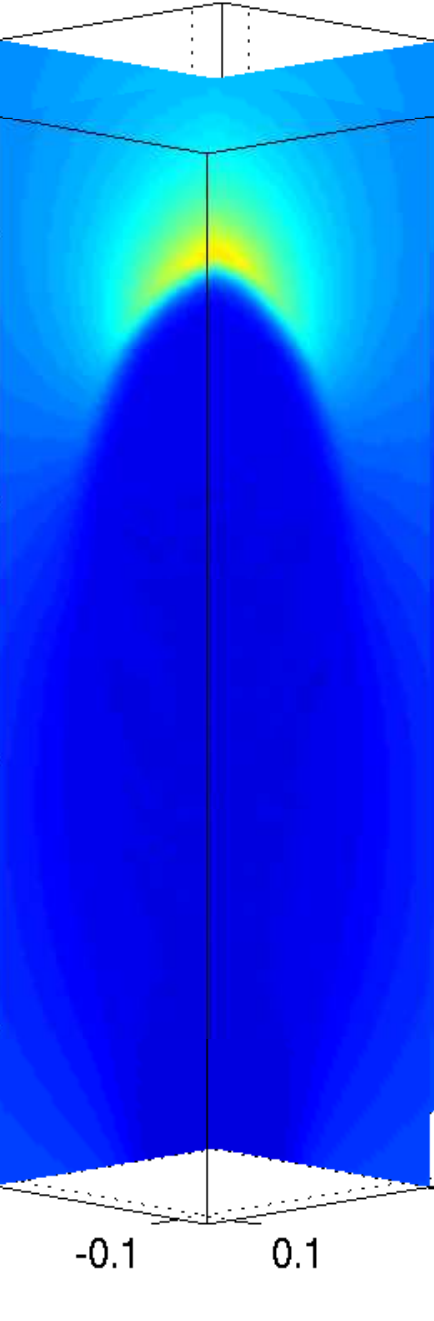}
\includegraphics[width=.12\textwidth,viewport=115 20 240 400, clip]{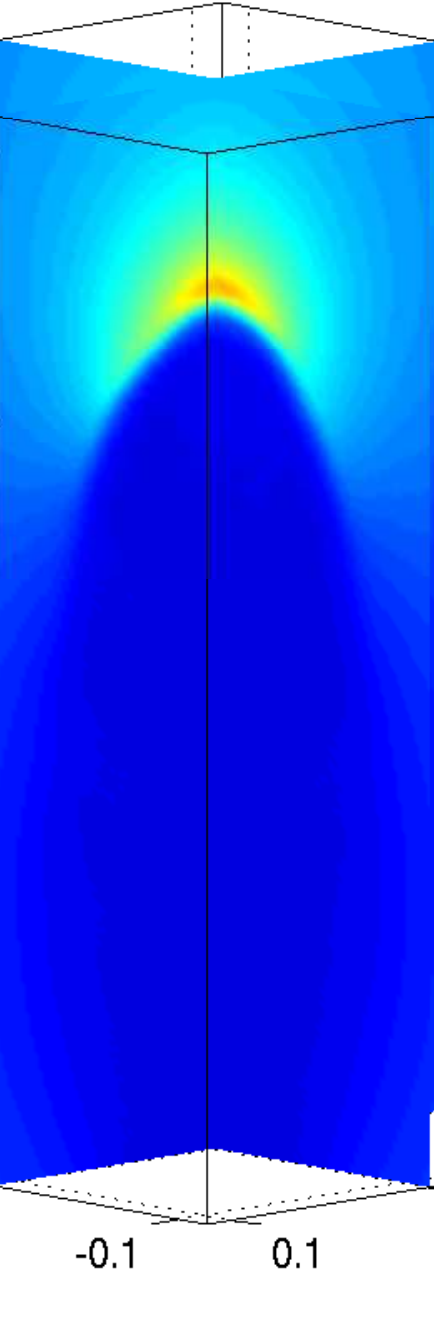}
\includegraphics[width=.12\textwidth,viewport=115 20 240 400, clip]{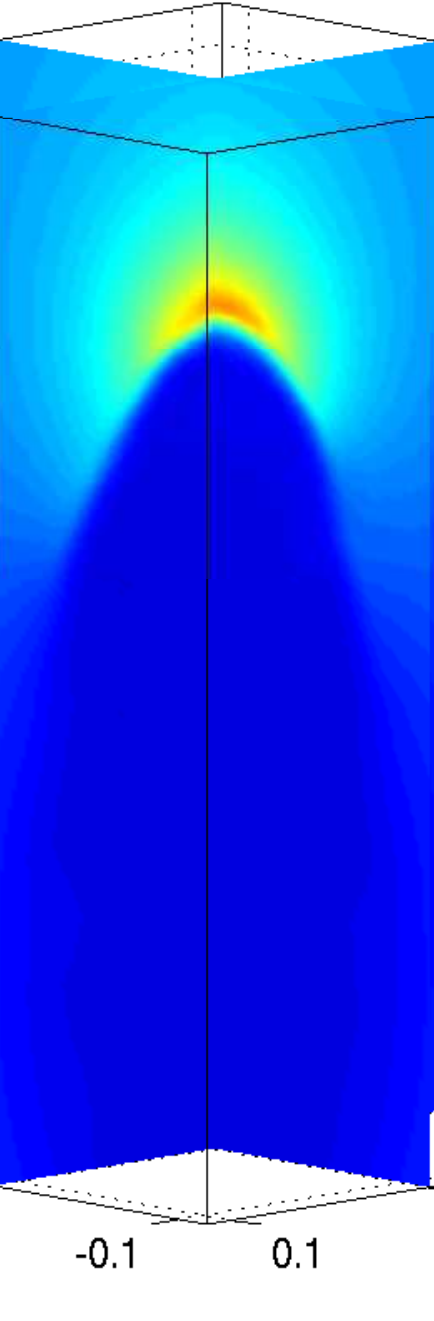}
\includegraphics[width=.12\textwidth,viewport=115 20 240 400, clip]{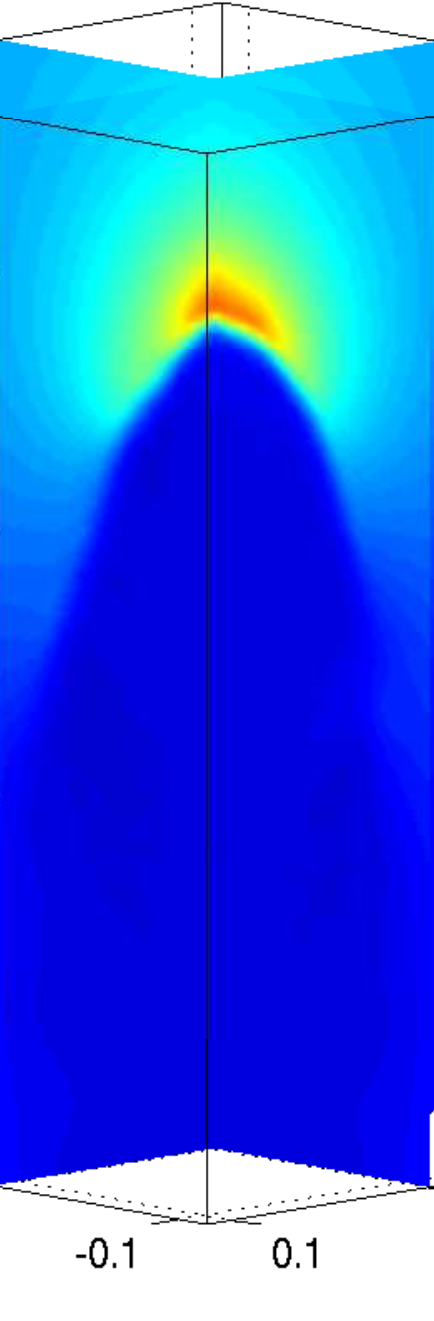}
\includegraphics[width=.12\textwidth,viewport=115 20 240 400, clip]{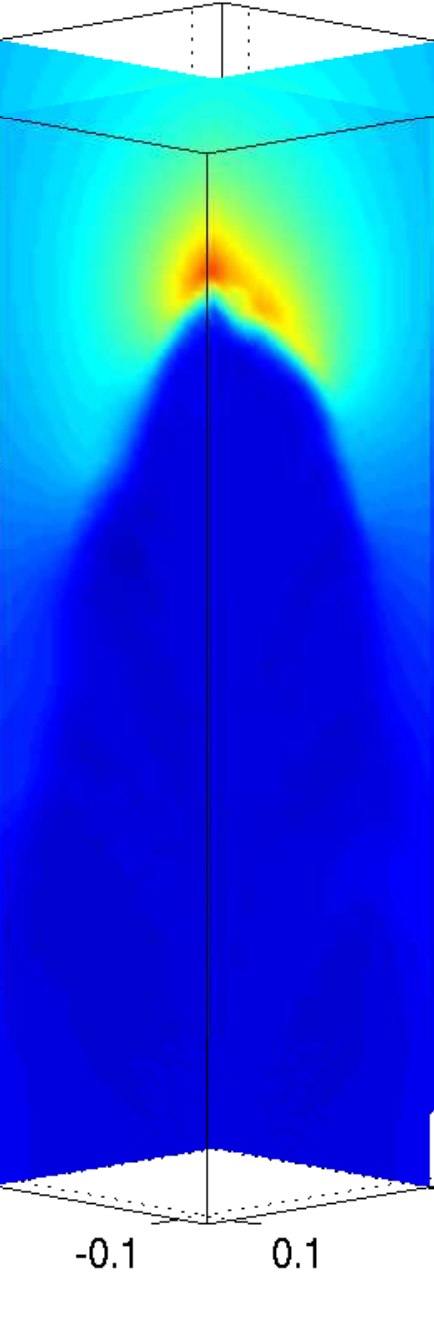}
\includegraphics[width=.12\textwidth,viewport=115 20 240 400, clip]{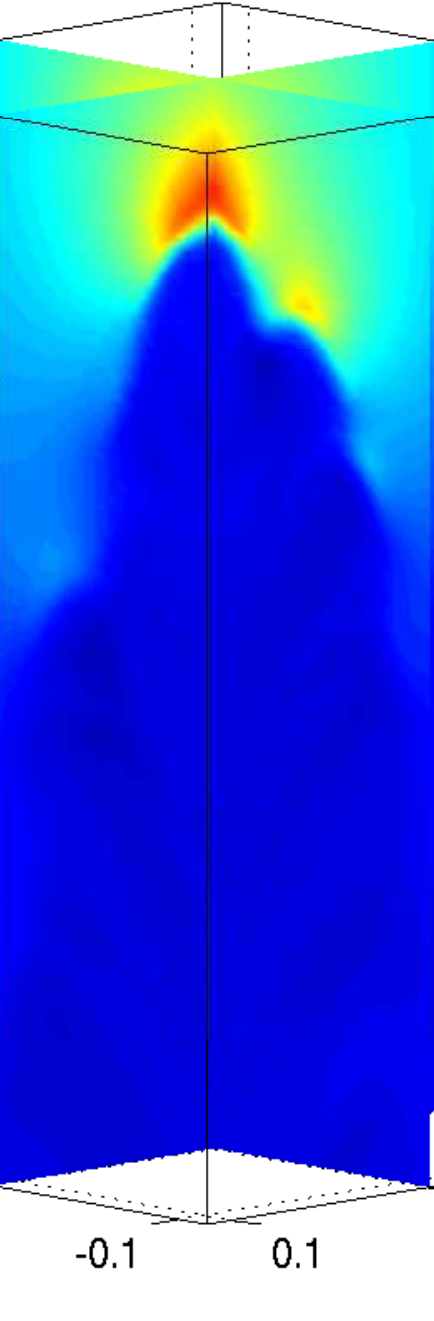}
\includegraphics[width=0.042\textwidth,viewport=314 0 364 432, clip]{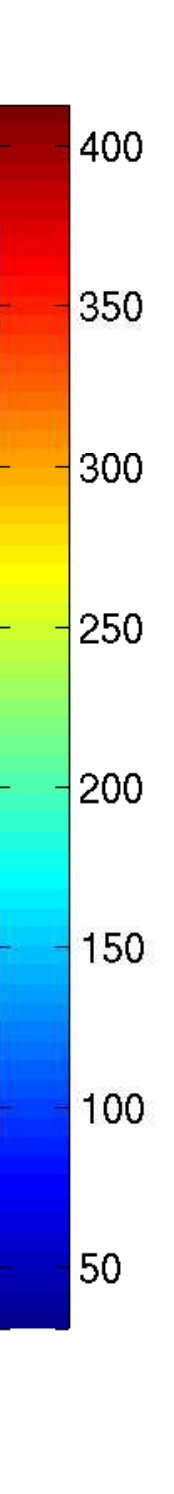}
\caption{The electric field in the vertical direction in the same models, at the same time steps and with the same spatial zoom as the electron densities in figure~\ref{fig:elecdenszoom} and the charge densities in figure~\ref{fig:chargedens}. The colours indicate electric fields from 0 (blue) to $450$~kV/cm (red).
}
\label{fig:efieldzoom}
\end{figure}

% FIG. 7
\begin{figure}
\centering
\includegraphics[width=0.6\textwidth]{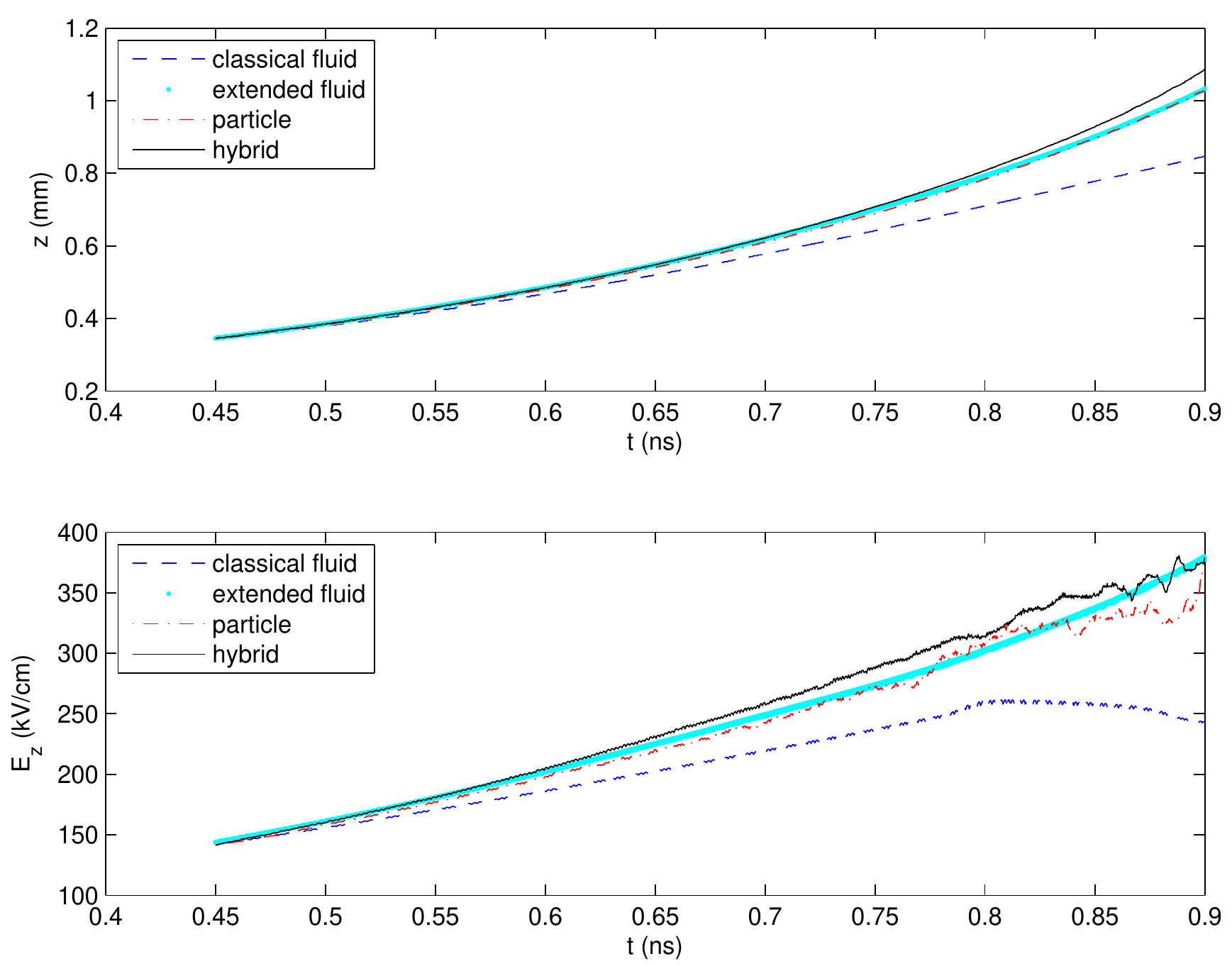}
\label{subfig:mm_z_Ez}
\caption{Upper panel: Front position as a function of time for the four models. The front position is here defined as the position of the maximum of the electric field on the axis within the ionization front; this maximum of the electric field as a function of time is plotted in the lower panel, also for the four models. }
\label{fig:frontposition}
\end{figure}

% FIG. 8
\begin{figure}
\centering
\includegraphics[width=0.6\textwidth]{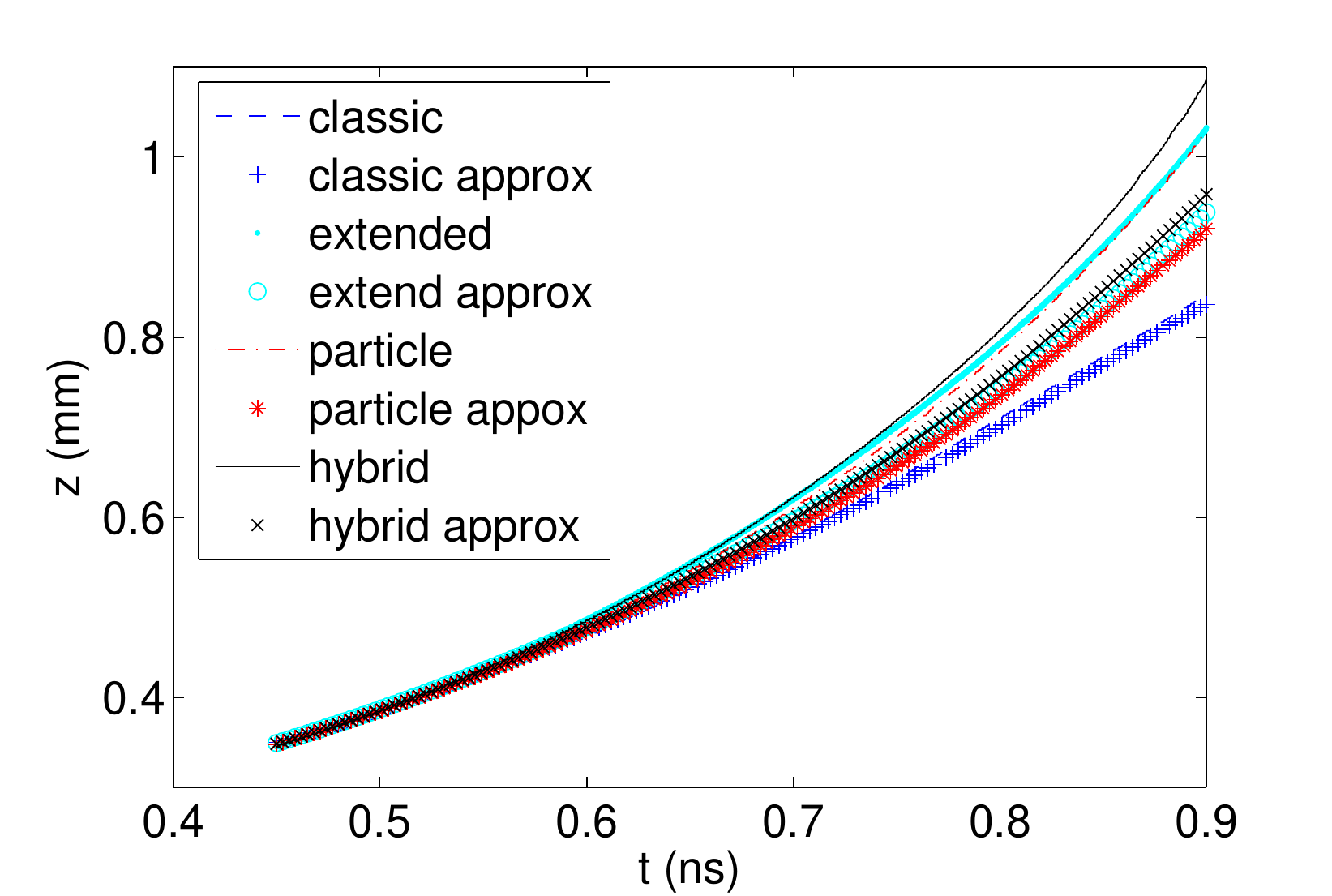}
\caption{Position of the streamer front as a function of time for the four models. Dashed, dotted, dashed-dotted and solid lines indicate the position determined from the simulations, as plotted previously in figure~\ref{fig:frontposition}a. The extended symbols (crosses and circles) indicate the positions for the four models determined through Eq.~(\ref{eq1}) where the maximal electric fields from figure~\ref{fig:frontposition}b were inserted and flux coefficients for $\mu_e(E)$, $D_e(E)$ and $\alpha(E)$ were used.
}
\label{fig:headposition}
\end{figure}

\end{document}